\documentclass[12pt,a4paper,JHEP3]{article}

\pdfoutput=1

\usepackage{jheppub}
\usepackage{graphicx}
\usepackage[dvipsnames]{xcolor}
\usepackage{dcolumn}
\usepackage{bm}
\usepackage{amstext}
\usepackage{amssymb}
\usepackage{subcaption}
\usepackage{amsmath}
\usepackage{caption}
\usepackage{ragged2e}
\usepackage{graphicx}
\usepackage{color}
\usepackage{bbold}
\usepackage{float}
\usepackage{delimset} 
\usepackage[colorlinks=true]{hyperref}
\usepackage{orcidlink}
\usetikzlibrary{
    math,
    decorations.pathmorphing,
    decorations.markings,
    positioning,
    patterns.meta,
    arrows.meta,
    shapes,
    backgrounds,
    calc
}
\usepackage{tikz-feynman}
\usepackage{extarrows}
\allowdisplaybreaks

\pgfdeclarelayer{foreground}
\pgfsetlayers{background,main,foreground}

\tikzset{ 
	graviton/.style={line width=.8pt, -latex,decorate, decoration={snake, segment length=4pt,amplitude=1.8pt, pre length=.1cm, post length=.25cm}},
	worldline/.style={gray, line width=1pt},
	worldlineBold/.style={black, line width=.6pt},
        background/.style={black,dotted,line width=1pt},
	zUndirected/.style={line width=1pt},
	zParticle/.style={line width=1pt,postaction={decorate},decoration={markings,mark=at position .6 with {\arrow[#1]{latex}}}},
	zParticleF/.style={line width=1pt,postaction={decorate}},
	cscalar/.style={line width=1pt,postaction={decorate},decoration={markings,mark=at position .6 with {\arrow[#1]{latex}}}},
	cscalar2/.style={line width=1pt,postaction={decorate},decoration={markings,mark=at position .8 with {\arrow[#1]{latex}}}},
	photon/.style={line width =.8pt, decorate, decoration={snake, segment length=3pt, amplitude=1.8pt,  pre length=.1cm, post length=.1cm}},
	 mid arrow/.style={postaction={decorate,decoration={
        markings,
        mark=at position .5 with {\arrow[#1]{latex}}}}} ,
        worlddot/.style={dotted, line width=.8pt},
	worlddot2/.style={dotted, line width=1pt},
  snake/.style={
      line width =.8pt,
      decorate,
      decoration={
        snake,
        segment length=5pt,
        amplitude=1.4pt,
        pre length=.0cm,
        post length=.0cm}
  },
  snakeDirected/.style={
    line width =.8pt,
    decorate,
    decoration={
      snake,
      segment length=5pt,
      amplitude=1.4pt,
      pre length=.0cm,
      post length=.1cm},
      postaction={decorate,decoration={markings,mark=at position 1 with {\arrow[xshift=4pt]{Triangle[length=6pt,width=5pt]}}}}},
  shadow/.style={
    draw=white,
    line width=7pt,
    shorten <=7pt,
    shorten >=7pt
  },
  dottedline/.style={
    dotted, line width=1.15
  },
    zParticle/.style={line width=1pt,postaction={decorate},decoration={markings,mark=at position .5 with {\arrow[xshift=3.5pt]{Triangle[length=6pt,width=5pt]}{Latex[length=5pt,width=4pt]}}}},
    loopContour/.style={
      line width=.8pt, draw=black, rounded corners=3pt,
      postaction={decorate},
      decoration={markings, mark=at position #1 with {\arrow{Stealth[length=4.5pt,width=3.5pt]}}}
  },
  loopContour/.default=.1,
  vertexCross/.style   ={cross out, draw, line width=1pt, minimum size=.16cm, inner sep=0pt},
  vertexPlus/.style    ={vertexCross, rotate=45},
  vertexOpen/.style    ={circle, draw, fill=white, line width=.8pt, minimum size=.14cm, inner sep=0pt},
  vertexSquare/.style  ={rectangle, draw, fill=black, minimum size=.11cm, inner sep=0pt},
  vertexDiamond/.style ={diamond, draw, fill=black, minimum size=.15cm, inner sep=0pt},
  vertexStar/.style    ={star, star points=5, star point ratio=2.3, draw, fill=black,
                         minimum size=.2cm, inner sep=0pt},
  vertexOTimes/.style  ={circle, draw, fill=white, line width=.8pt, minimum size=.2cm, inner sep=0pt,
                         append after command={
                           (\tikzlastnode.north west) edge[line width=.8pt] (\tikzlastnode.south east)
                           (\tikzlastnode.north east) edge[line width=.8pt] (\tikzlastnode.south west)}},
  markedVertex/.style={vertexSquare}
}

\newcommand{\drawMarkedVertex}[2][]{%
  \begin{pgfonlayer}{foreground}
    \node[markedVertex,#1] at (#2) {};
  \end{pgfonlayer}
}
\newcommand{\drawStaticBlob}[1]{%
  \draw[line width=.8pt,fill=white] (#1) circle (.14cm);%
  \draw[pattern=north east lines] (#1) circle (.14cm);}
\newcommand{\drawPerturbativeVertex}[1]{
  \begin{pgfonlayer}{foreground}
    \draw [fill=black, thick] (#1) circle (.045);
  \end{pgfonlayer}
}
\newcommand{\drawStaticPerturbativeVertex}[1]{
  \begin{pgfonlayer}{foreground}
    \draw [fill=Cerulean, thick,draw=Cerulean] (#1) circle (.045);
  \end{pgfonlayer}
}
\newcommand{\drawOnePointFunction}[1]{
    \begin{pgfonlayer}{foreground}
        \draw[line width=.8pt,fill=white] (#1) circle (.16cm);
        \draw[pattern=north east lines] (#1) circle (.16cm);
    \end{pgfonlayer}
}

\newcommand{\drawCurvedDottedLine}[5][1.5]{
    \draw [dottedline] (#2) to[in=#5,out=#4,looseness=#1] (#3) ;
}
\newcommand{\drawGravitonLine}[2]{
    \draw [snake] (#1) -- (#2) ;
}
\newcommand{\drawStaticGravitonLine}[2]{
    \draw [snake,Cerulean] (#1) -- (#2) ;
}

\newcommand{\drawDirectedGravitonLine}[2]{
    \draw [snakeDirected] (#1) -- (#2) ;
}
\newcommand{\drawDirectedStaticGravitonLine}[2]{
    \draw [snakeDirected,Cerulean] (#1) -- (#2) ;
}
\newcommand{\drawBHLineR}[2]{
    \draw [solid,thick] (#1) -- ($(#1)+(#2,0)$) ;
}
\newcommand{\drawStaticBHLineR}[2]{
    \draw [solid,thick,Cerulean] (#1) -- ($(#1)+(#2,0)$) ;
}
\newcommand{\drawDirectedBHLineR}[2]{
    \draw [zParticle] (#1) -- ($(#1)+(#2,0)$) ;
}
\newcommand{\drawStaticDirectedBHLineR}[2]{
    \draw [zParticle,Cerulean] (#1) -- ($(#1)+(#2,0)$) ;
}
\newcommand{\drawBHLine}[2]{
    \draw [solid,thick] (#1) -- (#2) ;
}

\newcommand{\drawStaticBHLine}[2]{
    \draw [solid,thick,Cerulean] (#1) -- (#2) ;
}
\newcommand{\drawStaticDirectedBHLine}[2]{
    \draw [zParticle,Cerulean] (#1) -- (#2) ;
}
\newcommand{\drawCurvedBHLine}[4]{
    \draw [solid,thick] (#1) to[in=#4,out=#3,looseness=1] (#2) ;
}
\newcommand{\drawStaticCurvedBHLine}[4]{
    \draw [solid,thick,Cerulean] (#1) to[in=#4,out=#3,looseness=1.2] (#2) ;
}
\newcommand{\drawDirectedBHLine}[2]{
    \draw [zParticle] (#1) -- (#2) ;
}
\newcommand{\drawDottedLineR}[2]{
    \draw [dottedline] (#1) -- ($(#1)+(#2,0)$) ;
}
\newcommand{\drawDottedLine}[2]{
    \draw [dottedline] (#1) -- (#2) ;
}
\newcommand{\drawActiveGravitonLine}[2]{
    \draw [snake,Plum,line width=1.2pt] (#1) -- (#2) ;
}
\newcommand{\drawRadiativeGravitonLine}[2]{
    \draw [snake,WildStrawberry,line width=1.2pt] (#1) -- (#2) ;
}
\newcommand{\drawDirectedActiveGravitonLine}[2]{
    \draw [snakeDirected,Plum,line width=1.2pt] (#1) -- (#2) ;
}
\newcommand{\drawDirectedRadiativeGravitonLine}[2]{
    \draw [snakeDirected,WildStrawberry,line width=1.2pt] (#1) -- (#2) ;
}
\newcommand{\drawShadowLine}[2]{
    \draw [shadow] (#1) -- (#2) ;
}
\newcommand{\drawCurvedGravitonLine}[5][1.5]{
    \draw [snake] (#2) to[in=#5,out=#4,looseness=#1] (#3) ;
}
\newcommand{\drawStaticCurvedGravitonLine}[5][1.5]{
    \draw [snake,Cerulean] (#2) to[in=#5,out=#4,looseness=#1] (#3) ;
}

\newcommand{\drawDirectedStaticCurvedGravitonLine}[5][1.5]{
    \draw [snakeDirected,Cerulean] (#2) to[in=#5,out=#4,looseness=#1] (#3) ;
}

\newcommand{\drawRadiativeCurvedGravitonLine}[4]{
    \draw [snake,WildStrawberry,line width=1pt] (#1) to[in=#4,out=#3,looseness=1.5] (#2) ;
}
\newcommand{\drawDirectedRadiativeCurvedGravitonLine}[4]{
    \draw [snakeDirected,WildStrawberry,line width=1pt] (#1) to[in=#4,out=#3,looseness=1.5] (#2) ;
}

\DeclareFontFamily{OT1}{pzc}{}
\DeclareFontShape{OT1}{pzc}{m}{it}{<-> s * [1.350] pzcmi7t}{}
\DeclareMathAlphabet{\mathpzc}{OT1}{pzc}{m}{it}

\def\sym{n'}

\def\dd{\delta\!\!\!{}^-\!}

\def\d{\mathrm{d}}
\def\i{\mathrm{i}}
\def\iZero{{\i 0^+}}
\def\eps{\epsilon}
\def\bell{{\bar\ell}}
\def\Ican{\vec I_{\rm c}}
\def\Jcan{\vec J_{\rm c}}

\def\nn{\nonumber}

\newcommand{\overbar}[1]{\mkern 1.5mu\overline{\mkern-1.5mu#1\mkern-1.5mu}\mkern 1.5mu}
\def\oneSFBar{\ensuremath{\overbar{\mathrm{1SF}}}}

\newcommand{\drawImpulseBlob}[2]{%
  \draw[line width=.8pt,fill=white] (#1) circle (.2cm);%
  \node[inner sep=0pt] at (#1) {\scalebox{.48}{\scriptsize #2PM}};}
  
\def\eqn#1{eq.~\eqref{#1}}
\def\eqns#1#2{eqs.~\eqref{#1} and~\eqref{#2}}

\def\fig#1{fig.~{\ref{#1}}}

\def\tab#1{table~{\ref{#1}}}
\def\tabs#1#2{tables~{\ref{#1}} and~{\ref{#2}}}

\def\app#1{appendix~{\ref{#1}}}
\newcommand{\secref}[1]{sec.~{\ref{#1}}}
\DeclareMathOperator{\arccosh}{arccosh}

\newcommand{\mn}{{\mu\nu}}
\newcommand{\ab}{{\alpha\beta}}
\newcommand{\rs}{{\rho\sigma}}
\newcommand{\pat}{\partial}

\newcommand*{\vct}[1]{\boldsymbol{#1}}

\usepackage{lmodern}

\begin{document}

\begin{flushright}
\begingroup\footnotesize\ttfamily
	HU-EP-26/32-RTG
\endgroup
\end{flushright}

\vspace{15mm}

\begin{center}
{\fontsize{19.6pt}{24.89pt}\selectfont\bfseries 
    Mechanical angular momentum loss at $\mathcal{O}(G^4)$ from static and radiative gravitons
\par}

\vspace{15mm}

\begingroup\scshape\large 
  Lara Bohnenblust${}^{1,2}$,
  Carl Jordan Eriksen${}^{1}$,
  Gustav Uhre Jakobsen${}^{1}$, 
  Kathrin Stoldt${}^{1}$, and 
  Andrii Yanovets${}^{1}$
\endgroup
\vspace{5mm}
					
\textit{${}^{1}$Institut f\"ur Physik, Humboldt-Universit\"at zu Berlin, 10099 Berlin, Germany} \\[0.25cm]
\textit{${}^{2}$Max-Planck-Institut f\"ur Gravitationsphysik
(Albert-Einstein-Institut), 14476 Potsdam, Germany } \\[0.25cm]

\bigskip
  
\texttt{\small\{lara.bohnenblust, carl.jordan.eriksen, gustav.uhre.jakobsen, kathrin.stoldt, andrii.yanovets\}@physik.hu-berlin.de}

\vspace{5mm}

\textbf{Abstract}\vspace{5mm}\par

\begin{minipage}{14.7cm}
{
    We determine the loss of mechanical angular momentum $L^{\mu}_{\rm loss}$ of two gravitationally scattering bodies at $\mathcal{O}(G^4)$ using worldline quantum field theory (WQFT).
    The gravitational memory effect, a phenomenon of the early- and late-time behaviour of the gravitational field, leads to static contributions to $L^{\mu}_{\rm loss}$.
    We illuminate how these arise: the linear memory contributes through so-called static correlators, while the nonlinear memory vanishes at this order due to symmetry considerations.
    The computation of the remaining, dynamical part resembles that of the $\mathcal{O}(G^4)$ impulse.
    We generate its integrand using a reformulation of the WQFT Feynman rules that conserves four-momentum at every vertex and plainly exhibits the classical loops.
    The three-loop integrals are reduced to master integrals, whose differential equations and boundary matching in the post-Newtonian limit we describe in some detail.
    In contrast to conservative effects at this order, only polylogarithms appear.
    While individual contributions to the dynamical part are divergent, these cancel in $L^{\mu}_{\rm loss}$.
    The quadratic-in-radiation part of our result reproduces earlier results, while the complete result agrees with existing post-Newtonian results.
}
\end{minipage}\par

\end{center}
\setcounter{page}{0}
\thispagestyle{empty}
\newpage
 
\tableofcontents
\clearpage\noindent
\newpage

\section{Introduction}
Accurate waveform models are indispensable for the analysis of gravitational wave data~\cite{LIGOScientific:2016aoc, LIGOScientific:2017vwq, KAGRA:2021vkt, LIGOScientific:2025slb, LISA:2017pwj, Punturo:2010zz, Ballmer:2022uxx,ET:2025xjr}, and their construction requires a detailed theoretical understanding of the dynamics of compact binaries. This need has driven intense research on the general-relativistic two-body problem, in which a variety of numerical~\cite{Pretorius:2005gq,Boyle:2019kee,Scheel:2025jct} and analytical methods, such as the post-Newtonian (PN)~\cite{Blanchet:2013haa,Porto:2016pyg,Levi:2018nxp,Porto:2026fsd,Brunello:2025gpf,Brunello:2026anu,Bini:2026dvn}, post-Minkowskian (PM)~\cite{Kosower:2022yvp,Bjerrum-Bohr:2022blt,Buonanno:2022pgc,DiVecchia:2023frv,Jakobsen:2023oow} and self-force (SF)~\cite{Mino:1996nk,Poisson:2011nh,Barack:2018yvs,Gralla:2021qaf,Warburton:2025ymy} expansions, are combined to describe the entire evolution, from the weak-field inspiral to the strong-field merger.

While the bound system is of direct significance to the observed gravitational wave signals, it is also worthwhile to analyse the unbound, scattering regime for multiple reasons.
First, local-in-time contributions to the dynamics can be mapped analytically from unbound to bound orbits~\cite{Kalin:2019rwq,Kalin:2019inp,Saketh:2021sri,Cho:2021arx,Gonzo:2023goe,Dlapa:2024cje,Dlapa:2025biy}.
Second, resummed models such as the effective-one-body approach~\cite{Buonanno:1998gg,Buonanno:2000ef} can be calibrated against scattering observables~\cite{Damour:2022ybd,Rettegno:2023ghr,Buonanno:2024vkx,Buonanno:2024byg,Damour:2025uka,Long:2025nmj,Clark:2025kvu,Barack:2026izc,Gamboa:2026ybr,Gamboa:2026jht,Albanesi:2025txj,Albanesi:2024xus}.
Third, for sufficiently large initial angular momentum, a scattering encounter remains in the weak-field regime throughout its evolution, both for low and high velocities, and is therefore particularly amenable to a perturbative treatment~\cite{Kovacs:1978eu,Westpfahl:1979gu,Bel:1981be,Damour:2017zjx}.

A convenient framework to describe compact objects at large separation is effective field theory (EFT). It treats black holes and neutron stars as point-like particles, specifically, as worldline fields, and is therefore named worldline effective field theory~\cite{Goldberger:2004jt,Levi:2015msa,Porto:2016pyg,Levi:2018nxp,Kalin:2020mvi}. This powerful approach formalises the inclusion of intrinsic spin~\cite{Levi:2015msa,Liu:2021zxr,Jakobsen:2021zvh,Haddad:2024ebn} and tidal deformability~\cite{Bini:2012gu,Kalin:2020lmz,Cheung:2020sdj,Haddad:2020que,Bern:2020uwk,Mandal:2023hqa,Jakobsen:2023pvx} in the point-particle description through higher-dimensional worldline operators.
Ignoring spins, such analyses show that black holes and neutron stars first differ at 6PM order in two-body scattering observables~\cite{Correia:2026utp,Brunello:2026lzf,Kalin:2020lmz}.
In the following we will therefore mostly refer to the compact bodies as black holes (BHs).

One important advantage of studying the perturbative scattering processes in an EFT language is that quantum field theory (QFT) offers an advanced toolbox designed for exactly this kind of problem~\cite{Kosower:2022yvp,Bjerrum-Bohr:2022blt,Buonanno:2022pgc}.
Namely, in this weak-coupling setting, the gravitational field may be described by the spin-$2$ graviton field.
Following this analogy, it is possible to set up the classical PM expansion in terms of Feynman diagrams and loop integrals~\cite{Bjerrum-Bohr:2013bxa,Bjerrum-Bohr:2018xdl,Cheung:2018wkq,Kosower:2018adc,Bern:2019nnu,Bern:2019crd,Cheung:2020gyp,Luna:2017dtq,Goldberger:2004jt,Kalin:2020mvi,Mogull:2020sak}.
At this point, the advanced and well-developed loop integration techniques of QFT come into play and allow for great progress at a relatively low cost~\cite{Bjerrum-Bohr:2002gqz,Holstein:2004dn,Bjerrum-Bohr:2018xdl}.
These are techniques such as integration-by-parts identities~\cite{Laporta:2000dsw}, dimensional regularisation, and the method of differential equations~\cite{Gehrmann:1999as,Henn:2013pwa}.

The synthesis of ideas from EFT, QFT, and the classical PM expansion is efficiently formalised in worldline QFT (WQFT)~\cite{Mogull:2020sak,Jakobsen:2021smu,Jakobsen:2022psy,Jakobsen:2023oow,Haddad:2024ebn,Haddad:2025cmw,benThesis,Bohnenblust:2025gir,Gonzo:2026yha}.
This framework naturally employs the language of QFT but targets, at the same time, directly classical physics.
By quantising not only the gravitational field, but also the deflection of the worldlines, one obtains a systematic way of computing perturbative corrections order by order in $G$ through Feynman diagrams and integrals.
The theory's backbone is that classical observables are derived by considering solely one-point functions and including \emph{tree-level} diagrams only.
Nevertheless, loop integrals still appear within the tree-level diagrams.
These are often referred to as classical loops, and for each additional order in the PM expansion, an additional such loop must be calculated.

At initial and final time, the black holes do not interact gravitationally and essentially correspond to free states.
A free black hole described by a worldline (WL) $x^\mu_i(\tau)$ moves in a straight line, $\tau p^\mu_i /m_i + b_i^\mu$, parametrised by the mass $m_i$, four-momentum $p^\mu_i$ and \emph{worldline offset} $b_i^\mu$.
While the worldline offsets by themselves are at best ambiguous, they are required to parametrise the physically relevant angular momentum content of each BH and, of greater significance, their change.

The angular momentum loss actually belongs to a larger class of observables where final and initial state variables are compared.
The simplest example is the change in four-momentum, the impulse $\Delta p^\mu_i$, which is fully known up to 4PM order \cite{Bern:2021yeh,Dlapa:2022lmu,Dlapa:2023hsl,Jakobsen:2023ndj,Jakobsen:2023hig,Jakobsen:2023pvx,Damgaard:2023ttc,Bern:2019nnu,Herrmann:2021lqe}, and at linear order in the SF expansion at 5PM~\cite{Driesse:2024feo,Driesse:2024xad,Dlapa:2025biy}. The conservative contribution at 5PM is additionally known at quadratic order in the SF expansion~\cite{Driesse:2026qiz,Bern:2025wyd,Dlapa:2026oyq}.
The change in the angular momentum of each BH or, equivalently, the changes in the worldline offsets $\Delta b_i^\mu$, has been studied less.
Here, complete results are known only at 3PM order~\cite{Damour:2020tta,Gralla:2021qaf,Jakobsen:2021smu,Manohar:2022dea,DiVecchia:2022piu} with partial results at 4PM order~\cite{Heissenberg:2024umh,Heissenberg:2025ocy}. 
In the PN expansion, the angular momentum loss is fully known to 3PN order~\cite{Bini:2021gat,Bini:2022enm} and to 7PN order at 1SF~\cite{Geralico:2025rof}.

The aim of the present work is to complete the knowledge of the angular momentum loss at 4PM order. We build on ref.~\cite{Jakobsen:2026ayw} where two of this work's authors have laid out a method for computing the change in the worldline offsets within the WQFT formalism and carried out the calculation explicitly at the known 3PM order.
This method is largely analogous to the WQFT approach of computing the impulse; in this work, we apply the known three-loop toolbox to the changes in worldline offsets and thereby derive directly the 4PM loss of angular momentum.

However, the changes in worldline offsets come with subtleties absent from the impulse.
For one, the worldline offsets depend both on translations of the coordinate centre and of the worldline proper times.
Further, although a truly isolated BH does follow a straight line, this is not true for the interacting system in four dimensions, where the long-range nature of the gravitational force implies that the velocity directions of the worldlines pick up logarithmic drifts~\cite{Bini:2024hme,Bini:2022wrq,Mogull:2025cfn}.
The changes in the worldline offsets therefore have infrared divergences and must be regulated.
Only when combined into a translation-invariant quantity, namely the (spatial) angular momentum, can one expect these divergences to vanish.

The total \emph{mechanical} angular momentum takes its familiar special-relativistic form, $L^\mn=2\sum_ib^{[\mu}_i p^{\nu]}_i$, composed of the worldline offsets and the four-momenta.
In addition, the gravitational field carries angular momentum, and it is only the sum of the mechanical angular momentum and the field content that is conserved.
Therefore, the balance law dictates that any mechanical loss of angular momentum must imply an increase in the field content of angular momentum and vice versa.
Computing the field content of angular momentum, however, is nontrivial because it comes from two sources: the radiative flux of gravitons and a reconfiguration of the static field related to the wave memory~\cite{Damour:2020tta,Veneziano:2022zwh,DiVecchia:2022owy,Riva:2023xxm}.
This distribution of the field content of angular momentum among radiative and static reconfigurations is not unique and changes under BMS translations~\cite{Veneziano:2022zwh,Riva:2023xxm}.
However, in the ``intrinsic'' frame, any mechanical loss of angular momentum is exactly balanced by a corresponding flux~\cite{Bini:2022wrq, Riva:2023xxm}.

The presented computation of the angular momentum loss at 4PM order comes in two parts: a \emph{static} contribution due to an on-shell static graviton and a \emph{dynamic} contribution due to conventional potential and radiative gravitons.

The static contributions to the angular momentum loss originate from the early- and late-time behaviour of the gravitational field, captured by the gravitational memory effect.
Linear memory contributions to the angular momentum loss were systematised in ref.~\cite{Jakobsen:2026ayw} as coming from so-called \emph{static correlators}.
This approach is easily extended and applied to the present 4PM order.
More generally, such linear memory effects to the angular momentum loss are understood and known to all orders in perturbation theory, parametrised in terms of the impulse~\cite{DiVecchia:2022owy}.
In the present work, we show how equivalent results are derived from the WQFT framework.

At 4PM order, the changes in the worldline offsets could in principle receive contributions from nonlinear memory in addition to linear memory. In \secref{sec:nonlinearmemory}, we derive nonlinear memory within the WQFT framework, recovering its known expression~\cite{PhysRevLett.67.1486,PhysRevD.45.520,PhysRevD.44.R2945}, and show that its 4PM contribution cancels completely in the spatial angular momentum loss, in agreement with ref.~\cite{Heissenberg:2024umh}.

The derivation of the 4PM dynamic contribution follows quite closely the analogous derivation of the 4PM impulse.
Most importantly, the odd-in-velocity sector of the same set of three-loop integrals is required and we discuss the structure of these integrals in some detail.
One interesting subtlety, however, is the intermediate cancellation of infrared divergences: The 4PM dissipative change in the worldline offsets comes with a $1/\eps$ pole which cancels against a 1PM conservative change.
This implies that the $\mathcal{O}(G^3,\eps^1)$ part of the 3PM impulse is required in the intermediate steps.

The computation of the dynamic part of the changes in worldline offsets requires a slight generalisation of the usual integrand construction of the impulse.
In order to systematically deal with this, we discuss the integrand generation in WQFT in detail.
In this process we systematically construct the off-shell integrand by introducing an alternative labelling of WQFT Feynman diagrams.
Importantly, with this perspective, one manifests the classical loops and imposes four-momentum conservation at each vertex.
The Feynman rules then mimic conventional QFT vertex rules and make the translation between the two perspectives even clearer.
From this perspective, the impulse and worldline offsets are derived from two-point functions while the waveform is derived from a three-point function.

The structure of the paper is as follows.
In \secref{sec:WEFTWQFT}, we review WQFT and the method and ingredients for computing changes in the worldline offsets and the angular momentum loss.
In \secref{sec:memory}, we discuss the computation of static contributions to the angular momentum loss and how the gravitational memory appears in WQFT diagrams.
Then, in \secref{sec:integrand}, we discuss the derivation of the 4PM integrand for dynamic contributions to $|L|_{\rm loss}$ and four-momentum conserving WQFT Feynman rules.
Continuing in \secref{sec:integrals}, we discuss the evaluation of the required three-loop integrals.
Finally, in \secref{section:Results}, we present the results for the angular momentum loss $|L|_{\rm loss}$.
In \secref{sec:conclusions}, we end with our conclusions.

We work with the mostly-minus metric $\eta_\mn=\text{diag}(1,-1,-1,-1)$ and set the speed of light to unity $c=1$.
Further, we use $D=4-2\eps$ for the space-time dimension in dimensional regularisation.
Throughout, we use the shorthand
\begin{align}
    \int_k
    =
    \int\frac{\d^Dk}{(2\pi)^D}
    \, ,
    \qquad
    \dd(\omega)
    =
    2\pi\delta(\omega)
    \, .
\end{align}

\paragraph{Note:} The \texttt{arXiv} submission of this work was coordinated with A. Georgoudis, C. Heissenberg, G. Mogull, and R. Russo \cite{GHMR}, and we confirm that our results for $|L|_{\rm loss}$ agree with theirs.

\section{Angular momentum loss from WQFT}
\label{sec:WEFTWQFT}
WQFT is situated within the framework of worldline effective field theory, where compact objects such as black holes and neutron stars are modelled as point particles described by effective worldline actions~\cite{Goldberger:2004jt,Porto:2016pyg,Levi:2018nxp}. 
The trajectories of the compact objects are coupled to gravity via the metric $g_{\mu\nu}$. In particular, we use the convenient Brink--Di Vecchia--Howe form of the point-particle action,
\begin{align}
    S_{{\rm BH}i}
    =
    -m_i
    \int \d\tau\,
    \big[
    \tfrac12
    g_\mn(x_i(\tau))
    \dot x_i^\mu(\tau)
    \dot x_i^\nu(\tau)
    +\tfrac12
    \big]
    \ .
\end{align}
Gravity in the bulk is governed by the Einstein--Hilbert action
\begin{align}
    S_{\rm EH}
    =
    -\frac{2}{\kappa^2}
    \int
    \mathrm{d}^D x\,
    \sqrt{-g}
    R
    \, ,
\end{align}
with $\kappa^2=32\pi G$, Newton's constant $G$ and the Ricci curvature $R$.
In dimensional regularisation, Newton's constant is technically lifted to $D$ dimensions.
The distinction is, however, immaterial for finite observables.
As we seek to describe a gravitationally interacting two-body system, we construct its total action using two copies of the worldline action plus the Einstein--Hilbert action:
\begin{align}
    S
    =
    S_{\rm BH1}
    +
    S_{\rm BH2}
    +
    S_{\rm EH}
    +
    S_{\rm gf}
    \ .
\end{align}
The gauge fixing term $S_{\rm gf} = \int \d^D x \,(\partial_{\nu}h^{\nu \mu}-\frac12 \partial^{\mu}h^{\nu}_{~\nu})^2+\mathcal{O}(h^3)$ enforces the generalised gauge of ref.~\cite{Driesse:2024feo}, which coincides with de Donder gauge at leading order in the graviton field.

In the case of absent gravitational interactions, the black holes move along straight lines which we parametrise by their four-velocities $v_i^\mu$ and \emph{worldline offsets} $b_i^\mu$ (i.e., the offset of each worldline from the observer). 
The full black hole trajectories then take the form
\begin{align}\label{eq:WLExp}
x_i^\mu(\tau)
=
\tau v_i^\mu+b_i^\mu
+
z_i^\mu(\tau)
\  ,
\end{align}
wherein $z_i^\mu(\tau)$ describes the deviation from straight-line motion. 
In the PM regime, we assume that the deviations are small and admit a Taylor series in the formal expansion parameter $\kappa$.
We further expand the metric $g_{\mu\nu}$ around the flat-space metric $\eta_{\mu\nu}$,
\begin{align}
    g_\mn(x)
    =
    \eta_\mn
    +
    \kappa h_\mn(x)
    \ .
\end{align}
Expanding the action in powers of the fluctuations $z_i^\mu$ and $h_{\mu\nu}$, the Feynman rules may be read off directly: the quadratic terms determine the propagators, while the higher-order terms give rise to interaction vertices. We represent the worldline and graviton propagators by solid and wiggly lines, respectively:
\begin{align}\label{WLprops}
    \begin{tikzpicture}[baseline=(anchor)]
        \coordinate (anchor) at (0,-.1);
        \coordinate (x) at (-.6,0);
        \coordinate (y) at (.6,0);
        \coordinate (in) at (-1.7,0);
        \coordinate (out) at (1.7,0);
        \coordinate (middle) at (0,0);
        \draw [white] (in) -- (out);
        \drawPerturbativeVertex{x}
        \drawPerturbativeVertex{y}
        \drawDirectedBHLine{x}{y}
        \node at (x) [above left=-.1cm and -.05cm,font=\footnotesize] {$\mu$};
        \node at (y) [above right=-.1cm and -.05cm,font=\footnotesize] {$\nu\vphantom{\mu}$};
        \node at (middle) [above=0cm] {$\omega$};
    \end{tikzpicture}
    \!\!\!\!&=\;\ 
    -\frac{\i\eta^{\mu\nu}}{m_j(\omega+\iZero)^{2}}
    \, ,
    \\
    \begin{tikzpicture}[baseline=(anchor)]
        \coordinate (anchor) at (0,-.1);
        \coordinate (x) at (-.6,0);
        \coordinate (y) at (.6,0);
        \coordinate (in) at (-1.7,0);
        \coordinate (out) at (1.7,0);
        \coordinate (middle) at (0,0);
        \draw [white] (in) -- (out);
        \drawPerturbativeVertex{x}
        \drawPerturbativeVertex{y}
        \drawDirectedGravitonLine{x}{y}
        \node at (x) [above left=-.1cm and -.05cm,font=\footnotesize] {$\mn$};
        \node at (y) [above right=-.1cm and -.05cm,font=\footnotesize] {$\ab$};
        \node at (middle) [above=0cm] {$k$};
    \end{tikzpicture}
    \!\!\!\!&=\;\ 
    \i
    \frac{
        \mathcal{P}_{\mn\ab}
    }{(k^0+\iZero)^2-\vct{k}^2}
    \, .
\end{align}
Here, $\mathcal P_{\mn\ab}=\eta_{\mu(\alpha}\eta_{\beta)\nu} -\tfrac1{D-2}  \eta_\mn \eta_\ab$ is the tensor structure of the graviton propagator in de Donder gauge.
To ensure causality in our results, we exclusively use retarded propagators \cite{Galley:2009px,Jakobsen:2022psy,Kalin:2022hph}. 
Since we can choose a time-like vector $v^\mu$ so that $\mathrm{sgn}(k\cdot v)=\mathrm{sgn}(k^0)$, the graviton propagator admits the covariant notation $(k^0+\iZero)^2-\vct{k}^2=k^2+\iZero\,k\cdot v$, where we may take either $v^\mu=v_1^\mu$ or $v^{\mu}=v_2^\mu$.

The retarded prescription of the graviton implies the physical boundary condition of no incoming gravitational radiation.
Instead, the retarded $\iZero$ prescription of the worldline fields is rather practical: it simply implies that our results are expressed in terms of initial state variables.

Before diving into explicit calculations, let us discuss the kinematics and the quantity we want to compute in more detail.
The four-velocities of each worldline satisfy
\begin{align}
    v_i^2=1
    \ ,\qquad
    v_1\cdot v_2=\gamma
    =
    \frac1{\sqrt{1-v^2}}
    \ .
\end{align}
Here, $\gamma$ and $v$ are the relative Lorentz factor and relative velocity, respectively.
Further, the four-momentum of each BH is $p_i^\mu=m_i v_i^\mu$.
Together with the worldline offsets, we may form the angular momentum vector of the two scattering black holes,
\begin{align}
    L^\mu
    =
    -\frac1{E} \epsilon^{\mu}{}_{\alpha\beta\rho}\,
    b^\alpha\,
    p_1^\beta\,
    p_2^\rho\,,
    \qquad
    b^\mu=b_2^\mu-b_1^\mu\,.
\end{align}
Here, $E$ is the total energy of the system $E^2=(p_1+p_2)^2$.
The angular momentum vector is invariant under both translations of the coordinate centre and proper time translations, where the variations of the offset parameters obey $\delta b_i^\mu\propto p_i^\mu$.
We may therefore freely fix the initial worldline offsets to be orthogonal to the velocities $b_i\cdot v_j=0$ for $i,j=1,2$.
The vector $b^\mu$ is then the physical impact parameter with absolute value $|b|=\sqrt{-b^2}$.

Our goal is to calculate the asymptotic change of $L^\mu$, where for any asymptotic worldline variable $\mathcal{O}$ we define
\begin{align}
    \Delta\mathcal{O}=\mathcal{O}_{+\infty}-\mathcal{O}_{-\infty}\,,
\end{align}
which is simply the difference between its value in the infinite future and the infinite past.
In practice, we focus on
\begin{align}\label{eq:deltaEL}
    \Delta(EL^\mu)
    =
    -\epsilon^{\mu}{}_{\alpha\beta\rho}\,
    \Delta\big(b^\alpha\,p_1^\beta\,p_2^\rho\big)\ ,
\end{align}
from which $\Delta L^\mu$ may easily be inferred assuming knowledge of $\Delta E$.
Note that since spinless scattering is planar, the angular momentum vector does not change its direction such that
\begin{align}\label{eq:sclarchange}
    \Delta (E L^\mu)
    =
    \Delta (E|L|) \hat L^\mu
    \ ,
\end{align}
where $\hat L^\mu=L^\mu/|L|$ and $|L|=\sqrt{-L^2}$.
The changes in both $E$ and $L^\mu$ are entirely dissipative.
For such dissipative observables we will often use the notation
\begin{align}
    \mathcal{O}_{\rm loss}
    =
    -\Delta \mathcal{O}
    \ ,
\end{align}
which measures the amount of $\mathcal{O}$ lost by the system during the scattering event.

It is useful to understand the relation of the angular momentum vector to the angular momentum tensor $L^\mn$ defined by,
\begin{align}
    L^\mn
    =
    2 \sum_i 
    b^{[\mu}_i
    p_i^{\nu]}
    =
    -2
    b^{[\mu}
    p^{\nu]}
    +
    2  b_{\rm CoE}^{[\mu}
    P^{\nu]}
    \ ,
\end{align}
with total momentum $P^\mu=p^\mu_1+p^\mu_2=(E,\vct{0})$ and relative momentum $p^\mu=(0,\vct{p}_1)=(0,-\vct{p}_2)$ in the centre-of-mass (CoM) frame.
The centre-of-energy parameter (CoE),
\begin{align}
    b_{\rm CoE}^\mu
    = 
    \sum_i 
    \frac{p_i\cdot P}{E^2}
    \,
    b_i^\mu
    \ ,
\end{align}
parametrises the dependence of $L^\mn$ on the coordinate centre.
The angular momentum vector is,
\begin{align}
    L^\mu = \tfrac12 E^{-1} \eps^\mu{}_{\ab\nu}
    L^\ab P^\nu 
    \ .
\end{align}
Thus, in order to compute the change in $L^\mu$ to 4PM order from knowledge of the change in the tensor, $\Delta L^\mn$, one gets two terms:
\begin{align}
    \Delta L^\mu 
    = 
    \tfrac1{2} E^{-1}
    \eps^\mu{}_{\ab\nu}
    \Delta L^\ab  P^\nu 
    +
    \tfrac12 
    \eps^\mu{}_{\ab\nu}
    L^\ab \Delta(E^{-1} P^\nu )
    +\mathcal O(G^5)
    \, .
\end{align}
The change in the unit vector $P^\mu/E$ is proportional only to $b^\mu$ or $p^\mu$ to order $G^6$ (i.e. orthogonal to itself).
The second term therefore couples to the centre-of-energy parameter and must be included unless one is in the centre-of-energy frame defined by $b_{\rm CoE}^\mu=0$.
Further, the mixed term is neglected because it would be at least of order $G^5$.

Having established the setup of our formalism and the quantity we wish to compute, we turn our attention to the calculation of the change in the worldline offsets $\Delta b_i^\mu$ which we will also refer to as the \emph{offset shifts}.
This shall form the backbone of our calculation. 
The offset shift can be calculated using the formula~\cite{Jakobsen:2026ayw}
\begin{align}
    \label{eq:our-fav-formula}
    \Delta b_i^\mu
    =
    \lim_{\omega_{\rm IR}\to0}
    \frac\d{\d\omega_{\rm IR}}
    \i\omega_{\rm IR}^2\langle
        z_i^\mu(\omega_{\rm IR})
    \rangle
    \,
    =
    \lim_{\omega_{\rm IR}\to0}
    \frac{\d}{\d \omega_{\rm IR}}
    \Big[
    \i \omega_{\rm IR}^2
    \,
    \begin{tikzpicture}[baseline={(0,-.1)}]
        \coordinate (cL) at (-1,0);
        \coordinate (cM) at (0,0);
        \coordinate (cMP) at (-.1,-.1);
        \drawDirectedBHLine{cL}{cM}
        \drawPerturbativeVertex{cM}
        \draw[line width=.8pt,fill=white] (cL) circle (.23cm) ;
        \draw[pattern=north east lines] (cL) circle (.23cm) ;
        \node[below right=.1cm, font=\footnotesize] at (cL) {$\ \ \omega_{\rm IR}$} ;
        \node[above right=.1cm, font=\scriptsize] at (cMP) {$\mu$} ;
    \end{tikzpicture}
    \Big]\ ,
\end{align}
where we have given the soft energy of the outgoing worldline its own symbol $\omega_{\rm IR}$. 
This formula differs only from the calculation of $\Delta p_i^\mu$,
\begin{align}
    \Delta p^\mu_i
    =
    \lim_{\omega_{\rm IR}\to0}
    \Big[
    -m_i \omega_{\rm IR}^2
    \,
    \begin{tikzpicture}[baseline={(0,-.1)}]
        \coordinate (cL) at (-1,0);
        \coordinate (cM) at (0,0);
        \coordinate (cMP) at (-.1,-.1);
        \drawDirectedBHLine{cL}{cM}
        \drawPerturbativeVertex{cM}
        \draw[line width=.8pt,fill=white] (cL) circle (.23cm) ;
        \draw[pattern=north east lines] (cL) circle (.23cm) ;
        \node[below right=.1cm, font=\footnotesize] at (cL) {$\ \ \omega_{\rm IR}$} ;
        \node[above right=.1cm, font=\scriptsize] at (cMP) {$\mu$} ;
    \end{tikzpicture}
    \Big]\ ,
\end{align}
by the derivative with respect to $\omega_{\rm IR}$ and a factor $-\i/m_i$.
The derivative corresponds to taking the subleading term in the expansion in small $\omega_{\mathrm{IR}}$ of the one-point function of the deflection.
Namely, by analysing the limit $\omega_{\rm IR}\to0$ more carefully, one finds,
\begin{align}\label{eq:softExpansion}
	-
    \omega_{\rm IR}^2 \ 
    \begin{tikzpicture}[baseline={(0,-.1)}]
        \coordinate (cL) at (-1,0);
        \coordinate (cM) at (0,0);
        \coordinate (cMP) at (-.1,-.1);
        \drawDirectedBHLine{cL}{cM}
        \drawPerturbativeVertex{cM}
        \draw[line width=.8pt,fill=white] (cL) circle (.23cm) ;
        \draw[pattern=north east lines] (cL) circle (.23cm) ;
        \node[below right=.1cm, font=\footnotesize] at (cL) {$\ \ \omega_{\rm IR}$} ;
        \node[above right=.1cm, font=\scriptsize] at (cMP) {$\mu$} ;
    \end{tikzpicture}
	&=
	\frac{\Delta p_i^\mu}{m_i}
    +
	\i\omega_{\rm IR} (
        \Delta_{\rm dyn} b_i^\mu
        +
        \omega_{\rm IR}^{-2\eps}
        \Delta_{\rm stat} b_i^\mu
    )
	+
	\dots\ .
\end{align}
In the subleading term of this expansion, we distinguish between two kinds of contributions: dynamical and static.
As introduced in \cite{Jakobsen:2026ayw}, a loop momentum can scale either with the modulus of the momentum transfer $q^\mu$ or with the frequency $\omega_{\mathrm{IR}}\ll |q|$, the latter of which can be identified with contributions due to zero-frequency gravitons \cite{DiVecchia:2022owy}:
\begin{align}\label{eq:scalingsStatDyn}
    \begin{aligned}
    \ell_{\mathrm{dyn}}^\mu
    &=
    (\ell^0_{\mathrm{dyn}},\vct{\ell}_{\mathrm{dyn}})
    \sim (|q|,|q|)
    \ ,
    \\
    \ell_{\mathrm{stat}}^\mu
    &=
    (\ell^0_{\mathrm{stat}},\vct{\ell}_{\mathrm{stat}})
    \sim (\omega_{\mathrm{IR}},\omega_{\mathrm{IR}})
    \ .
    \end{aligned}
\end{align}
The static contributions are peculiar to four dimensions, where gravity is a long-range force. 
In higher dimensions $D\ge5$ ($-2\eps\ge1$), the gravitational force falls off faster and these contributions no longer appear. Therefore, this static contribution comes with its subtle $\omega^{-2\epsilon}$ dependence: for $\eps=0$, this is a genuine contribution to $\Delta b_i^\mu$, while for $-2\eps\ge1$ it is negligible. 
In higher dimensions, then, $\Delta_{\rm stat} b_i^\mu$ would be the kick of a higher moment with units different from length.
To determine the offset shift in each region, we employ different methods.
Interestingly, since the static region is concerned with the long-range nature of gravity, it couples to the gravitational memory, both linear and nonlinear.

\subsection{Angular momentum loss in the PM expansion} 
In this section we study the angular momentum loss as it appears in the PM expansion.
In many respects, it mimics the PM expansion of the impulse except in its static contributions.

Let us first note that the offset shifts $\Delta b_i^\mu$ have the same simple mass polynomiality as the impulse~\cite{Damour:2019lcq,Jakobsen:2023oow,Driesse:2026qiz} since both are derived from the worldline one-point function.
In particular, the mass expansion of each worldline offset $\Delta b_i^\mu$ and of each (rescaled) impulse $\Delta p_i^\mu/m_i$ is exactly the same at each PM order.
For example, for the offset shift of WL1 at 4PM order:
\begin{subequations}
    \label{eqs:eqseqs}
\begin{align}
    \Delta^{(\rm 4PM)} b_1^{\mu}
    =
    \frac{G^4m_2}{|b|^3}\big[
    m_2^3 c^{\rm (4PM\,0SF)\mu}_b
    &+
    m_1m_2^2 c^{\rm (4PM\,1SF)\mu}_b
    \\[-6pt]
    &+
    m_1^2m_2 c^{\rm (4PM\,\overbar{1{SF}})\mu}_b
    +
    m_1^3 c^{\rm (4PM\,\overbar{0{SF}})\mu}_b
    \big]
    \ ,
    \nn
    \\
    \Delta^{(\rm 4PM)} p_1^{\mu}
    =
    \frac{G^4m_1m_2}{|b|^4}
    \big[
    m_2^3 c^{(\rm 4PM\,0SF)\mu}_p
    &+
    m_1m_2^2 c^{(\rm 4PM\,1SF)\mu}_p
    \\[-6pt]
    &+
    m_1^2m_2^1 c^{(\rm 4PM\,\overbar{1SF})\mu}_p
    +
    m_1^3 c^{(\rm 4PM\,\overbar{0SF})\mu}_p
    \big]
    \ .
    \nn
\end{align}
\end{subequations}
Here, we denote the different mass sectors $n$SF by their self-force order taken from the perspective of $m_2$ larger than $m_1$.
This perspective, however, is arbitrary and the bar notation $\overbar{n{\rm SF}}$ takes the perspective of $m_1$ larger than $m_2$.
In this sense, at 4PM one never encounters effects beyond the first self-force order.

The self-force counting in \eqn{eqs:eqseqs} is particularly relevant for the dynamical region where $n$SF and $\overbar{n{\rm SF}}$ contributions are computed from exactly the same kinds of integrals.
In this case, dissipative effects appear only at 1SF and $\overbar{1{\rm SF}}$.
Instead, in the static region, one encounters dissipative effects for $\Delta b_i^\mu$ also in the $\overbar{0{\rm SF}}$ contributions.
Indeed, this ``dissipative nature'' of $\overbar{0{\rm SF}}$ is known in the trajectory~\cite{Gralla:2021qaf,Bini:2024hme,Mogull:2025cfn}.

The change in the energy-scaled angular momentum vector $E L^\mu$ is composed linearly of the impulse and offset shifts, see \eqn{eq:deltaEL}, and therefore inherits a polynomial mass dependence.
Following \eqn{eq:sclarchange}, so does the change in the scalar $E |L|$.
The loss therefore takes the form,
\begin{align}
    (E|L|)_{\rm loss}
    &=
    \frac{G^4 m_1^2 m_2^2}{|b|^3} 
    \Big(
        m_1^2 c^{(1{\rm SF})} 
        +
        m_1 m_2 \big(
            c^{\rm (\overbar{1SF})}
            +2c^{(1{\rm SF})}
            \big)
        +
        m_2^2 c^{(1{\rm SF})} 
    \Big)
    \nn
    \\
    &=    
    \frac{G^4 M^6 \nu^2}{|b|^3} 
    \Big(
        c^{(1{\rm SF})} 
        +
        \nu 
        c^{\rm (\overbar{1SF})}
    \Big)
    \ .
    \label{eq:nuparamet}
\end{align}
In the first line we used the symmetry of $(E|L|)_{\rm loss}$ in the two bodies to put the same coefficient in front of $m_1^2$ and $m_2^2$ which allows for the rewriting of the second line in terms of the total mass and the symmetric mass ratio,
\begin{align}
    M=m_1+m_2\ ,\qquad \nu=\frac{m_1 m_2}{M^2}
    \ .
\end{align}
The $\overbar{1{\rm SF}}$ notation may seem a bit odd in \eqn{eq:nuparamet} as it now genuinely looks like a 2SF correction.
The coefficient $c^{\rm (1SF)}$, however, is derivable with the knowledge only of the 1SF coefficients of \eqn{eqs:eqseqs} (including, though, the static dissipation in $c_b^{\rm (4PM\,\overbar{0SF})}$) while the $c^{\rm (\overbar{1 SF})}$ instead requires the $\overbar{1{\rm SF}}$ coefficients of \eqn{eqs:eqseqs} as well.

In the dynamic region, dissipative effects may generally be associated with the presence of \emph{active} gravitons by which we mean propagating gravitons whose four-momentum may go on-shell.
Generally, not all gravitons in a given diagram will be active.
For example, any graviton emitted from a one-point worldline emission vertex has a vanishing energy component and is therefore \emph{passive}, i.e., it never goes on-shell.
In the dynamical region, one has at most two active gravitons at the 4PM order.

The categorisation of active gravitons is particularly useful in the low-velocity PN limit where $v\to0$.
Here, one finds two kinds of gravitons, potential (P) and radiative (R), well-known from PN theory~\cite{Goldberger:2004jt}.
Potential gravitons are passive while radiative gravitons are active and therefore sensitive to the retarded pole prescription.
For dynamical gravitons these regions are characterised as
\begin{align}
    \begin{aligned}\label{eq:scalingsDyn}
    \ell_{\rm P}^\mu
    &=
    (\ell^0_{\rm P},\vct{\ell}_{\rm P})
    \sim 
    |q|(v,1)
    \ ,
    \\
    \ell_{\rm R}^\mu
    &=
    (\ell^0_{\rm R},\vct{\ell}_{\rm R})
    \sim 
    |q|(v,v)
    \ .
    \end{aligned}
\end{align}
Having at most two active gravitons at 4PM order, we may at most have two radiative gravitons.

The notion of potential gravitons may easily be lifted to the PM context.
Namely, gravitons whose energy component is forced to vanish in either worldline frame are potential and cannot go on shell.
All other gravitons are generally active, and their retarded prescription matters.
For some active gravitons, the potential region is scaleless in the PN limit, such that only the radiation region contributes.
We refer to these as radiative and highlight them in red as a subclass of the more general active gravitons.
The radiative gravitons typically describe self-interactions of each body, i.e., subdiagrams that involve only a single WL and gravitons, or emitted gravitons that are put on shell.
In WQFT diagrams we distinguish them by colours:
\begin{align}
    \begin{tikzpicture}[baseline=(anchor)]
        \coordinate (anchor) at (0,-.1);
        \coordinate (x) at (-.6,0);
        \coordinate (y) at (.6,0);
        \drawPerturbativeVertex{x}
        \drawPerturbativeVertex{y}
        \drawDirectedGravitonLine{x}{y}
    \end{tikzpicture}
    =
    \text{Potential}
    \ ,
    \qquad
    \begin{tikzpicture}[baseline=(anchor)]
        \coordinate (anchor) at (0,-.1);
        \coordinate (x) at (-.6,0);
        \coordinate (y) at (.6,0);
        \drawPerturbativeVertex{x}
        \drawPerturbativeVertex{y}
        \drawDirectedRadiativeGravitonLine{x}{y}
    \end{tikzpicture}
    =
    \text{Radiative}
    \ ,
    \qquad
    \begin{tikzpicture}[baseline=(anchor)]
        \coordinate (anchor) at (0,-.1);
        \coordinate (x) at (-.6,0);
        \coordinate (y) at (.6,0);
        \drawPerturbativeVertex{x}
        \drawPerturbativeVertex{y}
        \drawDirectedActiveGravitonLine{x}{y}
    \end{tikzpicture}
    =
    \text{Active}
    \ .
\end{align}
The different colours serve solely to highlight different physical properties of the gravitons -- their Feynman rules are independent of it.
For example, the causal arrow for potential gravitons is immaterial.

Static gravitons (or zero-frequency gravitons) form another class defined by their scaling with $\omega_{\rm IR}$ (see \eqn{eq:scalingsStatDyn}).
As in ref.~\cite{Jakobsen:2026ayw}, we draw these in blue.
In the PN limit, one again finds two types of static gravitons whose scalings mimic the conventional potential and radiative ones:
 \begin{align}
    \begin{aligned}\label{eq:scalingsStat}
    \ell_{\rm stat.\,P}^\mu
    &=
    (\ell^0_{\rm stat.\, P},\vct{\ell}_{\rm stat.\, P})
    \sim 
    \frac{\omega_{\rm IR}}{v}
    (v,1)
    \ ,
    \\
    \ell_{\rm stat.\,R}^\mu
    &=
    (\ell^0_{\rm stat.\,R},\vct{\ell}_{\rm stat.\,R})
    \sim 
    \frac{\omega_{\rm IR}}{v}
    (v,v)
    \ .
    \end{aligned}
 \end{align}
 In other words, they are potential and radiative at the scale $\omega_{\rm IR}/v$ instead of the scale $|q|$.

Contributions involving only potential gravitons are purely conservative and we can discard them for computing the dissipative losses of $E$ and $|L|$. 
At the leading PM orders, one may then generally separate out contributions from one (R) or two (RR) radiative gravitons (two being the maximal up to 4PM order).
This clean separation would get less trivial at 5PM where R and RRR can have $1/\epsilon$ pole cancellations between them, i.e., ``tail'' effects~\cite{Dlapa:2021vgp,Bern:2021yeh,Driesse:2024feo}.

Contributions with two radiative gravitons appear first at 4PM order and their contribution to $\Delta(E L^\mu)$ is therefore clean:
\begin{align}\label{eq:ELdecomposedRR}
    \Delta_{\rm RR}^{\rm (4PM)} (EL^\mu)
    =
    -\epsilon^\mu{}_{\ab\gamma}
    \Big[
        \Delta_{\rm RR}^{\rm (4PM)}b^\alpha
    p_1^\beta
    p_2^\gamma
    +
    b^\alpha
    \Delta_{\rm RR}^{\rm (4PM)}p_1^\beta
    p_2^\gamma
    +
    b^\alpha
    p_1^\beta
    \Delta_{\rm RR}^{\rm (4PM)}p_2^\gamma
    \Big]
    \ .
\end{align}
This is the contribution first computed by C. Heissenberg in ref.~\cite{Heissenberg:2025ocy}.
Here, the subscript RR on $\Delta$ is meant to imply the parts of the change due to two radiative gravitons.

In the case of contributions from only one radiative graviton, one gets 4PM, 3PM and static 2PM contributions:
\begin{align}\label{eq:ELdecomposedR}
    \Delta_{\rm R}^{\!\rm(4PM)} (EL^\mu)
    =
    -\epsilon^\mu{}_{\ab\gamma}
    \Big[&
    \Delta_{\rm R}^{\!\rm(4PM)} b^\alpha
    p_1^\beta
    p_2^\gamma
    +
    b^\alpha
    \Delta_{\rm R}^{\!\rm(4PM)}p_1^\beta
    p_2^\gamma
    +
    b^\alpha
    p_1^\beta
    \Delta_{\rm R}^{\!\rm(4PM)}p_2^\gamma
    \nn\\
    &+
    \Delta_{\rm R}^{\!\rm(3PM)} b^\alpha
    (\Delta^{\!\rm(1PM)}_{\rm C}p_1^\beta
    p_2^\gamma
    +
    p_1^\beta
    \Delta^{\!\rm(1PM)}_{\rm C}p_2^\gamma)
    \nn\\
    &
    +
    \Delta^{\!\rm(1PM)}_{\rm C} b^\alpha
    (
    \Delta_{\rm R}^{\!\rm(3PM)}p_1^\beta
    p_2^\gamma
    +
    p_1^\beta
    \Delta_{\rm R}^{\!\rm(3PM)}p_2^\gamma
    )
    \nn\\
    &
    +
    \Delta^{\!\rm(2PM)}_{\rm R} b^\alpha
    (
    \Delta_{\rm C}^{\!\rm(2PM)}p_1^\beta
    p_2^\gamma
    +
    p_1^\beta
    \Delta_{\rm C}^{\!\rm(2PM)}p_2^\gamma
    )
    \Big]
    \ .
\end{align}
A subscript C here implies the conservative change which to the orders considered here can be identified with potential contributions only (i.e., $\rm R^0$).
Interestingly, both $\Delta_{\rm R}^{\!\rm(4PM)} b^\alpha$ and $\Delta^{\!\rm(1PM)}_{\rm C} b^\alpha$ come with $1/\eps$ poles which cancel each other.
Note, one would also have a term combining the 1PM conservative impulse and the 3PM R impulse.
This term, however, vanishes because the 1PM conservative impulse is parallel to $b^\mu$.

We conclude this section with reporting two of the lower PM-order results required for the R result of $\Delta(EL^\mu)$.
First, the 1PM dynamical, conservative change in the offset parameter:
\begin{align}
    \Delta_{\rm C}^{\!(1\rm PM)} b_1^\mu
    &=
    \frac{\d}{\d \omega_{\rm IR}}
    \Bigg[
    \i \omega_{\rm IR}^2
    \begin{tikzpicture}[baseline={(0,.4)}]
        \drawGravitonLine{.6,0}{.6,.9}
        \drawDottedLineR{.6,0}{.5}
        \drawDottedLineR{.6,.9}{-.5}
        \drawBHLineR{.6,.9}{.5}
        \drawDottedLineR{.6,0}{-.5}
        \drawPerturbativeVertex{.6,0}
        \drawPerturbativeVertex{.6,.9}
        \drawPerturbativeVertex{1.1,.9}
    \end{tikzpicture}
    \ \
    \Bigg]
    \Bigg|_{\omega_{\rm IR}\to0}
    \\
    &=
    e^{\epsilon\gamma_{E}}
    \pi^{\epsilon}
    \frac{m_2 G |b|^{2\epsilon}}{(\gamma^2-1)^{\frac32}}
    \Big[
        \frac{v_1^\mu-\gamma(3-2\gamma^2)v_2^\mu}{\epsilon}
        -(2\gamma^2-1) v_1^\mu
        +\gamma v_2^\mu
        +
        \mathcal{O}(\epsilon)
    \Big]
    \nn
    \ .
\end{align}
In the above equation, recall that $G$ is $D$-dimensional and units therefore check out correctly.
Indeed, this result has an infrared $1/\epsilon$ pole which is cancelled by a corresponding pole in $\Delta_{\rm R}^{\!\rm(4PM)} b^\alpha$.
Since $\Delta_{\rm C}^{\!(1\rm PM)} b_1^\mu$ multiplies the 3PM impulse, we require this observable to subleading order in $\eps$ in its dissipative $b$-direction:
\begin{subequations}
\begin{align}
    \hat b_\mu \Delta_{\rm R}^{(\rm 3PM)} p^\mu_1
    &=
    -\frac{G^3\nu^2M^4}{|b|^3}
    \bigg[
        \frac{e^{3\gamma_{E}}\pi^{3}\, |b|^{6}}{4(\gamma^2-1)}
    \bigg]^{\epsilon}
    \nn\\
    &\quad\times
    \Big[
        \frac{2(2\gamma^2-1)^2}{\gamma^2-1}\,\mathcal I(\gamma)
        +
        \epsilon\, \vec d^{\,(\rm 3PM)}(\gamma)\cdot\vec f^{\,(\rm 3PM)}(\gamma)
        +
        \mathcal{O}(\epsilon^2)
    \Big]
    \ ,
\end{align}
Here we have coefficients
\begin{align}\label{eq:labellabel}
    d^{(\rm 3PM)}_1
    &=
    \frac{16(-32+115\gamma^2-172\gamma^4+80\gamma^6)}{9(\gamma^2-1)^2}
    \ ,
    \\
    d^{(\rm 3PM)}_2
    &=
    -\frac{8\gamma(-62+183\gamma^2-210\gamma^4+80\gamma^6)}{3(\gamma^2-1)^2}
    \ ,
    \\
    d^{(\rm 3PM)}_3
    &=
    \frac{8(3+9\gamma^2-16\gamma^4+4\gamma^6)}{(\gamma^2-1)^2}
    \ ,
    \\
    d^{(\rm 3PM)}_4
    &=
    \frac{4\gamma(2\gamma^2-1)^2(2\gamma^2-3)}{(\gamma^2-1)^2}
    \ ,
\end{align}
and the basis functions
\begin{equation}
    \vec{f}^{\,(\rm 3PM)}(\gamma)
    =
    \begin{pmatrix}
        1
        \\[6pt]
        \dfrac{\arccosh(\gamma)}{\sqrt{\gamma^2-1}}
        \\[12pt]
        \arccosh^2(\gamma)
        \\[6pt]
        \dfrac{\arccosh^2(\gamma)+\operatorname{Li}_2\!\big(1-x^2\big)}{\sqrt{\gamma^2-1}}
    \end{pmatrix}
    \ ,
    \qquad
    x=\gamma-\sqrt{\gamma^2-1}
    \ .
\end{equation}
\end{subequations}
Finally, the function $\mathcal{I}(\gamma)$ is conventionally defined as~\cite{Damour:2020tta},
\begin{align}
    \mathcal{I}(\gamma)= \frac23\frac{8-5\gamma^2}{\gamma^2-1}
    -
    2\gamma\frac{3-2\gamma^2}{(\gamma^2-1)^{3/2}}
    \arccosh(\gamma)\, .
\end{align}

\section{Gravitational linear and nonlinear memory from diagrams}
\label{sec:memory}
In \secref{sec:WEFTWQFT}, we have seen that the offset shift receives contributions from two regions in the limit $\omega_{\rm IR}\to0$: a dynamical one, in which all loop momenta scale with $|q|$, and a static one, in which a graviton carries momentum of order $\omega_{\rm IR}$. In this section, we discuss those static contributions, first in general and then specifically at 4PM.

Static gravitons themselves may technically be potential or radiative, according to \eqn{eq:scalingsStat}. However, we observe that only radiative ones contribute. Those correspond to on-shell gravitons with vanishing frequency. We now proceed to show that these gravitons couple the worldline to the gravitational-wave memory.

We begin by considering a general diagram contributing to $\Delta b_i^\mu$ in the static region
\begin{align}\label{eq:generalDiagram}
    \begin{tikzpicture}[baseline={(0,-.2)}]
        \coordinate (LU) at (-.7,+.4) ;
        \coordinate (LD) at (-.7,-.4) ;
        \coordinate (M)  at (0 ,0  ) ;
        \coordinate (R)  at (.8 ,0  ) ;
        \coordinate (D)  at (0 ,-.8 ) ;
        \drawStaticGravitonLine{D}{M}
        \drawStaticBHLine{LU}{M}
        \drawStaticBHLine{LD}{M}
        \drawStaticDirectedBHLine{M}{R}
        \drawStaticPerturbativeVertex{R}
        \draw[line width=.8pt,fill=white] (LD) circle (.17cm) ;
        \draw[pattern=north east lines] (LD) circle (.17cm) ;
        \draw[line width=.8pt,fill=white] (LU) circle (.17cm) ;
        \draw[pattern=north east lines] (LU) circle (.17cm) ;
        \draw[line width=.8pt,fill=white] (D) circle (.17cm) ;
        \draw[pattern=north east lines] (D) circle (.17cm) ;
        \node[rotate=90] at ($(LU)!0.5!(LD)$) {\scriptsize ...} ;
        \drawStaticPerturbativeVertex{M}
        \node at ($(D)!.5!(M)$) [right=.0cm] {\footnotesize $k$} ;
        \node at ($(M)!.5!(R)$) [above=.08cm] {\footnotesize $\omega_{\rm IR}$} ;
        \node at (R) [right=.0cm] {\footnotesize $\mu$} ;
    \end{tikzpicture}
    \ \ ,
\end{align}
where we adopt the graphical representation of propagators and vertices that are dominated by static contributions as blue.
Due to the static graviton in this diagram being radiative, the blob connected to it represents an insertion of a waveform with on-shell momentum $k^2 =0$ and static energy component $k^0 \sim \omega_{\rm IR}$. Diagrammatically, the graviton insertion therefore is
\begin{align}
    \langle
        h_\mn(k)
    \rangle
    =
    \begin{tikzpicture}[baseline={(0,-.1)}]
        \coordinate (cL) at (-1,0);
        \coordinate (cM) at (0,0);
        \drawDirectedStaticGravitonLine{cL}{cM}
        \drawStaticPerturbativeVertex{cM}
        \draw[line width=.8pt,fill=white] (cL) circle (.23cm) ;
        \draw[pattern=north east lines] (cL) circle (.23cm) ;
        \node[below right, font=\footnotesize] at (cL) {$\ \ k$} ;
        \node[above right=-.1cm and -0.1cm, font=\scriptsize] at (cM) {$\mn$} ;
    \end{tikzpicture}
    \ .
\end{align}
This amounts to the small-frequency limit of the waveform, known as its \textit{soft} expansion. The static region of $\Delta b_i^\mu$ probes the soft limit of the waveform, which has been well studied~\cite{Weinberg:1965nx,Saha:2019tub,Sahoo:2021ctw,Alessio:2025flu} and is famously related to Weinberg's soft theorem in the gravitational context, as well as to gravitational memory~\cite{Strominger:2014pwa,He:2014laa}. 

To analyse this further, we express the frequency-domain waveform $f_\mn(k)$ by its definition as the leading coefficient in the far-field limit,
\begin{align}
    \lim_{|\vct{x}|\to\infty}|\vct{x}|
    \kappa h_\mn(x)
    =
    \int_\omega
    e^{-\i\omega u}
    f_{\mn}(\omega n)
    +
    \mathcal{O}(|\vct{x}|^{-1})
    \ ,
\end{align}
with retarded time $u=t-|\vct{x}|$.
This is in turn related to the one-point function of the metric perturbation $h_{\mu\nu}$ by amputating the external propagator as (see e.g.~\cite{Jakobsen:2021smu,Jakobsen:2021lvp,Jakobsen:2023oow,Bautista:2026qse}),
\begin{align}
    f_{\mu\nu}(k)
    =
    -\frac{\i\kappa
    }{4\pi}
    (-\i k^2)
    \langle h_\mn(k) \rangle
    \ .
\end{align}
Parametrising $k^\mu=\omega n^\mu$ with the null vector $n^\mu=(1,\hat{\vct{x}})$, the leading term of the soft expansion, i.e.\ the expansion around $\omega\to0$, reads
\begin{align}
    \i f^\mn(\omega n)
    =
    4 G
    \bigg[
    \frac{A_{(+)}^\mn(n)}{\omega+\iZero}
    -
    \frac{A_{(-)}^\mn(n)}{\omega-\iZero}
    \bigg]
    +
    \mathcal{O}(\omega^0)
    \ ,
\end{align}
where $A_{(+)\mn}$ and $A_{(-)\mn}$ describe the late- and early-time behaviour of the waveform, respectively.
In particular, the wave memory takes the form
\begin{align}
    \int_{-\infty}^{\infty}
    \d u
    \Big[
        \lim_{|\vct{x}|\to\infty}
        |\vct{x}|
        \kappa
        \frac{\d}{\d u}
        h_\mn(x)
    \Big]
    =
    -4G
    \big(
        A_{(+)\mn}(n)-A_{(-)\mn}(n)
    \big)
    \ ,
\end{align}
which makes the connection to the time-domain waveform manifest.

A simple power counting of the diagram~\eqref{eq:generalDiagram} (for details on the power counting see ref.~\cite{Jakobsen:2026ayw}) shows that exactly a behaviour of the waveform as $h\sim\omega^{-1}$ gives rise to nonzero contributions to $\Delta b_i^\mu$ and that all subleading terms in the soft expansion can be neglected. The (radiative) static contributions to the worldline offsets $\Delta b_i^{\mu}$ are therefore governed by the gravitational-wave memory.

The memory receives two kinds of contributions: linear (lm) and nonlinear (nlm). The linear memory, which is sourced by the change in the black-hole momenta, is given by
\begin{align}
    A_{(+)}^{\rm lm\,\mn}(n)
    =
    \sum_{i=1}^2
    \frac{p_{i,\infty}^\mu p_{i,\infty}^\nu}{
    p_{i,\infty}\cdot n}
    \ ,\qquad
    A_{(-)}^{\rm lm\,\mn}(n)
    =
    \sum_{i=1}^2
    \frac{p_{i}^\mu p_{i}^\nu}{
    p_{i}\cdot n}
    \ ,
\end{align}
where final momenta are defined as $p_{i,\infty}^\mu=p_i^\mu+\Delta p_i^\mu$.
These expressions are precisely given by Weinberg's soft theorem. 

The nonlinear memory, on the other hand, is sourced by the emitted gravitational radiation itself~\cite{PhysRevLett.67.1486,PhysRevD.45.520,PhysRevD.44.R2945,Heissenberg:2024umh,Georgoudis:2025vkk}. Being quadratic in the waveform, it can naively first contribute to $\Delta b_i^\mu$ at 4PM order. It reads
\begin{align}\label{eq:nonlinmem}
    A_{(+)}^{\rm nlm\,\mn}
    =
    \frac{\pi}{2G}
    \int_{\ell}
    \theta(v\cdot \ell)
    \dd(\ell^2)
    f^{\rs}(\ell)
    \mathcal P^\ab_\rs
    f^*_{\ab}(\ell)
    \frac{
        \ell^\mu \ell^\nu
        }{
            n\cdot \ell
        }
    \ ,
    \qquad
    A_{(-)}^{\rm nlm\,\mn}=0
    \ .
\end{align}
In the above equation, $\theta(v\cdot\ell)\,\dd(\ell^2)$ restricts $\ell^\mu$ to be on shell with positive energy, where $v^{\mu}$ may either be $v_1^{\mu}$ or $v_2^{\mu}$. Observe that the nonlinear memory has a structure similar to the linear one, where the outgoing/incoming black-hole momenta are replaced with on-shell graviton momenta.
In ref.~\cite{Heissenberg:2024umh}, it was shown that nonlinear memory contributions to the loss of the spatial components of the angular momentum tensor vanish at this order due to a symmetry argument.
In other words, nonlinear memory contributions to the change in $L^\mu$ can only appear from 5PM order onwards.

In the following, we first derive the linear memory at all orders and compute its contribution to the offset shift at 4PM from static correlators as introduced in ref.~\cite{Jakobsen:2026ayw}. 
We then show how nonlinear memory appears diagrammatically, and why its contribution to $L^\mu_{\rm loss}$ vanishes at the 4PM order~\cite{Heissenberg:2024umh}.

\subsection{Linear gravitational memory at all orders}
The linear gravitational wave memory in the context of the waveform can be isolated by the subsequent diagrams,
\begin{align}\label{eq:blobBrokenDown}
    \begin{tikzpicture}[baseline={(0,-.1)}]
        \coordinate (cL) at (-1,0);
        \coordinate (cM) at (0,0);
        \drawDirectedStaticGravitonLine{cL}{cM}
        \drawStaticPerturbativeVertex{cM}
        \draw[line width=.8pt,fill=white] (cL) circle (.23cm) ;
        \draw[pattern=north east lines] (cL) circle (.23cm) ;
        \node[below right, font=\footnotesize] at (cL) {$\ \ k$} ;
        \node[above right=-.1cm and -0.1cm, font=\scriptsize] at (cM) {$\mn$} ;
    \end{tikzpicture}
    \
    \big|_{\rm lm}
    =
    \sum_{n=0}^{\infty}
    \
    \frac1{n!}
    \
    \bigg[\
    \begin{tikzpicture}[baseline={([yshift=-.5ex]current bounding box.center)},yscale=-1]
        \drawDirectedStaticCurvedGravitonLine[1]{.6,0}{1.3,.7}{90}{180}
        \drawStaticPerturbativeVertex{1.3,.7}
        \node[right=1pt,font=\scriptsize] at (1.3,.7) {$\mn$} ;
        \node[font=\footnotesize] at (1.12,.42) {$k$} ;
        \drawDottedLineR{.6,0}{.5}
        \drawStaticBHLineR{.6,0}{-.6}
        \drawStaticCurvedBHLine{.6,0}{0,.7}{150}{0}
        \drawStaticPerturbativeVertex{.6,0}
        \drawStaticBlob{0,0}
        \drawStaticBlob{0,.7}
        \node[rotate=90] at (0,.35) {\scriptsize ...} ;
        \draw[decorate,decoration={brace,amplitude=2.5pt,raise=1pt}]
            (-.15,.8) -- (-.15,-.1)
            node[midway,left=3pt,inner sep=0] {\scriptsize $n\,$} ;
    \end{tikzpicture}
    \ +\
    \begin{tikzpicture}[baseline={([yshift=-.5ex]current bounding box.center)}]
        \drawDirectedStaticCurvedGravitonLine[1]{.6,0}{1.3,.7}{90}{180}
        \drawStaticPerturbativeVertex{1.3,.7}
        \node[right=1pt,font=\scriptsize] at (1.3,.7) {$\mn$} ;
        \node[font=\footnotesize] at (1.12,.42) {$k$} ;
        \drawDottedLineR{.6,0}{.5}
        \drawStaticBHLineR{.6,0}{-.6}
        \drawStaticCurvedBHLine{.6,0}{0,.7}{150}{0}
        \drawStaticPerturbativeVertex{.6,0}
        \drawStaticBlob{0,0}
        \drawStaticBlob{0,.7}
        \node[rotate=90] at (0,.35) {\scriptsize ...} ;
        \draw[decorate,decoration={brace,amplitude=2.5pt,raise=1pt}]
            (-.15,-.1) -- (-.15,.8)
            node[midway,left=3pt,inner sep=0] {\scriptsize $n\,$} ;
    \end{tikzpicture}
    \
    \bigg]
    \ .
\end{align}
We proceed to compute this infinite sum of diagrams at leading order in the soft limit $k^0\to0$.
We focus on the sum of emissions from worldline 1, with the ones of worldline 2 given by swapping particle labels.
The required tower of vertex rules is a simple result of WQFT \cite{Mogull:2020sak}:
\begin{align}
\label{eq:hejFarvel}
&
\begin{tikzpicture}[baseline={([yshift=-.5ex]current bounding box.center)},yscale=-1]
    \drawCurvedGravitonLine{.6,0}{1.3,.7}{90}{180}
    \drawDottedLineR{.6,0}{.5}
    \drawBHLineR{.6,0}{-.6}
    \drawCurvedBHLine{.6,0}{0,.7}{150}{0}
    \drawPerturbativeVertex{.6,0}
    \node[rotate=90] at (0,.35) {\scriptsize ...} ;
    \node at (1.05,.7) [below=2pt] {\scriptsize $k$} ;
    \node at (1.3,.7) [right=1pt] {\scriptsize $\mu\nu$} ;
    \node at (.3,0) [above=1pt] {\scriptsize $\omega_1$} ;
    \node at (.3,.64) [below=1pt] {\scriptsize $\omega_n$} ;
    \node at (0,0) [left=-2pt] {\scriptsize $\rho_1$} ;
    \node at (0,.7) [left=-2pt] {\scriptsize $\rho_n$} ;
    \draw[decorate,decoration={brace,amplitude=2.5pt,raise=1pt}]
        (-.5,.8) -- (-.5,-.1)
        node[midway,left=3pt,inner sep=0] {\scriptsize $n\,$} ;
\end{tikzpicture}
 =
\i V^{\mu\nu}_{\rho_1\cdots\rho_n}(k;\omega_1,\cdots,\omega_n)
={}
\i^{n-1}
m_1\kappa \,e^{\i k\cdot b_1}\,
\dd
\big(
    k\cdot v_1+
  {\textstyle \sum_{i=1}^{n}}
    \omega_i
\big)
\\
&
\qquad\qquad
\times\bigg[
    \frac{1}{2}
        \bigg(
        \prod_{i=1}^n k_{\rho_i}
        \bigg)
        v_1^\mu v_1^\nu
+\sum_{i=1}^n\omega_i
    \bigg(
    \prod_{j\neq i}^n k_{\rho_j}
    \bigg)
    v_1^{(\mu}\delta^{\nu)}_{\rho_i}
+\sum_{i<j}^n\omega_i\omega_j
    \bigg(
    \prod_{l\neq i,j}^n k_{\rho_l}
    \bigg)
    \delta^{(\mu}_{\rho_i}\delta^{\nu)}_{\rho_j}
\bigg].
\nn
\end{align}
Here, all frequencies and momenta are outgoing.
In this context, as $k^\mu$ is soft, we may omit the exponential factor $\text{exp}(\i k\cdot b_1)$ as the phase $k\cdot b_1\ll1$ is suppressed.
We also drop labellings on the diagrams going forward.
The integrand becomes,
\begin{align}
    &\begin{tikzpicture}[baseline={([yshift=-.5ex]current bounding box.center)},yscale=-1]
        \drawDirectedStaticCurvedGravitonLine{.6,0}{1.3,.7}{90}{180}
        \drawDottedLineR{.6,0}{.5}
        \drawStaticBHLineR{.6,0}{-.6}
        \drawStaticCurvedBHLine{.6,0}{0,.7}{150}{0}
        \drawStaticPerturbativeVertex{.6,0}
        \drawStaticBlob{0,0}
        \drawStaticBlob{0,.7}
        \node[rotate=90] at (0,.35) {\scriptsize ...} ;
        \draw[decorate,decoration={brace,amplitude=2.5pt,raise=1pt}]
            (-.15,.8) -- (-.15,-.1)
            node[midway,left=3pt,inner sep=0] {\scriptsize $n\,$} ;
    \end{tikzpicture}
    =
    \prod_{m=1}^n \bigg[
         -\int_{\omega_m}\frac{\Delta p_1^{\sigma_m}}{(\omega_m+\iZero)^2 m_1}
    \bigg]
    \i V_{\sigma_1\dots\sigma_n}^\mn(k;-\omega_1,\dots,-\omega_n)
    \ .
\end{align}
Here, we have used the ``soft'' or ``static'' expansion of the worldline fields~\eqref{eq:softExpansion}, which tells us that their leading order is proportional to the impulse.
Therefore, we have $n$ insertions of the impulse, each with a worldline propagator, and we integrate over all frequencies.
Note that we have amputated the graviton propagator (and therefore omit a dot on the external line).

Before carrying out the frequency integrals, we simplify the frequency numerators introduced by the Feynman vertex rule.
Since all indices are contracted with the impulse, the vertex splits into three tensor structures, $v_1^\mu v_1^\nu$, $v_1^{(\mu}\Delta p_1^{\nu)}$ and $\Delta p_1^\mu\Delta p_1^\nu$, carrying zero, one, and two frequencies, respectively. The frequency dependencies can be eliminated by using the delta function constraint.
First, the sum $\sum_i\omega_i$ can be replaced by $k\cdot v_1$.
Second, one may use the following replacement:
\begin{align}
    \sum_{i<j}\omega_i\omega_j\to \frac12 \sum_{ij}\omega_i\omega_j
    =
    \frac12 (k\cdot v_1)^2
    \ .
\end{align}
This is implied because the integral over frequencies vanishes with any insertion of $\omega_i^2$ for $n\ge2$.
In this case the insertion completely cancels a propagator and the frequency integrand collapses to a product with no simple poles.

With the frequency numerators eliminated, the relevant integral is a convolution of $n$ worldline propagators, which for any $n\ge0$ is easily computed:
\begin{align}\label{eq:integral1d}
    \int_{\omega_1\dots\omega_n}
         \frac{\dd(k\cdot v_1-\sum_{i=1}^n\omega_i)}{
            \prod_{i=1}^n (\omega_i+\iZero)^2}
    =
    \frac{(-\i)^{n-1}n!}{(k\cdot v_1+\iZero)^{n+1}}
    -
    \delta_{n0}
    \frac{\i}{k\cdot v_1-\iZero}
    \ .
\end{align}
For $n\ge1$, the delta function is integrated away and only the retarded poles of the worldline propagators remain.
For $n=0$, instead, there is no integral and the delta function remains, which we have written as the cut of the propagator,
\begin{align}\label{eq:cutsubstraction}
    \dd(k\cdot v_1)
    =
    \frac{\i}{k\cdot v_1+\iZero}
    -
    \frac{\i}{k\cdot v_1-\iZero}
    \ .
\end{align}
The first, retarded, term is simply the general result evaluated at $n=0$, while the second, advanced, term appears only for $n=0$.
As we will see, it is this advanced term that gives rise to the initial-state contribution to the memory.

Combining the simplified numerators with \eqn{eq:integral1d}, the general contribution is
\begin{align}\label{eq:OneDiagram}
        &\begin{tikzpicture}[baseline={([yshift=-.5ex]current bounding box.center)},yscale=-1]
        \drawDirectedStaticCurvedGravitonLine{.6,0}{1.3,.7}{90}{180}
        \drawDottedLineR{.6,0}{.5}
        \drawStaticBHLineR{.6,0}{-.6}
        \drawStaticCurvedBHLine{.6,0}{0,.7}{150}{0}
        \drawStaticPerturbativeVertex{.6,0}
        \drawStaticBlob{0,0}
        \drawStaticBlob{0,.7}
        \node[rotate=90] at (0,.35) {\scriptsize ...} ;
        \draw[decorate,decoration={brace,amplitude=2.5pt,raise=1pt}]
            (-.15,.8) -- (-.15,-.1)
            node[midway,left=3pt,inner sep=0] {\scriptsize $n\,$} ;
    \end{tikzpicture}
    =
    \frac{(-1)^n n!\,\kappa}{
        2 m_1^{n-1} (k\cdot v_1+\iZero)^{n+1}
    }
    \Big[
        (k\cdot \Delta p_1)^{n}
        v_1^\mu v_1^\nu
        -
        2 (k\cdot v_1)
        (k\cdot \Delta p_1)^{n-1}
        v_1^{(\mu} \Delta p^{\nu)}_1
    \nn\\
    &\hspace{3.5cm}
        +
        (k\cdot v_1)^2
        (k\cdot \Delta p_1)^{n-2}
        \Delta p_1^\mu \Delta p_1^\nu
    \Big]
    -
    \delta_{n0}
    \frac{
        \kappa
        m_1
        v_1^\mu v_1^\nu
    }{2 (k\cdot v_1-\iZero)}
    \ ,
\end{align}
where the second and third terms in the square brackets are present only for $n\ge1$ and $n\ge2$, respectively.

Summing over $n$, each of the three tensor structures gives rise to its own series, starting at $n=0$, $n=1$ and $n=2$, respectively.
We shift the summation index of the second and third series such that all three start at $n=0$, and cancel the factors of $k\cdot v_1$ against the retarded propagators.
The sum then reads,
\begin{align}
    &\sum_{n=0}^{\infty}
    \
    \frac1{n!}
    \
    \begin{tikzpicture}[baseline={([yshift=-.5ex]current bounding box.center)},yscale=-1]
        \drawDirectedStaticCurvedGravitonLine{.6,0}{1.3,.7}{90}{180}
        \drawDottedLineR{.6,0}{.5}
        \drawStaticBHLineR{.6,0}{-.6}
        \drawStaticCurvedBHLine{.6,0}{0,.7}{150}{0}
        \drawStaticPerturbativeVertex{.6,0}
        \drawStaticBlob{0,0}
        \drawStaticBlob{0,.7}
        \node[rotate=90] at (0,.35) {\scriptsize ...} ;
        \draw[decorate,decoration={brace,amplitude=2.5pt,raise=1pt}]
            (-.15,.8) -- (-.15,-.1)
            node[midway,left=3pt,inner sep=0] {\scriptsize $n\,$} ;
    \end{tikzpicture}
    \nn\\
    &=\sum_{n=0}^\infty
        \frac{(-1)^n \kappa (k\cdot \Delta p_1)^{n}}{
        2 m_1^{n+1} (k\cdot v_1+\iZero)^{n+1}
    }
    \Big(
        m_1^2 v_1^\mu v_1^\nu
        +
        2 m_1 v_1^{(\mu} \Delta p^{\nu)}_1
        +
        \Delta p_1^\mu \Delta p_1^\nu
    \Big)
    -
    \frac{
        \kappa
        m_1
        v_1^\mu v_1^\nu
    }{2 (k\cdot v_1-\iZero)}
    \nn\\
    &
    =
    \frac{
        \kappa
    (m_1v_1+\Delta p_1)^\mu
    (m_1v_1+\Delta p_1)^\nu
    }{2 m_1 (k\cdot v_1+\iZero)}\Big[
    \sum_{n=0}^\infty
        \frac{(-1)^n (k\cdot \Delta p_1)^{n}}{
        m_1^{n}  (k\cdot v_1+\iZero)^{n}
    }\Big]
    -
    \frac{
        \kappa
        m_1
        v_1^\mu v_1^\nu
    }{2 (k\cdot v_1-\iZero)}
    \ .
\end{align}
The three terms in the brackets of the second line correspond to the three terms of \eqn{eq:OneDiagram}, and they factorise into a product in the last line.

The infinite sum is now a geometric series, which we can resum to
\begin{align}\label{eq:softWaveformWL1}
    &\sum_{n=0}^{\infty}
    \
    \frac1{n!}
    \
    \begin{tikzpicture}[baseline={([yshift=-.5ex]current bounding box.center)},yscale=-1]
        \drawDirectedStaticCurvedGravitonLine{.6,0}{1.3,.7}{90}{180}
        \drawDottedLineR{.6,0}{.5}
        \drawStaticBHLineR{.6,0}{-.6}
        \drawStaticCurvedBHLine{.6,0}{0,.7}{150}{0}
        \drawStaticPerturbativeVertex{.6,0}
        \drawStaticBlob{0,0}
        \drawStaticBlob{0,.7}
        \node[rotate=90] at (0,.35) {\scriptsize ...} ;
        \draw[decorate,decoration={brace,amplitude=2.5pt,raise=1pt}]
            (-.15,.8) -- (-.15,-.1)
            node[midway,left=3pt,inner sep=0] {\scriptsize $n\,$} ;
    \end{tikzpicture}
    =
    \frac{
        \kappa
    (m_1v_1+\Delta p_1)^\mu
    (m_1v_1+\Delta p_1)^\nu
    }{2 (k\cdot( m_1v_1+\Delta p_1)+\iZero)}
    -
    \frac{
        \kappa
    m_1v_1^\mu
    m_1v_1^\nu
    }{2 (k\cdot m_1v_1-\iZero)}
    \ .
\end{align}
The impulse insertions thus resum into a retarded term depending only on the final momentum $m_1v_1^\mu+\Delta p_1^\mu$, while the advanced term from $n=0$ depends only on the initial momentum $m_1v_1^\mu$.

Summing the result with its partner by replacing the labellings of WL1 with WL2, the linear memory contribution to the waveform reads
\begin{align}
    \label{eq:fLin}
    \i f_{\rm lm}^\ab
    &=
    4 G
    \mathcal P^\ab_\mn
    \sum_{i=1}^2
    \Delta\bigg[
        \frac{p_{i}^\mu p_{i}^\nu}{
        p_{i}\cdot k-\iZero}
    \bigg]
    \\
    &
    =
    4 G
    \mathcal P^\ab_\mn
    \sum_{i=1}^2
    \bigg[
        \frac{p_{i,\infty}^\mu p_{i,\infty}^\nu}{
        p_{i,\infty}\cdot k+\iZero}
        -
        \frac{p_{i}^\mu p_{i}^\nu}{
        p_{i}\cdot k-\iZero}
    \bigg]
    \ .
    \nn
\end{align}
where $p_{i,\infty}^\mu=p_i^\mu+\Delta p_i^\mu$.
In the first line, we emphasise the simple interpretation as the difference of the final linearised metric from the initial linearised metric.

The result~\eqref{eq:fLin} is precisely Weinberg's soft factor, with the $\iZero$ prescription following here from the retarded propagators. The same steps could be followed for the upper part of the static diagram in \eqn{eq:generalDiagram}, which corresponds to the vertex absorbing the static graviton together with the outgoing deflection and its impulse insertions. Gluing the two resummed parts with the static graviton propagator, one could then compute the linear-memory contribution to $L_{\rm loss}^\mu$ as a function of the initial and final momenta only, and thereby reproduce the all-order result of ref.~\cite{DiVecchia:2022owy}. Since the soft waveform involves only on-shell gravitons, ref.~\cite{DiVecchia:2022owy} moreover shows that static \emph{potential} gravitons do not contribute to $L_{\rm loss}^\mu$ at any PM order, extending our observation at 4PM.

\subsection{Linear memory calculation at $\mathcal{O}(G^4)$ from static correlators}
\label{sec:staticCorrelators}
To obtain the linear memory contributions at 4PM order, we employ the formalism derived in ref.~\cite{Jakobsen:2026ayw}, which can be thought of as gluing two static waveforms from the previous section. As the formalism is described in detail there, we only summarise it briefly here and mainly focus on the features specific to the 4PM calculation.

In the generic static diagram of \eqn{eq:generalDiagram}, the blob to which the static graviton attaches can be broken down into the subdiagrams shown on the right-hand side of \eqn{eq:blobBrokenDown}.
This leaves us with diagrams of the kind

\begin{align}
\begin{tikzpicture}[baseline={(0,.4)}]
    \drawStaticGravitonLine{.6,0}{.6,.9}
    \drawDottedLineR{.6,0}{.5}
    \drawStaticBHLineR{.6,.9}{.5}
    \drawStaticBHLineR{.6,.9}{-.6}
    \drawStaticCurvedBHLine{.6,.9}{0,1.4}{160}{0}
    \drawStaticBHLineR{.6,0}{-.6}
    \drawStaticCurvedBHLine{.6,0}{0,-.5}{200}{0}
    \drawStaticPerturbativeVertex{.6,0}
    \drawStaticPerturbativeVertex{.6,.9}
    \drawStaticBlob{0,.9}
    \drawStaticBlob{0,1.4}
    \drawStaticBlob{0,0}
    \drawStaticBlob{0,-.5}
    \node[rotate=90] at (0,1.15) {\scriptsize ...} ;
    \node[rotate=90] at (0,-.25) {\scriptsize ...} ;
    \draw[decorate,decoration={brace,amplitude=2.5pt,raise=1pt}]
        (-.15,.8) -- (-.15,1.5) node[midway,left=3pt,inner sep=0] {\scriptsize $n\,$} ;
    \draw[decorate,decoration={brace,amplitude=2.5pt,raise=1pt}]
        (-.15,-.6) -- (-.15,.1) node[midway,left=3pt,inner sep=0] {\scriptsize $\sym\,$} ;
\end{tikzpicture}\ ,
\qquad\qquad
\begin{tikzpicture}[baseline={(0,.4)}]
    \drawDottedLine{0,.5}{.8,.5}
    \drawStaticBHLineR{.8,.5}{.5}
    \drawStaticCurvedGravitonLine{0,.5}{.8,.5}{-80}{-100}
    \drawStaticBHLineR{0,.5}{-.6}
    \drawStaticCurvedBHLine{0,.5}{-.6,0}{200}{0}
    \drawStaticCurvedBHLine{.8,.5}{.2,.8}{160}{0}
    \drawStaticCurvedBHLine{.8,.5}{.2,1.3}{120}{0}
    \drawStaticPerturbativeVertex{0,.5}
    \drawStaticPerturbativeVertex{.8,.5}
    \drawStaticBlob{-.6,.5}
    \drawStaticBlob{-.6,0}
    \drawStaticBlob{.2,.8}
    \drawStaticBlob{.2,1.3}
    \node[rotate=90] at (-.6,.25) {\scriptsize ...} ;
    \node[rotate=90] at (.2,1.05) {\scriptsize ...} ;
    \draw[decorate,decoration={brace,amplitude=2.5pt,raise=1pt}]
        (-.75,-.1) -- (-.75,.6) node[midway,left=3pt,inner sep=0] {\scriptsize $\sym\,$} ;
    \draw[decorate,decoration={brace,amplitude=2.5pt,raise=1pt}]
        (.05,.7) -- (.05,1.4) node[midway,left=3pt,inner sep=0] {\scriptsize $n\,$} ;
\end{tikzpicture}\ ,
\end{align}
with the blue skeletons introduced as \emph{static correlators} in ref.~\cite{Jakobsen:2026ayw}.
Each hatched blob now represents the insertion of lower-order impulses.
Whereas at the 2PM and 3PM orders linear impulse insertions suffice, at 4PM we must also consider correlators with multiple impulse insertions. All required static correlators are shown in \fig{fig:ElevenCorrelators} and we remind the reader that the blue colour marks gravitons in the static region and energies scaling with $\omega_{\mathrm{IR}}$.
\usetikzlibrary{patterns}

To obtain the change in the worldline offset at 4PM, we contract the correlators with the lower-order impulse insertions shown in \fig{fig:ElevenCorrelators}, where the number inside each blob gives the PM order of the inserted impulse.
\begin{figure}[t]
  \centering
  \begin{subfigure}[b]{0.15\textwidth}
    \centering
    \begin{tikzpicture}[baseline={(0,.4)}]
        \drawStaticGravitonLine{.6,0}{.6,.9}
        \drawDottedLineR{.6,0}{.5}
        \drawDottedLineR{.6,.9}{-.6}
        \drawStaticBHLineR{.6,.9}{.5}
        \drawStaticBHLineR{.6,0}{-.6}
        \drawStaticPerturbativeVertex{.6,0}
        \drawStaticPerturbativeVertex{.6,.9}
        \drawImpulseBlob{0,0}{3}
    \end{tikzpicture}
    \caption{}
    \label{fig:ElevenCorrelators-a}
  \end{subfigure}
  \hfill
  \begin{subfigure}[b]{0.15\textwidth}
    \centering
    \begin{tikzpicture}[baseline={(0,.4)}]
        \drawDottedLine{0,.5}{.8,.5}
        \drawStaticBHLineR{0,.5}{-.6}
        \drawStaticBHLineR{.8,.5}{.5}
        \drawStaticPerturbativeVertex{0,.5}
        \drawStaticPerturbativeVertex{.8,.5}
        \drawStaticCurvedGravitonLine{0,.5}{.8,.5}{-80}{-100}
        \drawImpulseBlob{-.6,.5}{3}
    \end{tikzpicture}
    \caption{}
    \label{fig:ElevenCorrelators-b}
  \end{subfigure}
  \hfill
  \begin{subfigure}[b]{0.15\textwidth}
    \centering
    \begin{tikzpicture}[baseline={(0,.4)}]
        \drawStaticGravitonLine{.6,0}{.6,.9}
        \drawDottedLineR{.6,0}{.5}
        \drawStaticBHLineR{.6,.9}{-.6}
        \drawStaticBHLineR{.6,.9}{.5}
        \drawStaticBHLineR{.6,0}{-.6}
        \drawStaticPerturbativeVertex{.6,0}
        \drawStaticPerturbativeVertex{.6,.9}
        \drawImpulseBlob{0,0}{2}
        \drawImpulseBlob{0,.9}{1}
    \end{tikzpicture}
    \caption{}
    \label{fig:ElevenCorrelators-c}
  \end{subfigure}
  \hfill
  \begin{subfigure}[b]{0.15\textwidth}
    \centering
    \begin{tikzpicture}[baseline={(0,.4)}]
        \drawStaticGravitonLine{.6,0}{.6,.9}
        \drawDottedLineR{.6,0}{.5}
        \drawDottedLineR{.6,.9}{-.6}
        \drawStaticBHLineR{.6,.9}{.5}
        \drawStaticBHLineR{.6,0}{-.6}
        \drawStaticCurvedBHLine{.6,0}{0,.5}{160}{0}
        \drawStaticPerturbativeVertex{.6,0}
        \drawStaticPerturbativeVertex{.6,.9}
        \drawImpulseBlob{0,0}{2}
        \drawImpulseBlob{0,.5}{1}
    \end{tikzpicture}
    \caption{}
    \label{fig:ElevenCorrelators-d}
  \end{subfigure}
  \hfill
  \begin{subfigure}[b]{0.15\textwidth}
    \centering
    \begin{tikzpicture}[baseline={(0,.4)}]
        \drawDottedLine{0,.5}{.8,.5}
        \drawStaticBHLineR{0,.5}{-.6}
        \drawStaticBHLineR{.8,.5}{.5}
        \drawStaticCurvedBHLine{.8,.5}{.1,.8}{160}{0}
        \drawStaticPerturbativeVertex{0,.5}
        \drawStaticPerturbativeVertex{.8,.5}
        \drawStaticCurvedGravitonLine{0,.5}{.8,.5}{-80}{-100}
        \drawImpulseBlob{-.6,.5}{2}
        \drawImpulseBlob{.1,.8}{1}
    \end{tikzpicture}
    \caption{}
    \label{fig:ElevenCorrelators-e}
  \end{subfigure}
  \hfill
  \begin{subfigure}[b]{0.15\textwidth}
    \centering
    \begin{tikzpicture}[baseline={(0,.4)}]
        \drawDottedLine{0,.5}{.8,.5}
        \drawStaticBHLineR{0,.5}{-.6}
        \drawStaticBHLineR{.8,.5}{.5}
        \drawStaticCurvedBHLine{0,.5}{-.6,1}{160}{0}
        \drawStaticPerturbativeVertex{0,.5}
        \drawStaticPerturbativeVertex{.8,.5}
        \drawStaticCurvedGravitonLine{0,.5}{.8,.5}{-80}{-100}
        \drawImpulseBlob{-.6,.5}{2}
        \drawImpulseBlob{-.6,1}{1}
    \end{tikzpicture}
    \caption{}
    \label{fig:ElevenCorrelators-f}
  \end{subfigure}

  \vspace{1em}

  \begin{subfigure}[b]{0.15\textwidth}
    \centering
    \begin{tikzpicture}[baseline={(0,.4)}]
        \drawStaticGravitonLine{.6,0}{.6,.9}
        \drawDottedLineR{.6,0}{.5}
        \drawStaticBHLineR{.6,.9}{-.6}
        \drawStaticBHLineR{.6,.9}{.5}
        \drawStaticBHLineR{.6,0}{-.6}
        \drawStaticCurvedBHLine{.6,.9}{0,1.4}{160}{0}
        \drawStaticPerturbativeVertex{.6,0}
        \drawStaticPerturbativeVertex{.6,.9}
        \drawImpulseBlob{0,0}{1}
        \drawImpulseBlob{0,.9}{1}
        \drawImpulseBlob{0,1.4}{1}
    \end{tikzpicture}
    \caption{}
    \label{fig:ElevenCorrelators-g}
  \end{subfigure}
  \hfill
  \begin{subfigure}[b]{0.15\textwidth}
    \centering
    \begin{tikzpicture}[baseline={(0,.4)}]
        \drawStaticGravitonLine{.6,0}{.6,.9}
        \drawDottedLineR{.6,0}{.5}
        \drawStaticBHLineR{.6,.9}{-.6}
        \drawStaticBHLineR{.6,.9}{.5}
        \drawStaticBHLineR{.6,0}{-.6}
        \drawStaticCurvedBHLine{.6,0}{0,-.5}{200}{0}
        \drawStaticPerturbativeVertex{.6,0}
        \drawStaticPerturbativeVertex{.6,.9}
        \drawImpulseBlob{0,0}{1}
        \drawImpulseBlob{0,-.5}{1}
        \drawImpulseBlob{0,.9}{1}
    \end{tikzpicture}
    \caption{}
    \label{fig:ElevenCorrelators-h}
  \end{subfigure}
  \hfill
  \begin{subfigure}[b]{0.15\textwidth}
    \centering
    \begin{tikzpicture}[baseline={(0,.4)}]
        \drawStaticGravitonLine{.6,0}{.6,.9}
        \drawDottedLineR{.6,0}{.5}
        \drawDottedLineR{.6,.9}{-.6}
        \drawStaticBHLineR{.6,.9}{.5}
        \drawStaticBHLineR{.6,0}{-.6}
        \drawStaticCurvedBHLine{.6,0}{0,-.5}{200}{0}
        \drawStaticCurvedBHLine{.6,0}{0,.5}{160}{0}
        \drawStaticPerturbativeVertex{.6,0}
        \drawStaticPerturbativeVertex{.6,.9}
        \drawImpulseBlob{0,0}{1}
        \drawImpulseBlob{0,-.5}{1}
        \drawImpulseBlob{0,.5}{1}
    \end{tikzpicture}
    \caption{}
    \label{fig:ElevenCorrelators-i}
  \end{subfigure}
  \hfill
  \begin{subfigure}[b]{0.15\textwidth}
    \centering
    \begin{tikzpicture}[baseline={(0,.4)}]
        \drawDottedLine{0,.5}{.8,.5}
        \drawStaticBHLineR{0,.5}{-.6}
        \drawStaticBHLineR{.8,.5}{.5}
        \drawStaticCurvedBHLine{.8,.5}{.2,.85}{160}{0}
        \drawStaticCurvedBHLine{.8,.5}{.2,1.35}{120}{0}
        \drawStaticPerturbativeVertex{0,.5}
        \drawStaticPerturbativeVertex{.8,.5}
        \drawStaticCurvedGravitonLine{0,.5}{.8,.5}{-80}{-100}
        \drawImpulseBlob{-.6,.5}{1}
        \drawImpulseBlob{.2,.85}{1}
        \drawImpulseBlob{.2,1.35}{1}
    \end{tikzpicture}
    \caption{}
    \label{fig:ElevenCorrelators-j}
  \end{subfigure}
  \hfill
  \begin{subfigure}[b]{0.15\textwidth}
    \centering
    \begin{tikzpicture}[baseline={(0,.4)}]
        \drawDottedLine{0,.5}{.8,.5}
        \drawStaticBHLineR{0,.5}{-.6}
        \drawStaticBHLineR{.8,.5}{.5}
        \drawStaticCurvedBHLine{0,.5}{-.6,1}{160}{0}
        \drawStaticCurvedBHLine{.8,.5}{.3,.85}{160}{0}
        \drawStaticPerturbativeVertex{0,.5}
        \drawStaticPerturbativeVertex{.8,.5}
        \drawStaticCurvedGravitonLine{0,.5}{.8,.5}{-80}{-100}
        \drawImpulseBlob{-.6,.5}{1}
        \drawImpulseBlob{-.6,1}{1}
        \drawImpulseBlob{.3,.85}{1}
    \end{tikzpicture}
    \caption{}
    \label{fig:ElevenCorrelators-k}
  \end{subfigure}
  \hfill
  \begin{subfigure}[b]{0.15\textwidth}
    \centering
    \begin{tikzpicture}[baseline={(0,.4)}]
        \drawDottedLine{0,.5}{.8,.5}
        \drawStaticBHLineR{0,.5}{-.6}
        \drawStaticBHLineR{.8,.5}{.5}
        \drawStaticCurvedBHLine{0,.5}{-.6,0}{200}{0}
        \drawStaticCurvedBHLine{0,.5}{-.6,1}{160}{0}
        \drawStaticPerturbativeVertex{0,.5}
        \drawStaticPerturbativeVertex{.8,.5}
        \drawStaticCurvedGravitonLine{0,.5}{.8,.5}{-80}{-100}
        \drawImpulseBlob{-.6,.5}{1}
        \drawImpulseBlob{-.6,0}{1}
        \drawImpulseBlob{-.6,1}{1}
    \end{tikzpicture}
    \caption{}
    \label{fig:ElevenCorrelators-l}
  \end{subfigure}
  \caption{
    Static correlators contracted with lower-order impulse insertions, contributing to the 4PM offset shift.
    The label inside each blob gives the PM order of the inserted impulse. These represent all contributions up to swapping the 1PM and 2PM insertions if both occur in the same diagram.
  }
  \label{fig:ElevenCorrelators}
\end{figure}
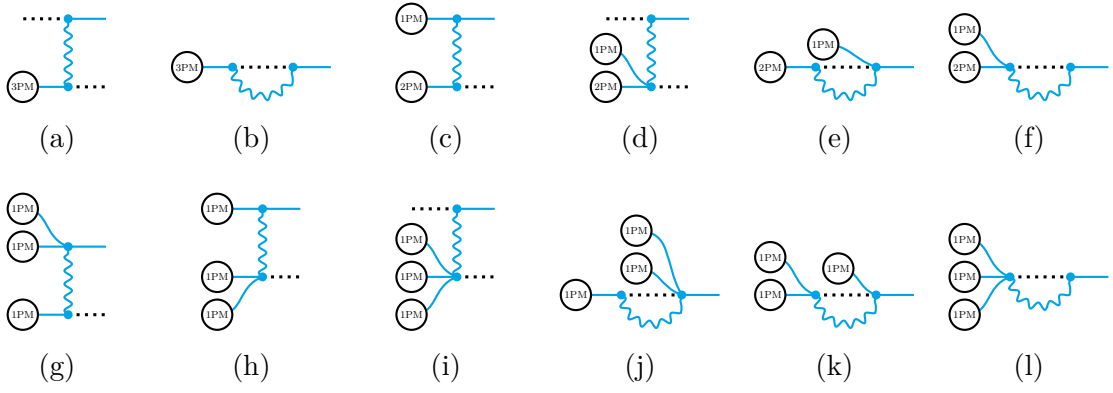
In this construction, we include symmetry factors of the resulting diagrams after impulse insertion (e.g.\ $1/2$ for two identical insertions at the same vertex).
Since we must only consider \emph{radiative} static gravitons, contractions with the potential-region impulse contribute to the PR region of the offset kick and each diagram in \fig{fig:ElevenCorrelators} can contribute there. 
Contrariwise, since the radiative-region impulse starts at 3PM, only contractions with the 3PM impulse yield the RR part of the linear memory.

\begin{figure}[t]
  \centering
  \begin{subfigure}[b]{0.15\textwidth}
    \centering
    \begin{tikzpicture}[baseline={(0,.4)}]
        \drawStaticGravitonLine{.6,0}{.6,.9}
        \drawDottedLineR{.6,0}{.5}
        \drawDottedLineR{.6,.9}{-.6}
        \drawStaticBHLineR{.6,.9}{.5}
        \drawStaticBHLineR{.6,0}{-.6}
        \drawStaticPerturbativeVertex{.6,0}
        \drawStaticPerturbativeVertex{.6,.9}
        \drawImpulseBlob{0,0}{2}
    \end{tikzpicture}
    \caption{}
    \label{fig:LessCorrelators3PM-a}
  \end{subfigure}
  \hfill
  \begin{subfigure}[b]{0.15\textwidth}
    \centering
    \begin{tikzpicture}[baseline={(0,.4)}]
        \drawDottedLine{0,.5}{.8,.5}
        \drawStaticBHLineR{0,.5}{-.6}
        \drawStaticBHLineR{.8,.5}{.5}
        \drawStaticPerturbativeVertex{0,.5}
        \drawStaticPerturbativeVertex{.8,.5}
        \drawStaticCurvedGravitonLine{0,.5}{.8,.5}{-80}{-100}
        \drawImpulseBlob{-.6,.5}{2}
    \end{tikzpicture}
    \caption{}
    \label{fig:LessCorrelators3PM-b}
  \end{subfigure}
  \hfill
  \begin{subfigure}[b]{0.15\textwidth}
    \centering
    \begin{tikzpicture}[baseline={(0,.4)}]
        \drawStaticGravitonLine{.6,0}{.6,.9}
        \drawDottedLineR{.6,0}{.5}
        \drawStaticBHLineR{.6,.9}{-.6}
        \drawStaticBHLineR{.6,.9}{.5}
        \drawStaticBHLineR{.6,0}{-.6}
        \drawStaticPerturbativeVertex{.6,0}
        \drawStaticPerturbativeVertex{.6,.9}
        \drawImpulseBlob{0,0}{1}
        \drawImpulseBlob{0,.9}{1}
    \end{tikzpicture}
    \caption{}
    \label{fig:LessCorrelators3PM-c}
  \end{subfigure}
  \hfill
  \begin{subfigure}[b]{0.15\textwidth}
    \centering
    \begin{tikzpicture}[baseline={(0,.4)}]
        \drawStaticGravitonLine{.6,0}{.6,.9}
        \drawDottedLineR{.6,0}{.5}
        \drawDottedLineR{.6,.9}{-.6}
        \drawStaticBHLineR{.6,.9}{.5}
        \drawStaticBHLineR{.6,0}{-.6}
        \drawStaticCurvedBHLine{.6,0}{0,.5}{160}{0}
        \drawStaticPerturbativeVertex{.6,0}
        \drawStaticPerturbativeVertex{.6,.9}
        \drawImpulseBlob{0,0}{1}
        \drawImpulseBlob{0,.5}{1}
    \end{tikzpicture}
    \caption{}
    \label{fig:LessCorrelators3PM-d}
  \end{subfigure}
  \hfill
  \begin{subfigure}[b]{0.15\textwidth}
    \centering
    \begin{tikzpicture}[baseline={(0,.4)}]
        \drawDottedLine{0,.5}{.8,.5}
        \drawStaticBHLineR{0,.5}{-.6}
        \drawStaticBHLineR{.8,.5}{.5}
        \drawStaticCurvedBHLine{.8,.5}{.1,.8}{160}{0}
        \drawStaticPerturbativeVertex{0,.5}
        \drawStaticPerturbativeVertex{.8,.5}
        \drawStaticCurvedGravitonLine{0,.5}{.8,.5}{-80}{-100}
        \drawImpulseBlob{-.6,.5}{1}
        \drawImpulseBlob{.1,.8}{1}
    \end{tikzpicture}
    \caption{}
    \label{fig:LessCorrelators3PM-e}
  \end{subfigure}
  \hfill
  \begin{subfigure}[b]{0.15\textwidth}
    \centering
    \begin{tikzpicture}[baseline={(0,.4)}]
        \drawDottedLine{0,.5}{.8,.5}
        \drawStaticBHLineR{0,.5}{-.6}
        \drawStaticBHLineR{.8,.5}{.5}
        \drawStaticCurvedBHLine{0,.5}{-.6,1}{160}{0}
        \drawStaticPerturbativeVertex{0,.5}
        \drawStaticPerturbativeVertex{.8,.5}
        \drawStaticCurvedGravitonLine{0,.5}{.8,.5}{-80}{-100}
        \drawImpulseBlob{-.6,.5}{1}
        \drawImpulseBlob{-.6,1}{1}
    \end{tikzpicture}
    \caption{}
    \label{fig:LessCorrelators3PM-f}
  \end{subfigure}
  \caption{
    Static correlators contracted with lower-order impulse insertions, yielding the velocity components of the 3PM offset shift.
    The label inside each blob gives the PM order of the inserted impulse.
  }
  \label{fig:LessCorrelators3PM}
\end{figure}
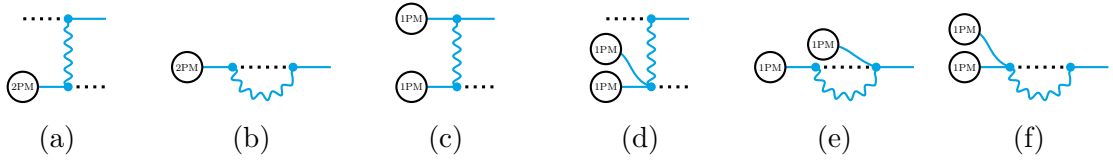

Finally, we obtain the result below, where we split the expression into its $\mathrm{RR}$ region and $\mathrm{R}$ region, respectively.
\begin{align}
    \hat b_\mu\Delta^{\!\rm(4PM)}_{\rm stat} b^{\mu}=\frac{G^4M^4\nu}{|b|^3}\Big[\hat b_\mu\Delta^{\!\rm(4PM)}_{\rm stat,RR}b^\mu+\hat b_\mu\Delta^{\!\rm(4PM)}_{\rm stat,R} b^\mu\Big]\, .
\end{align}
The $\mathrm{RR}$ region can be cast in the very compact form
\begin{align}
    \hat b_\mu\Delta^{\!\rm(4PM)}_{\rm stat,RR} b^\mu=-\frac{2\nu(2\gamma^2-1)^2}{(\gamma^2-1)}\mathcal{I}(\gamma)^2\, .
\end{align}
In the $\mathrm{PR}$ region the $4\mathrm{PM}$ impact parameter kick is given by a 1SF contribution,
\begin{align}
    \hat b_\mu\Delta^{\!\rm(4PM,1SF)}_{\rm stat,R} b^\mu&=\frac{8(8-349\gamma^2+1260\gamma^4-1884\gamma^6+1304\gamma^8-264\gamma^{10})}{15(\gamma^2-1)^{7/2}}\\
    &\qquad-\frac{8\gamma\,(-3+2\gamma^2+10\gamma^4-20\gamma^6+16\gamma^8)}{(\gamma^2-1)^{4}}\mathrm{arccosh}(\gamma)\, ,&&
    \nonumber
\end{align}
and a \oneSFBar{} contribution,
\begin{flalign}
    \hat b_\mu\Delta^{\!\rm(4PM,\oneSFBar{})}_{\rm stat,R} b^\mu&=\frac{8\nu}{45(\gamma^2-1)^{7/2}}\Bigl(-48+1670\gamma+2094\gamma^2-3245\gamma^3-7560\gamma^4\\*
&\qquad\qquad+120\gamma^5+11304\gamma^6+2710\gamma^7-7824\gamma^8-940\gamma^9+1584\gamma^{10}\Bigr)\notag\\*
&\hspace{.3cm}-\frac{8\nu}{3(\gamma^2-1)^{4}}\Bigl(30+18\gamma+333\gamma^2-12\gamma^3-1142\gamma^4-60\gamma^5\notag\\*
&\qquad\qquad+1224\gamma^6+120\gamma^7-504\gamma^8-96\gamma^9+80\gamma^{10}\Bigr)\operatorname{arccosh}(\gamma)\notag\\*
&\hspace{.3cm}+\frac{16\nu\gamma(2\gamma^2-3)(4\gamma^4-12\gamma^2-3)}{(\gamma^2-1)^{5/2}}\operatorname{arccosh}^2(\gamma)
\, .
\notag
\end{flalign}
To capture all constituents of the static angular momentum flux at 4PM, we need to consider not only contractions with correlators that yield the 4PM offset shift, since terms which could be discarded at lower orders may now become relevant.
Contrary to the calculation at 3PM, pieces of the impact parameter proportional to the velocities now have to be taken into account, as they may be combined with the component of an impulse along the $b$-direction. They are calculated from the diagrams in \fig{fig:LessCorrelators3PM}.


The required velocity components of the 3PM offset parameter kick read
\begin{subequations}
\label{eq:vel-3pm-off-kick}
\begin{align}
    \check v^1_\mu\Delta^{\!\rm(3PM)}_{\rm stat} b^{\mu}
    &=
    \frac{
        -
        (2\gamma^2-1)^2
        G^3M^2\nu 
    }{
        (2\gamma^2-3)
        (\gamma^2-1)^2
        \gamma |b|^2
    }
    \bigg[
        24
        (2-3\gamma^2+\gamma^4)
        (m_1+\gamma m_2)
        \\
        &\hspace{4.8cm}+
        \big(
            (9-6\gamma^2+4\gamma^4)m_1
            +
            (3+4\gamma^4)m_2
        \big)\mathcal{I}(\gamma)
    \bigg]
    \, ,
    \notag\\
    \check v^2_\mu\Delta^{\!\rm(3PM)}_{\rm stat} b^{\mu}
    &=
    \frac{
        (2\gamma^2-1)^2
        G^3M^2\nu 
    }{
        (2\gamma^2-3)
        (\gamma^2-1)^2
        \gamma |b|^2
    }
    \bigg[
        24
        (2-3\gamma^2+\gamma^4)
        (m_2+\gamma m_1)
        \\
        &\hspace{4.8cm}+
        \big(
            (9-6\gamma^2+4\gamma^4)m_2
            +
            (3+4\gamma^4)m_1
        \big)\mathcal{I}(\gamma)
    \bigg]
    \, ,
    \notag
\end{align}
\end{subequations}
where we used the dual velocities,
\begin{align}
    \label{eq:dual-vels}
    \check v^{1\mu}
    =
    \frac{
        \gamma v_2^\mu-v_1^\mu
    }{\gamma^2-1}
    \, ,
    \qquad
    \check v^{2\mu}
    =
    \frac{
        \gamma v_1^\mu-v_2^\mu
    }{\gamma^2-1}
    \, ,
\end{align}
to pick out the components along $v_1^\mu$ and $v_2^\mu$.
Using \eqns{eq:ELdecomposedRR}{eq:ELdecomposedR}, as well as contributions involving $\Delta^{\!\rm(2PM)}_{\rm stat} b^\mu$ from \cite{Jakobsen:2026ayw}, we immediately obtain the static contribution to $(E|L|)_{\rm loss}$ at $\mathcal{O}(G^4)$, which we provide in \secref{section:Results}.

\subsection{Nonlinear gravitational memory}
\label{sec:nonlinearmemory}
In the linear memory, the soft graviton is emitted directly from one of the worldlines, and the memory is sourced by the change in the black hole momenta.
The soft graviton may, however, also be emitted from the gravitational field itself via a three-graviton vertex. This gives rise to the nonlinear memory.
From a classical perspective, this is because the gravitational waves emitted during the scattering source part of the memory themselves.
In contrast to the soft graviton, these emitted gravitons are ``hard'', meaning that their momenta scale with $|q|$, as in the dynamical region of \secref{sec:WEFTWQFT}.
Since there is no incoming radiation, this source is only present at late times, and the nonlinear memory therefore contributes only to $A_{(+)}^\mn$ in \eqn{eq:nonlinmem}, which we will derive in the following.

For this purpose, we analyse the diagram in which the soft graviton with momentum $k^\mu$ is emitted from a three-graviton vertex connected to two graviton one-point functions,
\begin{align}\label{eq:nonlinearmem}
    f_{{\rm nl},\,\mn}(k)
    \;
    =
    \;
    -\frac{\i\kappa}{4\pi}
    (-\i k^2)
    \;
    \frac12
    \;\,
    \begin{tikzpicture}[baseline={(0,0)}]
        \coordinate (cR) at (+1,0);
        \coordinate (cL) at (-1,0);
        \coordinate (cM) at (0,0);
        \coordinate (cU) at (0,.8);
        \drawGravitonLine{cL}{cM}
        \drawGravitonLine{cR}{cM}
        \drawDirectedGravitonLine{cM}{cU}
        \draw[line width=.8pt,fill=white] (cL) circle (.23cm) ;
        \draw[pattern=north east lines] (cL) circle (.23cm) ;
        \draw[line width=.8pt,fill=white] (cR) circle (.23cm) ;
        \draw[pattern=north east lines] (cR) circle (.23cm) ;
        \drawPerturbativeVertex{cM}
        \drawPerturbativeVertex{cU}
        \node[below right, font=\footnotesize] at (cL) {$\ \ \ell_1$} ;
        \node[below left, font=\footnotesize] at (cR) {$\ell_2\ \ $} ;
        \node[below right, font=\footnotesize] at (cU) {$k$} ;
        \node[above=-.05cm, font=\scriptsize] at (cU) {$\mu\nu$} ;
    \end{tikzpicture}
    \ .
\end{align}
The factor of $(-\i k^2)$ amputates the diagram, and we include the symmetry factor $1/2$.
The two hatched blobs in this diagram are the full one-point graviton fields, and the diagram above therefore has contributions at any order in $G$.
Later, for contributions relevant to $\Delta L^\mu$ at 4PM, these two hatched blobs will be specialised to their leading-order $\sim\kappa^3$ contributions. In principle, the soft graviton may also be emitted from higher-point graviton vertices with additional blobs attached. However, these do not contribute at 4PM, and, as we will see, the diagram above is sufficient to reproduce the known nonlinear memory.
All momentum labellings follow the causality flow and we therefore have,
\begin{align}
    \ell^\mu_1+\ell^\mu_2=k^\mu
    \ .
\end{align}

We amputate each graviton insertion as,
\begin{align}
    \begin{tikzpicture}[baseline={(0,-.1)}]
        \coordinate (cL) at (-1,0);
        \coordinate (cM) at (0,0);
        \drawDirectedGravitonLine{cL}{cM}
        \draw[line width=.8pt,fill=white] (cL) circle (.23cm) ;
        \draw[pattern=north east lines] (cL) circle (.23cm) ;
        \node[below right, font=\footnotesize] at (cL) {$\ \ \ell$} ;
        \node[above right=-.1cm and -0.1cm, font=\scriptsize] at (cM) {$\mn$} ;
        \drawPerturbativeVertex{cM}
    \end{tikzpicture}
    =
    -\frac{4\pi}{\kappa}\frac1{\ell^2}
    f_\mn(\ell)
    \, ,
\end{align}
For on-shell $\ell^\mu$, the insertion $f_\mn(\ell)$ is then the waveform (and as we will see, $\ell^\mu$ will become on-shell).

We find the integrand,
\begin{align}
    f_{\rm nl}^\mn(k)
    =
    \frac{2\pi}{\kappa}
    \int_{\ell_1\ell_2}
    \frac{
        V^{\mn\ab\rs}_{\rm GR}(k,-\ell_1,-\ell_2)
        f_\ab(\ell_1)
        f_\rs(\ell_2)
        }{
            (\ell_1^2+\iZero \ell_1\!\cdot\! v) (\ell_2^2+\iZero \ell_2\!\cdot\! v)
        }
        \dd^{(D)}(k-\ell_1-\ell_2)
    \, .
\end{align}
Here, $\i V^{\mn\ab\rs}_{\rm GR}(k,\ell_1,\ell_2)$ is the three-graviton vertex with all momenta outgoing.
Further, we explicitly use retarded propagators clearly indicated by the $\iZero$ prescription multiplied by the sign of the time component of the loop momenta.
In principle, we should also insert the de Donder numerator $\mathcal P^\mn_\ab$, but, as we will see, $f^\mn_{\rm nl}(k)$ is traceless at leading order in $\omega\to0$, so this insertion is immaterial.

We are interested in the limit where $k^0=\omega\to0$.
For this purpose, it is essential to employ a partial-fraction identity (also employed in ref.~\cite{Jakobsen:2021lvp}),
\begin{align}
    \frac1{\ell_1^2 \ell_2^2}
    =
    -
    \frac1{2}
    \sum_{i=1}^2
    \frac1{k \cdot \ell_i \, \ell_i^2}
    \ .
\end{align}
This identity holds true because we impose on-shellness of $k^\mu$ and momentum conservation. Both taken together imply $\ell_1^2-\ell_2^2=2k\cdot\ell_1=-2k\cdot\ell_2$.
Naturally, one must be careful with $\iZero$ prescriptions in this identity.
The linear propagator arises from the difference of the loop momenta,
\begin{align}
    (\ell_1^2+\iZero\, \ell_1\cdot v)-(\ell_2^2+\iZero\, \ell_2\cdot v)
    &=
    2 k\cdot \ell_1
    +\iZero v\cdot(\ell_1-\ell_2)
    \\
    &=
    -2 k\cdot \ell_2
    -\iZero v\cdot(\ell_2-\ell_1)
    \nn
    \ .
\end{align}
This informs us of the correct $\iZero$ prescription of the linear propagator.
The quadratic ones all have a retarded prescription.

The nonlinear memory now splits itself into two terms:
\begin{align}
    f_{\rm nl}^\mn(k)
    =
    -\frac{\pi
    }{\kappa}
    \sum_{i=1}^2
    \int_{\ell_1\ell_2}
    \frac{
        V^{\mn\ab\rs}_{\rm GR}(k,-\ell_1,-\ell_2)
        f_\ab(\ell_1)
        f_\rs(\ell_2)
        }{
            k\cdot \ell_i \,\ell_i^2
        }
        \dd^{(D)}(k-\ell_1-\ell_2)
    \ .
\end{align}
We omit explicit $\iZero$ prescriptions.
In the first term $i=1$, it is natural to integrate out $\ell_2^\mu$ using the delta function, and in the second term it is natural to integrate out $\ell_1^\mu$.
At the same time, we want to take the $\omega\to0$ limit, which we may attempt by simply expanding the integrand.
Strictly speaking, this corresponds to the hard region, $\ell_i^\mu\sim|q|$, in a method-of-regions expansion, and soft regions with $\ell_i^\mu\sim\omega$ could in principle also contribute. As we will see, however, the hard region alone is sufficient to reproduce the known all-order nonlinear memory.
In the limit $\omega\to0$, the delta constraint simply reads $\ell_1^\mu+\ell_2^\mu=0$, and both terms are equal.
Also, we see explicitly that our contribution scales as the expected $\omega^{-1}$ for memory effects (since $k^\mu\sim k^0=\omega$).

At leading order in the $\omega\to0$ limit, we then find:
\begin{align}
    f_{\rm nlm}^\mn(k)
    =
    -\frac{2\pi
    }{\kappa}
    \int_{\ell}
    \frac{
        V^{\mn\ab\rs}_{\rm GR}(0,\ell,-\ell)
        f_\ab(\ell)
        f_\rs(-\ell)
        }{
            (k\cdot \ell+\iZero v\cdot \ell) \,(\ell^2 +\iZero v\cdot \ell)
        }
    \ .
\end{align}
Here, we have labelled the leading-order contribution in $\omega$ as (nonlinear) memory, i.e.\ ``nlm''.
The numerator of this expression is even in $\ell$.
In the denominator, the first factor is odd in $\ell$.
We may therefore pick out the odd part of the last remaining quadratic retarded propagator:
\begin{align}
    \frac1{\ell^2+\iZero v\cdot \ell}
    \to
    -\frac{\i}2 \,\text{sign}(v\cdot \ell)
    \dd(\ell^2)
    \ .
\end{align}
This forces $\ell^\mu$ to be on-shell and the three-graviton vertex simplifies drastically. Further, the graviton insertions also localise on their on-shell contributions, which obey the de Donder condition when contracted with $\ell^\mu$.
We find
\begin{align}
    f_{\rm nlm}^\mn(k)
    &=
    \frac{\pi\i
    }{\kappa}
    \int_{\ell}
    \text{sign}(v\cdot \ell)
    \dd(\ell^2)
    \frac{
        V^{\mn\ab\rs}_{\rm GR}(0,\ell,-\ell)
        f_\ab(\ell)
        f_\rs(-\ell)
        }{
            k\cdot \ell+\iZero v\cdot \ell
        }\\
        &=
    -2\pi\i
    \int_{\ell}
    \theta(v\cdot \ell)
    \dd(\ell^2)
    f^{\rs}(\ell)
    \mathcal P^{-1}_{\rs\ab}
    f^{*\ab}(\ell)
    \frac{
        \ell^\mu \ell^\nu
        }{
            k\cdot \ell+\iZero
        }
    \ .\label{eq:nonlinearMemFinal}
\end{align}
Here, we also specialised to the positive-frequency phase space using the symmetry under $\ell\to-\ell$, and used that $f_\mn(-\ell)=f_\mn^*(\ell)$.
Further, we use $\mathcal P^{-1}_{\ab\mn}=\eta_{\mu(\alpha}\eta_{\beta)\nu}-\frac12 \eta_\mn\eta_\ab$ which is the inverse operator to $\mathcal P^{\mn\ab}$ (and incidentally also its four-dimensional limit).
Identifying $k^\mu=\omega n^\mu$ we reach agreement with the nonlinear memory part in eqs. (4) through (6)
in \cite{Georgoudis:2025vkk}. We can identify 
\begin{align}
    \rho(\ell)
    =
    \frac{\pi}{2G} \theta(v\cdot \ell)
    \dd(\ell^2)
    f^{\rs}(\ell)
    \mathcal P^{-1}_{\rs\ab}
    f^{*\ab}(\ell)\ ,
\end{align}
with the graviton emission densities of \cite{Heissenberg:2024umh, Georgoudis:2025vkk}, and may also write
\begin{align}
\i f_{\rm nlm}^\mn(k)
    &=
    4G
    \int_{\ell}
    \rho(\ell)
    \frac{
        \ell^\mu \ell^\nu
        }{
            k\cdot \ell+\iZero
        }\ .
\end{align}
Specifically, $\rho(\ell)$ is the density of gravitons where, e.g., the total radiation of four momentum is given as~\cite{Damour:2020tta,Jakobsen:2021smu},
\begin{align}
    P^\mu_{\rm loss}
    =
    \int_\ell\rho(\ell)
    \ell^\mu
    \ ,
\end{align}
such that, also,
\begin{align}
    \i f^\mn_{\rm nlm}(k) k_\nu
    =
    4 G P_{\rm loss}^\mu
    \ .
\end{align}
Notice that this is essential for the gravitational memory in its entirety obeying the de Donder gauge.
Our expression for the nonlinear memory agrees with those given in refs.~\cite{Heissenberg:2024umh,Georgoudis:2025vkk}.

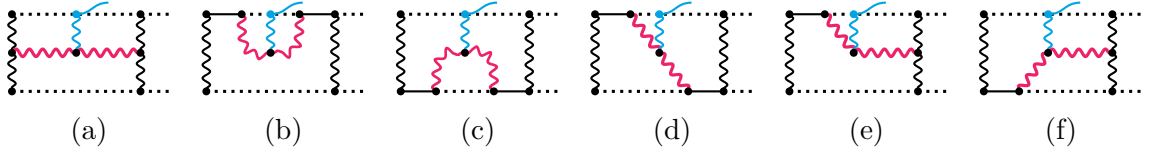
\begin{figure}[t]
    \centering
    \begin{subfigure}[b]{0.15\textwidth}
        \centering
        \begin{tikzpicture}[scale=.85]
            \coordinate (t1) at (0,1.2); \coordinate (t2) at (1,1.2); \coordinate (t3) at (2,1.2);
            \coordinate (m1) at (0,.6);  \coordinate (m2) at (1,.6);  \coordinate (m3) at (2,.6);
            \coordinate (b1) at (0,0);   \coordinate (b2) at (2,0);
            \coordinate (out) at (1.5,1.38);
            \drawDottedLine{t1}{t3} \drawDottedLineR{t3}{.45}
            \drawDottedLine{b1}{b2} \drawDottedLineR{b2}{.45}
            \drawGravitonLine{t1}{m1} \drawGravitonLine{m1}{b1}
            \drawGravitonLine{t3}{m3} \drawGravitonLine{m3}{b2}
            \drawRadiativeGravitonLine{m1}{m2} \drawRadiativeGravitonLine{m2}{m3}
            \drawStaticGravitonLine{t2}{m2}
            \drawStaticCurvedBHLine{t2}{out}{0}{180}
            \foreach \v in {t1,t3,m1,m2,m3,b1,b2} {\drawPerturbativeVertex{\v}}
            \drawStaticPerturbativeVertex{t2}
        \end{tikzpicture}
        \caption{}
        \label{fig:wqft-a}
    \end{subfigure}
    \hfill
    \begin{subfigure}[b]{0.15\textwidth}
        \centering
        \begin{tikzpicture}[scale=.85]
            \coordinate (t1) at (0,1.2);   \coordinate (t2) at (.55,1.2); \coordinate (t3) at (1,1.2);
            \coordinate (t4) at (1.45,1.2);\coordinate (t5) at (2,1.2);
            \coordinate (m1) at (1,.6);
            \coordinate (b1) at (0,0);     \coordinate (b2) at (2,0);
            \coordinate (out) at (1.5,1.38);
            \drawDottedLine{t2}{t4} \drawDottedLineR{t5}{.45}
            \drawDottedLine{b1}{b2} \drawDottedLineR{b2}{.45}
            \drawBHLine{t1}{t2} \drawBHLine{t4}{t5}
            \drawGravitonLine{t1}{b1} \drawGravitonLine{t5}{b2}
            \drawRadiativeCurvedGravitonLine{t2}{m1}{-90}{180}
            \drawRadiativeCurvedGravitonLine{t4}{m1}{-90}{0}
            \drawStaticGravitonLine{t3}{m1}
            \drawStaticCurvedBHLine{t3}{out}{0}{180}
            \foreach \v in {t1,t2,t4,t5,m1,b1,b2} {\drawPerturbativeVertex{\v}}
            \drawStaticPerturbativeVertex{t3}
        \end{tikzpicture}
        \caption{}
        \label{fig:wqft-b}
    \end{subfigure}
    \hfill
    \begin{subfigure}[b]{0.15\textwidth}
        \centering
        \begin{tikzpicture}[scale=.85]
            \coordinate (t1) at (0,1.2);   \coordinate (t2) at (1,1.2); \coordinate (t3) at (2,1.2);
            \coordinate (m1) at (1,.6);
            \coordinate (b1) at (0,0);     \coordinate (b2) at (.55,0);
            \coordinate (b3) at (1.45,0);  \coordinate (b4) at (2,0);
            \coordinate (out) at (1.5,1.38);
            \drawDottedLine{t1}{t3} \drawDottedLineR{t3}{.45}
            \drawDottedLine{b2}{b3} \drawDottedLineR{b4}{.45}
            \drawBHLine{b1}{b2} \drawBHLine{b3}{b4}
            \drawGravitonLine{t1}{b1} \drawGravitonLine{t3}{b4}
            \drawRadiativeCurvedGravitonLine{b2}{m1}{90}{180}
            \drawRadiativeCurvedGravitonLine{b3}{m1}{90}{0}
            \drawStaticGravitonLine{t2}{m1}
            \drawStaticCurvedBHLine{t2}{out}{0}{180}
            \foreach \v in {t1,t3,m1,b1,b2,b3,b4} {\drawPerturbativeVertex{\v}}
            \drawStaticPerturbativeVertex{t2}
        \end{tikzpicture}
        \caption{}
        \label{fig:wqft-c}
    \end{subfigure}
    \hfill
    \begin{subfigure}[b]{0.15\textwidth}
        \centering
        \begin{tikzpicture}[scale=.85]
            \coordinate (t1) at (0,1.2);   \coordinate (t2) at (.55,1.2);
            \coordinate (t3) at (1,1.2);   \coordinate (t4) at (2,1.2);
            \coordinate (m1) at (1,.6);
            \coordinate (b1) at (0,0);     \coordinate (b2) at (1.45,0);  \coordinate (b3) at (2,0);
            \coordinate (out) at (1.5,1.38);
            \drawDottedLine{t2}{t4} \drawDottedLineR{t4}{.45}
            \drawDottedLine{b1}{b2} \drawDottedLineR{b3}{.45}
            \drawBHLine{t1}{t2} \drawBHLine{b2}{b3}
            \drawGravitonLine{t1}{b1} \drawGravitonLine{t4}{b3}
            \drawRadiativeGravitonLine{t2}{m1} \drawRadiativeGravitonLine{m1}{b2}
            \drawStaticGravitonLine{t3}{m1}
            \drawStaticCurvedBHLine{t3}{out}{0}{180}
            \foreach \v in {t1,t2,t4,m1,b1,b2,b3} {\drawPerturbativeVertex{\v}}
            \drawStaticPerturbativeVertex{t3}
        \end{tikzpicture}
        \caption{}
        \label{fig:wqft-d}
    \end{subfigure}
    \hfill
    \begin{subfigure}[b]{0.15\textwidth}
        \centering
        \begin{tikzpicture}[scale=.85]
            \coordinate (t1) at (0,1.2);   \coordinate (t2) at (.55,1.2);
            \coordinate (t3) at (1,1.2);   \coordinate (t4) at (2,1.2);
            \coordinate (m1) at (1,.6);    \coordinate (m2) at (2,.6);
            \coordinate (b1) at (0,0);     \coordinate (b2) at (2,0);
            \coordinate (out) at (1.5,1.38);
            \drawDottedLine{t2}{t4} \drawDottedLineR{t4}{.45}
            \drawDottedLine{b1}{b2} \drawDottedLineR{b2}{.45}
            \drawBHLine{t1}{t2}
            \drawGravitonLine{t1}{b1}
            \drawRadiativeGravitonLine{t2}{m1} \drawRadiativeGravitonLine{m1}{m2}
            \drawStaticGravitonLine{t3}{m1}
            \drawGravitonLine{t4}{m2} \drawGravitonLine{m2}{b2}
            \drawStaticCurvedBHLine{t3}{out}{0}{180}
            \foreach \v in {t1,t2,t4,m1,m2,b1,b2} {\drawPerturbativeVertex{\v}}
            \drawStaticPerturbativeVertex{t3}
        \end{tikzpicture}
        \caption{}
        \label{fig:wqft-e}
    \end{subfigure}
    \hfill
    \begin{subfigure}[b]{0.15\textwidth}
        \centering
        \begin{tikzpicture}[scale=.85]
            \coordinate (t1) at (0,1.2);   \coordinate (t2) at (1,1.2);  \coordinate (t3) at (2,1.2);
            \coordinate (m1) at (1,.6);    \coordinate (m2) at (2,.6);
            \coordinate (b1) at (0,0);     \coordinate (b2) at (.55,0);  \coordinate (b3) at (2,0);
            \coordinate (out) at (1.5,1.38);
            \drawDottedLine{t1}{t3} \drawDottedLineR{t3}{.45}
            \drawDottedLine{b2}{b3} \drawDottedLineR{b3}{.45}
            \drawBHLine{b1}{b2}
            \drawGravitonLine{t1}{b1}
            \drawRadiativeGravitonLine{b2}{m1} \drawRadiativeGravitonLine{m1}{m2}
            \drawStaticGravitonLine{t2}{m1}
            \drawGravitonLine{t3}{m2} \drawGravitonLine{m2}{b3}
            \drawStaticCurvedBHLine{t2}{out}{0}{180}
            \foreach \v in {t1,t3,m1,m2,b1,b2,b3} {\drawPerturbativeVertex{\v}}
            \drawStaticPerturbativeVertex{t2}
        \end{tikzpicture}
        \caption{}
        \label{fig:wqft-f}
    \end{subfigure}
    \caption{
        The six diagrams representing nonlinear-memory contributions to the changes in the worldline offsets.
        Ignoring the static graviton insertion, these diagrams are the six 3PM vacuum diagrams with an active graviton.
        The first three graphs have a symmetry factor $1/2$; the last three have none.
        As it turns out, these diagrams evaluate to a nonzero contribution only in the velocity direction, which means they do not contribute to $\Delta L^\mu$ at 4PM.
    }
    \label{fig:wqft-diagrams}
\end{figure}

\subsection{Vanishing of nonlinear memory contributions at $\mathcal{O}(G^4)$}
The nonlinear memory contributions to the change in worldline offsets at 4PM order come from diagrams with the leading-order waveform inserted for $\langle h_\mn \rangle$ in \eqn{eq:nonlinearMemFinal}.
Pictorially, the nonlinear memory contributions to the change in the worldline offsets are given by the following diagram,
\begin{align}\label{eq:connonlin}
    \begin{tikzpicture}[baseline={(0,0)}]
        \coordinate (cR) at (+1,0);
        \coordinate (cL) at (-1,0);
        \coordinate (cM) at (0,0);
        \coordinate (cU) at (0,.8);
        \drawRadiativeGravitonLine{cL}{cM}
        \drawRadiativeGravitonLine{cR}{cM}
        \drawStaticGravitonLine{cM}{cU}
        \drawStaticDirectedBHLineR{cU}{.5}
        \drawStaticPerturbativeVertex{.5,.8}
        \drawDottedLineR{cU}{-.5}
        \draw[line width=.8pt,fill=white] (cL) circle (.23cm) ;
        \draw[pattern=north east lines] (cL) circle (.23cm) ;
        \draw[line width=.8pt,fill=white] (cR) circle (.23cm) ;
        \draw[pattern=north east lines] (cR) circle (.23cm) ;
        \drawPerturbativeVertex{cM}
        \drawStaticPerturbativeVertex{cU}
        \node[below right, font=\footnotesize] at (cL) {$\ \ \ell_1$} ;
        \node[below right=-.05cm and -.2cm, font=\footnotesize] at (cU) {$\ \ \omega_{\rm IR}$} ;
        \node[above right=-.05cm and 0cm, font=\footnotesize] at (cU) {$\ \ \sigma$} ;
        \node[below left, font=\footnotesize] at (cR) {$\ell_2\ \ $} ;
        \node[left, font=\footnotesize] at ($(cM)!.5!(cU)$) {$k$} ;
    \end{tikzpicture}
    =
    -\i\frac1{\omega_{\rm IR}^{1+2\eps}}
    \Delta_{\rm nlm} b_{1}^\sigma
    +
    \dots
    \ .
\end{align}
where at 4PM, the waveform insertions are at leading $\mathcal{O}(\kappa^3)$ order and the dots denote subleading order in $\omega_{\rm IR}$.
Inserting the waveform gives rise to the six diagrams shown in \fig{fig:wqft-diagrams}.
To compute the corresponding contribution to $\Delta L^\mu$, we need only the $\hat b^\mu$ direction of these diagrams.
This is true because the nonlinear contributions to $\Delta b^\mu$ start at 4PM order and to obtain the change of $L^\mu$ the velocity components are projected out in the formula,
\begin{align}\label{eq:ld}
   \Delta (E L)^\mu
    =
    -\eps^{\mu}_{\ \nu\ab}\Delta(
     (b_2^\nu-b_1^\nu) p_1^\alpha p_2^\beta)
    \ .
\end{align}

The contribution to $\Delta b_{1}^\mu$ from \eqn{eq:connonlin} reads,
\begin{align}\label{eq:DeltaNlmDef}
    \Delta_{\rm nlm} b^\sigma_1
    =
    -\frac{4\pi}{\kappa\, m_1}
    \lim_{\omega_{\rm IR}\to0}
    \omega_{\rm IR}^{-(1-2\eps)}
    \int_k
    f_{\rm nlm}^\mn(k)
    \frac{\i\mathcal P_{\mn\ab}}{k^2+\iZero k\cdot v_1}
    V^{\ab}{}_\sigma(-k;\omega_{\rm IR})
    \ ,
\end{align}
with static $k^\mu\sim\omega_{\rm IR}$ and the worldline interaction $V^{\ab}{}_\sigma(-k;\omega_{\rm IR})$ from eq.~\eqref{eq:hejFarvel}. 
Insertion of the Feynman vertex results in
\begin{align}\label{eq:fjfjfjfjf}
    \Delta_{\rm nlm} b^\sigma_1
    =
    16\pi \i\, G
    \int_{\tilde k\ell}
    \rho(\ell)\,
    \dd(\tilde k\cdot v_1-1)\,(\ell\cdot v_1)
    \frac{
        \ell^\sigma-\tfrac12(\ell\cdot v_1)\,\tilde k^\sigma
    }{
        (\tilde k\cdot \ell+\iZero)
        (\tilde k^2+\iZero\, \tilde k\cdot v_1)
    }
    \ ,
\end{align}
where $k=\omega_{\rm IR} \tilde k$ has been rescaled with $\omega_{\rm IR}$.
As we are interested in the $\hat b^\mu$ component, the index $\sigma$ may be projected onto $\hat b^\mu$.
This invites us to consider the behaviour of the integrand under flipping the sign of the transverse components of the loop momenta $\ell^\mu$ and $\tilde k^\mu$.
In this context, by transverse components, we mean the components orthogonal to $v_i^\mu$.

Naturally, any scalar product of $\ell^\mu$ or $\tilde k^\mu$ with the velocities is independent of the transverse components.
Further, any inner product of the momenta ($\ell$ and $\tilde k$) among themselves is even under flipping the sign of the transverse components.
Therefore, apart from the factor $\rho(\ell)$, the $\hat b$-component of the integrand in eq.~\eqref{eq:fjfjfjfjf} is odd under this sign flip.

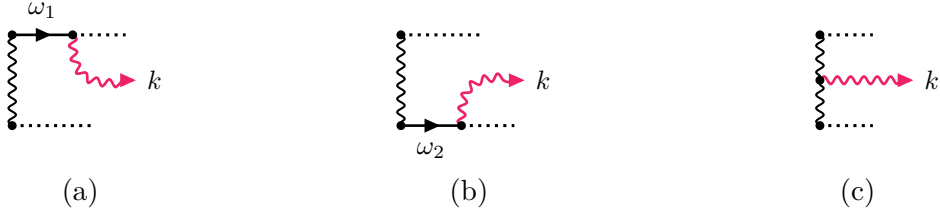
\begin{figure}[t]
  \centering
  \begin{subfigure}[b]{0.32\textwidth}
    \centering
    \begin{tikzpicture}
      \path[use as bounding box] (-.1,-.4) rectangle (1.9,1.6);
      \coordinate (t1) at (0,1.2);  \coordinate (t2) at (.8,1.2);
      \coordinate (b1) at (0,0);
      \coordinate (k)  at (1.6,.6);
      \drawDottedLine{t2}{1.5,1.2}
      \drawDottedLine{b1}{1.1,0}
      \drawDirectedBHLine{t1}{t2}
      \drawGravitonLine{b1}{t1}
      \drawDirectedRadiativeCurvedGravitonLine{t2}{k}{-90}{180}
      \foreach \v in {t1,t2,b1} {\drawPerturbativeVertex{\v}}
      \node at ($(t1)!.5!(t2)$) [above=.08cm] {\footnotesize $\omega_1$};
      \node at (k) [right=.04cm] {\footnotesize $k$};
    \end{tikzpicture}
    \caption{}
    \label{fig:brem-a}
  \end{subfigure}
  \hfill
  \begin{subfigure}[b]{0.32\textwidth}
    \centering
    \begin{tikzpicture}
      \path[use as bounding box] (-.1,-.4) rectangle (1.9,1.6);
      \coordinate (t1) at (0,1.2);
      \coordinate (b1) at (0,0);    \coordinate (b2) at (.8,0);
      \coordinate (k)  at (1.6,.6);
      \drawDottedLine{t1}{1.1,1.2}
      \drawDottedLine{b2}{1.5,0}
      \drawDirectedBHLine{b1}{b2}
      \drawGravitonLine{t1}{b1}
      \drawDirectedRadiativeCurvedGravitonLine{b2}{k}{90}{180}
      \foreach \v in {t1,b1,b2} {\drawPerturbativeVertex{\v}}
      \node at ($(b1)!.5!(b2)$) [below=.08cm] {\footnotesize $\omega_2$};
      \node at (k) [right=.04cm] {\footnotesize $k$};
    \end{tikzpicture}
    \caption{}
    \label{fig:brem-b}
  \end{subfigure}
  \hfill
  \begin{subfigure}[b]{0.32\textwidth}
    \centering
    \begin{tikzpicture}
      \path[use as bounding box] (-.1,-.4) rectangle (1.9,1.6);
      \coordinate (t1) at (.4,1.2);
      \coordinate (m)  at (.4,.6);
      \coordinate (b1) at (.4,0);
      \coordinate (k)  at (1.6,.6);
      \drawDottedLine{t1}{1.1,1.2}
      \drawDottedLine{b1}{1.1,0}
      \drawGravitonLine{t1}{m} \drawGravitonLine{b1}{m}
      \drawDirectedRadiativeCurvedGravitonLine{m}{k}{0}{180}
      \foreach \v in {t1,m,b1} {\drawPerturbativeVertex{\v}}
      \node at (k) [right=.04cm] {\footnotesize $k$};
    \end{tikzpicture}
    \caption{}
    \label{fig:brem-c}
  \end{subfigure}
  \caption{The three diagrams contributing to gravitational bremsstrahlung at
    leading $\kappa^3$ order, after fig.~1 of ref.~\cite{Jakobsen:2021smu}.}
  \label{fig:waveformkappa3}
\end{figure}

This forces us to consider the behaviour of $\rho(\ell)$ under sign-flipping of the transverse components of $\ell$.
If $\rho(\ell)$ were even under this sign flip, the integrand would be odd in its entirety and the integral would therefore vanish.
Since $\rho(\ell)$ is composed of two copies of the $\kappa^3$ waveform, this in turn requires that we understand its behaviour under sign-flipping.
The three diagrams contributing to the waveform at $\kappa^3$ are drawn for reference in \fig{fig:waveformkappa3}~\cite{Jakobsen:2021smu}.
Generally, the waveform may be written as a Fourier transform from the ``amplitude'', or $q$-space result, as,~\cite{Jakobsen:2021smu,Bohnenblust:2025gir}
\begin{align}\label{eq:paraFour}
    f_\mn(\ell)
    =
    \int_{q_1q_2}
    e^{\i q_1\cdot b_1+\i q_2\cdot b_2}
    \dd(\ell-q_1-q_2)
    f_\mn(q_1,q_2)
    \ ,
\end{align}
with $f_\mn(q_1,q_2)$ independent of $b_i^\mu$.
If 
\begin{equation}\label{eq:reality}
    f_\mn^*(q_1,q_2)=f_\mn(q_1,q_2)
    \ ,
\end{equation}
then it is also the case that,
\begin{align}
    f_\mn(\ell) \big|_{\ell\cdot b\to -\ell\cdot b}
    =
    f^*_\mn(\ell)
    \ .
\end{align} 
This flip essentially inverts the sign in the exponential in \eqn{eq:paraFour} which corresponds to complex conjugation (because $f_\mn(q_1,q_2)$ is assumed real).

The $\kappa^3$ waveform does obey eq.~\eqref{eq:reality} except for its memory term.
Namely, failures to obey this identity come from the $\iZero$ prescription of WL or graviton propagators.
In the diagrams of \fig{fig:waveformkappa3}, only the $\iZero$ of the  linear propagators in \fig{fig:brem-a} and \fig{fig:brem-b} matter and give rise to the memory effect which is proportional to $1/(\ell\cdot v+\iZero)$.
More specifically, we may split the waveform into a real part that fulfils \eqn{eq:reality} and its imaginary part which in turn is proportional to $\dd(\ell\cdot v)$.
Furthermore, since $\ell^\mu$ is on shell, its overall scale is set by its energy in any frame $\ell^0$.
Dimensional analysis of the integrand \eqn{eq:fjfjfjfjf} shows that it may be considered as quadratic in $\ell^0 f(\ell)$ (in analogy, for example, to the integrand of $E_{\rm rad}$) and therefore the imaginary part disappears at leading order.

In other words, it is in fact the case that,
\begin{align}
    \ell^0 f(\ell)
    \big|_{\ell\cdot b\to -\ell\cdot b}
    =
    (\ell^0 f(\ell))^*
    \ ,
\end{align}
and therefore the $\hat b$-projected integrand in \eqn{eq:fjfjfjfjf}, which includes $|\ell^0 f(\ell)|^2$, is overall odd in the transverse components.
With this knowledge, we finally understand why the nonlinear memory contribution to $|L|_{\rm loss}$ vanishes at this order as observed in ref.~\cite{Heissenberg:2024umh}.

\section{Integrand generation}
\label{sec:integrand}
With the static contribution taken care of, we now turn to the dynamical contribution to the dissipative worldline offset kicks, i.e., the contribution from diagrams whose gravitons all scale dynamically.
We address its computation using \eqn{eq:our-fav-formula}, which, like the impulse, is obtained from the deflection one-point function.
As such, the computation follows the impulse one closely, mainly differing through the presence of the derivative with respect to the energy of the outgoing deflection in \eqn{eq:our-fav-formula}.
This derivative forces us to generate the integrand off-shell, mirroring the computation of the trajectory~\cite{Mogull:2025cfn}, albeit the on-shell limit is taken afterwards.
Additionally, we need only compute the 1SF and \oneSFBar{} contributions, since the probe contributions are purely conservative.
Hence, we require the off-shell 4PM integrand of the deflection one-point function at these two SF orders.

Below, we first describe how this integrand is generated.
Here, an important ingredient, to which we devote some attention, is a slight reframing of our Feynman rules which manifests the general structure of our WQFT Feynman diagrams including the classical loop integrals within them.

\begin{figure}[t]
    \centering
    \begin{subfigure}[b]{0.19\textwidth}
        \centering
        \begin{tikzpicture}[baseline={(0,-.1)}]
            \coordinate (t1) at (0,.65);
            \coordinate (t2) at (.55,.65);
            \coordinate (t3) at (1.1,.65);
            \coordinate (t4) at (1.65,.65);
            \coordinate (out) at (2.1,.65);
            \coordinate (b1) at (0,-.65);
            \coordinate (b2) at (.55,-.65);
            \coordinate (b3) at (1.1,-.65);
            \coordinate (b4) at (1.65,-.65);
            \drawDottedLine{t2}{t3}
            \drawDottedLine{b1}{b2}
            \drawDottedLine{b3}{b4}
            \drawDottedLine{b4}{out|-b4}
            \drawBHLine{t1}{t2}
            \drawBHLine{t3}{t4}
            \drawBHLine{b2}{b3}
            \drawDirectedBHLine{t4}{out}
            \drawGravitonLine{t1}{b1}
            \drawActiveGravitonLine{t2}{b2}
            \drawActiveGravitonLine{b3}{t3}
            \drawGravitonLine{t4}{b4}
            \foreach \v in {t1,t2,t3,t4,b1,b2,b3,b4} {\drawPerturbativeVertex{\v}}
        \end{tikzpicture}
        \caption{}
        \label{fig:sample-diagrams-a}
    \end{subfigure}
    \hfill
    \begin{subfigure}[b]{0.19\textwidth}
        \centering
        \begin{tikzpicture}[baseline={(0,-.1)}]
            \coordinate (t1) at (0,.65);
            \coordinate (t2) at (.55,.65);
            \coordinate (t3) at (1.1,.65);
            \coordinate (t4) at (1.65,.65);
            \coordinate (out) at (2.1,.65);
            \coordinate (b1) at (0,-.65);
            \coordinate (b2) at (.55,-.65);
            \coordinate (b3) at (1.1,-.65);
            \coordinate (b4) at (1.65,-.65);
            \drawDottedLine{t3}{t4}
            \drawDottedLine{b1}{b3}
            \drawDottedLine{b4}{out|-b4}
            \drawBHLine{t1}{t2}
            \drawBHLine{t2}{t3}
            \drawBHLine{b3}{b4}
            \drawDirectedBHLine{t4}{out}
            \drawGravitonLine{t1}{b1}
            \drawGravitonLine{t2}{b2}
            \drawActiveGravitonLine{t3}{b3}
            \drawActiveGravitonLine{b4}{t4}
            \foreach \v in {t1,t2,t3,t4,b1,b2,b3,b4} {\drawPerturbativeVertex{\v}}
        \end{tikzpicture}
        \caption{}
        \label{fig:sample-diagrams-b}
    \end{subfigure}
    \hfill
    \begin{subfigure}[b]{0.19\textwidth}
        \centering
        \begin{tikzpicture}[baseline={(0,-.1)}]
            \coordinate (t1) at (0,.65);
            \coordinate (v2) at (.55,.25);
            \coordinate (t3) at (1.1,.65);
            \coordinate (t4) at (1.65,.65);
            \coordinate (out) at (2.1,.65);
            \coordinate (b1) at (0,-.65);
            \coordinate (b2) at (.55,-.65);
            \coordinate (b3) at (1.1,-.65);
            \coordinate (b4) at (1.65,-.65);
            \drawDottedLine{t3}{t4}
            \drawDottedLine{b1}{b3}
            \drawDottedLine{b4}{out|-b4}
            \drawBHLine{t1}{t3}
            \drawCurvedBHLine{v2}{t3}{0}{180}
            \drawBHLine{b3}{b4}
            \drawDirectedBHLine{t4}{out}
            \drawGravitonLine{t1}{b1}
            \drawGravitonLine{v2}{b2}
            \drawActiveGravitonLine{t3}{b3}
            \drawActiveGravitonLine{b4}{t4}
            \foreach \v in {t1,v2,t3,t4,b1,b2,b3,b4} {\drawPerturbativeVertex{\v}}
        \end{tikzpicture}
        \caption{}
        \label{fig:sample-diagrams-c}
    \end{subfigure}
    \hfill
    \begin{subfigure}[b]{0.19\textwidth}
        \centering
        \begin{tikzpicture}[baseline={(0,-.1)}]
            \coordinate (tl) at (0,.65);
            \coordinate (tm) at (.8,.65);
            \coordinate (tr) at (1.65,.65);
            \coordinate (out) at (2.1,.65);
            \coordinate (g1) at (0,0);
            \coordinate (g2) at (.8,0);
            \coordinate (bl) at (0,-.65);
            \coordinate (bm) at (.8,-.65);
            \coordinate (br) at (1.65,-.65);
            \drawDottedLine{tl}{tm}
            \drawDottedLine{bl}{br}
            \drawDottedLine{b4}{out|-br}
            \drawBHLine{tm}{tr}
            \drawDirectedBHLine{tr}{out}
            \drawGravitonLine{tl}{g1}
            \drawGravitonLine{g1}{bl}
            \drawActiveGravitonLine{g1}{g2}
            \drawActiveGravitonLine{tm}{g2}
            \drawGravitonLine{g2}{bm}
            \drawGravitonLine{tr}{br}
            \foreach \v in {tl,tm,tr,g1,g2,bl,bm,br} {\drawPerturbativeVertex{\v}}
        \end{tikzpicture}
        \caption{}
        \label{fig:sample-diagrams-d}
    \end{subfigure}
    \hfill
    \begin{subfigure}[b]{0.19\textwidth}
        \centering
        \begin{tikzpicture}[baseline={(0,-.1)}]
            \coordinate (tl) at (0,.65);
            \coordinate (tr) at (1.65,.65);
            \coordinate (out) at (2.1,.65);
            \coordinate (g1) at (0,0);
            \coordinate (g2) at (.8,0);
            \coordinate (g3) at (1.65,0);
            \coordinate (bl) at (0,-.65);
            \coordinate (bm) at (.8,-.65);
            \coordinate (br) at (1.65,-.65);
            \drawDottedLine{tl}{tr}
            \drawDottedLine{bl}{br}
            \drawDottedLine{b4}{out|-br}
            \drawDirectedBHLine{tr}{out}
            \drawGravitonLine{bl}{g1}
            \drawGravitonLine{g1}{tl}
            \drawGravitonLine{bm}{g2}
            \drawActiveGravitonLine{g1}{g2}
            \drawActiveGravitonLine{g2}{g3}
            \drawGravitonLine{br}{g3}
            \drawActiveGravitonLine{g3}{tr}
            \foreach \v in {tl,tr,g1,g2,g3,bl,bm,br} {\drawPerturbativeVertex{\v}}
        \end{tikzpicture}
        \caption{}
        \label{fig:sample-diagrams-e}
    \end{subfigure}
    \\[1em]
    \begin{subfigure}[b]{0.19\textwidth}
        \centering
        \begin{tikzpicture}[baseline={(0,-.1)}]
            \coordinate (t1) at (0,.65);
            \coordinate (t2) at (.55,.65);
            \coordinate (t3) at (1.1,.65);
            \coordinate (t4) at (1.65,.65);
            \coordinate (out) at (2.1,.65);
            \coordinate (b1) at (0,-.65);
            \coordinate (b2) at (.55,-.65);
            \coordinate (b3) at (1.1,-.65);
            \coordinate (b4) at (1.65,-.65);
            \drawDottedLine{t1}{t2}
            \drawDottedLine{t3}{t4}
            \drawDottedLine{b2}{b3}
            \drawDottedLine{b4}{out|-b4}
            \drawBHLine{t2}{t3}
            \drawBHLine{b1}{b2}
            \drawBHLine{b3}{b4}
            \drawDirectedBHLine{t4}{out}
            \drawGravitonLine{t1}{b1}
            \drawActiveGravitonLine{t2}{b2}
            \drawActiveGravitonLine{t3}{b3}
            \drawActiveGravitonLine{t4}{b4}
            \foreach \v in {t1,t2,t3,t4,b1,b2,b3,b4} {\drawPerturbativeVertex{\v}}
        \end{tikzpicture}
        \caption{}
        \label{fig:sample-diagrams-f}
    \end{subfigure}
    \hfill
    \begin{subfigure}[b]{0.19\textwidth}
        \centering
        \begin{tikzpicture}[baseline={(0,-.1)}]
            \coordinate (t1) at (0,.65);
            \coordinate (t2) at (.55,.65);
            \coordinate (t3) at (1.1,.65);
            \coordinate (t4) at (1.65,.65);
            \coordinate (out) at (2.1,.65);
            \coordinate (b1) at (0,-.65);
            \coordinate (b2) at (.55,-.65);
            \coordinate (b3) at (1.1,-.65);
            \coordinate (b4) at (1.65,-.65);
            \drawDottedLine{t1}{t3}
            \drawDottedLine{b3}{b4}
            \drawDottedLine{b4}{out|-b4}
            \drawBHLine{t3}{t4}
            \drawBHLine{b1}{b2}
            \drawBHLine{b2}{b3}
            \drawDirectedBHLine{t4}{out}
            \drawGravitonLine{t1}{b1}
            \drawGravitonLine{t2}{b2}
            \drawActiveGravitonLine{t3}{b3}
            \drawGravitonLine{t4}{b4}
            \foreach \v in {t1,t2,t3,t4,b1,b2,b3,b4} {\drawPerturbativeVertex{\v}}
        \end{tikzpicture}
        \caption{}
        \label{fig:sample-diagrams-g}
    \end{subfigure}
    \hfill
    \begin{subfigure}[b]{0.19\textwidth}
        \centering
        \begin{tikzpicture}[baseline={(0,-.1)}]
            \coordinate (t1) at (0,.65);
            \coordinate (t2) at (.55,.65);
            \coordinate (t3) at (1.1,.65);
            \coordinate (t4) at (1.65,.65);
            \coordinate (out) at (2.1,.65);
            \coordinate (b1) at (0,-.65);
            \coordinate (v2) at (.55,-.25);
            \coordinate (b3) at (1.1,-.65);
            \coordinate (b4) at (1.65,-.65);
            \drawDottedLine{t1}{t3}
            \drawDottedLine{b3}{b4}
            \drawDottedLine{b4}{out|-b4}
            \drawBHLine{t3}{t4}
            \drawBHLine{b1}{b3}
            \drawCurvedBHLine{v2}{b3}{0}{180}
            \drawDirectedBHLine{t4}{out}
            \drawGravitonLine{t1}{b1}
            \drawGravitonLine{t2}{v2}
            \drawActiveGravitonLine{t3}{b3}
            \drawGravitonLine{t4}{b4}
            \foreach \v in {t1,t2,t3,t4,b1,v2,b3,b4} {\drawPerturbativeVertex{\v}}
        \end{tikzpicture}
        \caption{}
        \label{fig:sample-diagrams-h}
    \end{subfigure}
    \hfill
    \begin{subfigure}[b]{0.19\textwidth}
        \centering
        \begin{tikzpicture}[baseline={(0,-.1)}]
            \coordinate (tl) at (0,.65);
            \coordinate (tm) at (.8,.65);
            \coordinate (tr) at (1.65,.65);
            \coordinate (out) at (2.1,.65);
            \coordinate (g1) at (0,0);
            \coordinate (g2) at (.8,0);
            \coordinate (bl) at (0,-.65);
            \coordinate (bm) at (.8,-.65);
            \coordinate (br) at (1.65,-.65);
            \drawDottedLine{tl}{tr}
            \drawDottedLine{bl}{bm}
            \drawDottedLine{br}{out|-br}
            \drawBHLine{bm}{br}
            \drawDirectedBHLine{tr}{out}
            \drawGravitonLine{bl}{g1}
            \drawGravitonLine{g1}{tl}
            \drawActiveGravitonLine{g1}{g2}
            \drawActiveGravitonLine{bm}{g2}
            \drawGravitonLine{g2}{tm}
            \drawActiveGravitonLine{tr}{br}
            \foreach \v in {tl,tm,tr,g1,g2,bl,bm,br} {\drawPerturbativeVertex{\v}}
        \end{tikzpicture}
        \caption{}
        \label{fig:sample-diagrams-i}
    \end{subfigure}
    \hfill
    \begin{subfigure}[b]{0.19\textwidth}
        \centering
        \begin{tikzpicture}[baseline={(0,-.1)}]
            \coordinate (tl) at (0,.65);
            \coordinate (tm) at (.8,.65);
            \coordinate (tr) at (1.65,.65);
            \coordinate (out) at (2.1,.65);
            \coordinate (g1) at (0,.1);
            \coordinate (r) at (1.65,.1);
            \coordinate (g3) at (.8,-.3);
            \coordinate (bl) at (0,-.65);
            \coordinate (bm) at (.8,-.65);
            \drawDottedLine{tl}{tr}
            \drawDottedLine{bl}{bm}
            \drawDottedLine{bm}{out|-bm}
            \drawDirectedBHLine{tr}{out}
            \drawGravitonLine{tl}{g1}
            \drawGravitonLine{g1}{bl}
            \drawActiveGravitonLine{tr}{r}
            \drawGravitonLine{g3}{bm}
            \drawGravitonLine{tm}{g3}
            \drawShadowLine{g1}{r}
            \drawActiveGravitonLine{r}{g3}
            \drawActiveGravitonLine{g1}{r}
            \foreach \v in {tl,tm,tr,g1,r,g3,bl,bm} {\drawPerturbativeVertex{\v}}
        \end{tikzpicture}
        \caption{}
        \label{fig:sample-diagrams-j}
    \end{subfigure}
    \caption{
        A representative sample of the amputated 4PM diagrams at 1SF (top row) and \oneSFBar{} (bottom row) with active gravitons highlighted.
        In \subref{fig:sample-diagrams-b}, \subref{fig:sample-diagrams-c}, \subref{fig:sample-diagrams-e}, \subref{fig:sample-diagrams-f}, \subref{fig:sample-diagrams-i} and \subref{fig:sample-diagrams-j}, the right-most graviton would on shell be passive, but off shell it is active, since $q\cdot v_1\neq 0$.
        It is the curved deflection line in diagrams like \subref{fig:sample-diagrams-c} and \subref{fig:sample-diagrams-h} that necessitates the introduction of the planar-2 families.
        The \oneSFBar{} diagrams resemble their 1SF counterparts, but the off-shell asymmetry between the worldlines is apparent. For instance, the right-most graviton in \subref{fig:sample-diagrams-g}, the counterpart of \subref{fig:sample-diagrams-b}, is passive, while in \subref{fig:sample-diagrams-b} it is active.
        Also, the crossing gravitons in \subref{fig:sample-diagrams-j} make it non-planar, which is why we need the $\overbar{\rm NP}$ family at \oneSFBar{}.
    }
    \label{fig:sample-diagrams}
\end{figure}
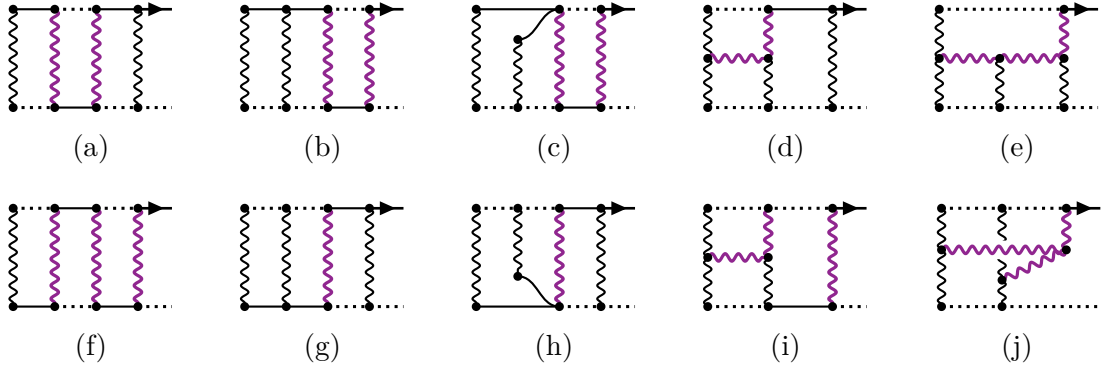

\subsection{Construction of off-shell integrand}
To generate the necessary diagrams, we use a version of Berends--Giele recursion~\cite{Berends:1987me,Jakobsen:2023ndj,Jakobsen:2023oow}.
Stated briefly, this involves recursing the classical Dyson--Schwinger equations for the WQFT one-point functions, which in the spinless case we are considering, read
\begin{align}\label{eq:wl-ds-eq}
    \begin{tikzpicture}[baseline={(0,-.1)}]
        \coordinate (x) at (0,0);
        \coordinate (y) at (.6,0);
        \drawOnePointFunction{$(x)-(.1,0)$}
        \drawDirectedBHLine{x}{y}
        \drawPerturbativeVertex{y}
    \end{tikzpicture}
    \;&=\;
    \begin{tikzpicture}[baseline={(0,-.1)}]
        \coordinate (x) at (.5,-.5);
        \coordinate (y) at (.5,0);
        \coordinate (z) at (1.1,0);
        \coordinate (in) at (.1,0);
        \drawDottedLine{in}{y}
        \drawOnePointFunction{x}
        \drawGravitonLine{x}{y}
        \drawPerturbativeVertex{y}
        \drawDirectedBHLine{y}{z}
        \drawPerturbativeVertex{z}
    \end{tikzpicture}
    \;+\;
    \begin{tikzpicture}[baseline={(0,-.1)}]
        \coordinate (x1) at (.5,-.5);
        \coordinate (x2) at (.1,0);
        \coordinate (y) at (.5,0);
        \coordinate (z) at (1.1,0);
        \drawOnePointFunction{x1}
        \drawOnePointFunction{x2}
        \drawGravitonLine{x1}{y}
        \drawBHLine{x2}{y}
        \drawPerturbativeVertex{y}
        \drawDirectedBHLine{y}{z}
        \drawPerturbativeVertex{z}
    \end{tikzpicture}
    \;+\;
    \frac12\;
    \begin{tikzpicture}[baseline={(0,-.1)}]
        \coordinate (x1) at (.5,-.5);
        \coordinate (x2) at (.1,0);
        \coordinate (x3) at (.1,-.4);
        \coordinate (y) at (.5,0);
        \coordinate (z) at (1.1,0);
        \drawOnePointFunction{x1}
        \drawOnePointFunction{x2}
        \drawOnePointFunction{x3}
        \drawGravitonLine{x1}{y}
        \drawBHLine{x2}{y}
        \drawCurvedBHLine{x3}{y}{0}{180}
        \drawPerturbativeVertex{y}
        \drawDirectedBHLine{y}{z}
        \drawPerturbativeVertex{z}
    \end{tikzpicture}
    \;+\;
    \frac1{3!}\;
    \begin{tikzpicture}[baseline={(0,-.1)}]
        \coordinate (x1) at (.5,-.5);
        \coordinate (x2) at (.1,0);
        \coordinate (x3) at (.1,-.4);
        \coordinate (x4) at (.1,.4);
        \coordinate (y) at (.5,0);
        \coordinate (z) at (1.1,0);
        \drawOnePointFunction{x1}
        \drawOnePointFunction{x2}
        \drawOnePointFunction{x3}
        \drawOnePointFunction{x4}
        \drawGravitonLine{x1}{y}
        \drawBHLine{x2}{y}
        \drawCurvedBHLine{x3}{y}{0}{180}
        \drawCurvedBHLine{x4}{y}{0}{180}
        \drawPerturbativeVertex{y}
        \drawDirectedBHLine{y}{z}
        \drawPerturbativeVertex{z}
    \end{tikzpicture}
    +
    \cdots
    \, ,
    \\[.5cm]\label{eq:graviton-ds-eq}
    \begin{tikzpicture}[baseline={(0,-.1)}]
        \coordinate (x) at (0,0);
        \coordinate (y) at (.7,0);
        \drawOnePointFunction{x}
        \drawDirectedGravitonLine{x}{y}
        \drawPerturbativeVertex{y}
    \end{tikzpicture}
    \;&=\;
    \begin{tikzpicture}[baseline={(0,-.1)}]
        \coordinate (x) at (0,.4);
        \coordinate (y) at (.4,0);
        \coordinate (inT) at (-.3,.4);
        \coordinate (outT) at (.3,.4);
        \coordinate (inB) at (-.2,-.4);
        \coordinate (outB) at (.2,-.4);
        \drawDottedLine{inT}{x}
        \drawDottedLine{x}{outT}
        \drawDottedLine{inB}{outB}
        \drawPerturbativeVertex{x}
        \draw [snakeDirected] (x) to[out=270,in=-180,looseness=1.4] (y);
        \drawPerturbativeVertex{y}
    \end{tikzpicture}
    \;+\;
    \begin{tikzpicture}[baseline={(0,-.1)}]
        \coordinate (x) at (-.4,.4);
        \coordinate (y) at (0,.4);
        \coordinate (z) at (.4,0);
        \coordinate (outT) at (.4,.4);
        \coordinate (inB) at (-.3,-.4);
        \coordinate (outB) at (.2,-.4);
        \drawDottedLine{y}{outT}
        \drawDottedLine{inB}{outB}
        \drawOnePointFunction{x}
        \drawBHLine{x}{y}
        \drawPerturbativeVertex{y}
        \draw [snakeDirected] (y) to[out=270,in=-180,looseness=1.4] (z);
        \drawPerturbativeVertex{z}
    \end{tikzpicture}
    \;+\;
    \frac12\;
    \begin{tikzpicture}[baseline={(0,-.1)}]
        \coordinate (x1) at (-.4,0);
        \coordinate (x2) at (-.4,.4);
        \coordinate (y) at (0,.4);
        \coordinate (z) at (.4,0);
        \coordinate (outT) at (.4,.4);
        \coordinate (inB) at (-.3,-.4);
        \coordinate (outB) at (.2,-.4);
        \drawDottedLine{y}{outT}
        \drawDottedLine{inB}{outB}
        \drawOnePointFunction{x1}
        \drawOnePointFunction{x2}
        \drawCurvedBHLine{x1}{y}{0}{180}
        \drawBHLine{x2}{y}
        \drawPerturbativeVertex{y}
        \draw [snakeDirected] (y) to[out=270,in=-180,looseness=1.4] (z);
        \drawPerturbativeVertex{z}
    \end{tikzpicture}
    \;+
    \cdots
    +\;
    \begin{tikzpicture}[baseline={(0,-.1)}]
        \coordinate (x) at (0,-.4);
        \coordinate (y) at (.4,0);
        \coordinate (inB) at (-.3,-.4);
        \coordinate (outB) at (.3,-.4);
        \coordinate (inT) at (-.2,.4);
        \coordinate (outT) at (.2,.4);
        \drawDottedLine{inB}{x}
        \drawDottedLine{x}{outB}
        \drawDottedLine{inT}{outT}
        \drawPerturbativeVertex{x}
        \draw [snakeDirected] (x) to[out=90,in=-180,looseness=1.4] (y);
        \drawPerturbativeVertex{y}
    \end{tikzpicture}
    \;+\;
    \begin{tikzpicture}[baseline={(0,-.1)}]
        \coordinate (x) at (-.4,-.4);
        \coordinate (y) at (0,-.4);
        \coordinate (z) at (.4,0);
        \coordinate (outB) at (.4,-.4);
        \coordinate (inT) at (-.3,.4);
        \coordinate (outT) at (.2,.4);
        \drawDottedLine{y}{outB}
        \drawDottedLine{inT}{outT}
        \drawOnePointFunction{x}
        \drawBHLine{x}{y}
        \drawPerturbativeVertex{y}
        \draw [snakeDirected] (y) to[out=90,in=-180,looseness=1.4] (z);
        \drawPerturbativeVertex{z}
    \end{tikzpicture}
    \;+\;
    \frac12\;
    \begin{tikzpicture}[baseline={(0,-.1)}]
        \coordinate (x1) at (-.4,0);
        \coordinate (x2) at (-.4,-.4);
        \coordinate (y) at (0,-.4);
        \coordinate (z) at (.4,0);
        \coordinate (outB) at (.4,-.4);
        \coordinate (inT) at (-.3,.4);
        \coordinate (outT) at (.2,.4);
        \drawDottedLine{y}{outB}
        \drawDottedLine{inT}{outT}
        \drawOnePointFunction{x1}
        \drawOnePointFunction{x2}
        \drawCurvedBHLine{x1}{y}{0}{180}
        \drawBHLine{x2}{y}
        \drawPerturbativeVertex{y}
        \draw [snakeDirected] (y) to[out=90,in=-180,looseness=1.4] (z);
        \drawPerturbativeVertex{z}
    \end{tikzpicture}
    \;+
    \cdots
    \notag\\[.2cm]
    &\hspace{5.4cm}+\;
    \frac12\;
    \begin{tikzpicture}[baseline={(0,-.1)}]
        \coordinate (x1) at (0,-.2);
        \coordinate (x2) at (0,.2);
        \coordinate (y) at (.5,0);
        \coordinate (z) at (1.1,0);
        \drawOnePointFunction{x1}
        \drawOnePointFunction{x2}
        \drawGravitonLine{x1}{y}
        \drawGravitonLine{x2}{y}
        \drawPerturbativeVertex{y}
        \drawDirectedGravitonLine{y}{z}
        \drawPerturbativeVertex{z}
    \end{tikzpicture}
    \;+\;
    \frac1{3!}\;
    \begin{tikzpicture}[baseline={(0,-.1)}]
        \coordinate (x1) at (0,-.4);
        \coordinate (x2) at (0,0);
        \coordinate (x3) at (0,.4);
        \coordinate (y) at (.5,0);
        \coordinate (z) at (1.1,0);
        \drawOnePointFunction{x1}
        \drawOnePointFunction{x2}
        \drawOnePointFunction{x3}
        \drawGravitonLine{x1}{y}
        \drawGravitonLine{x2}{y}
        \drawGravitonLine{x3}{y}
        \drawPerturbativeVertex{y}
        \drawDirectedGravitonLine{y}{z}
        \drawPerturbativeVertex{z}
    \end{tikzpicture}
    +
    \cdots
    \, .
\end{align}
Here, a hatched blob denotes a one-point function, identified through which line emanates from it.
In this recursive relation, all hatched blobs are off-shell.
The dotted lines indicate to which worldline the one-point function corresponds to, and \eqn{eq:wl-ds-eq} holds for both WL1 and WL2.
Repeated application of these equations yields all relevant diagrams with symmetry factors~\cite{Jakobsen:2023ndj}; in the present computation, we find 75 diagrams at 1SF and 83 diagrams at \oneSFBar, of which we display a representative sample in \fig{fig:sample-diagrams}.

\begin{table}[t]
    \centering
    \small
    \setlength{\tabcolsep}{4pt}
    \renewcommand{\arraystretch}{1.25}
    \begin{tabular}{c|cc|cc|c}
        & M1 & M2 & $\overbar{\rm M1}$ & $\overbar{\rm M2}$ & $\overbar{\rm NP}$ \\
        \hline
        $D_1$ & \multicolumn{2}{c|}{$\ell_1\cdot v_2$} & \multicolumn{2}{c|}{$\bell_1\cdot v_1$} & $\bell_1\cdot v_1$ \\
        $D_2$ & \multicolumn{2}{c|}{$\ell_2\cdot v_2$} & \multicolumn{2}{c|}{$\bell_2\cdot v_1$} & $\bell_2\cdot v_1$ \\
        $D_3$ & \multicolumn{2}{c|}{$\bell_1\cdot v_1$} & \multicolumn{2}{c|}{$\ell_1\cdot v_2$} & $\ell_1\cdot v_2$ \\
        \hline
        $D_4$ & \multicolumn{2}{c|}{$\ell_1\cdot v_1$} & \multicolumn{2}{c|}{$\bell_1\cdot v_2$} & $\bell_1\cdot v_2$ \\
        $D_5$ & $\ell_2\cdot v_1$ & $(\ell_1-\ell_2)\cdot v_1$ & $\bell_2\cdot v_2$ & $(\bell_1-\bell_2)\cdot v_2$ & $\bell_2\cdot v_2$ \\
        $D_6$ & \multicolumn{2}{c|}{$\bell_1\cdot v_2$} & \multicolumn{2}{c|}{$\ell_1\cdot v_1$} & $\ell_1\cdot v_1$ \\
        \hline
        $\hat D_5$ & $(\ell_2-q)\cdot v_1$ & $(\ell_1-\ell_2+q)\cdot v_1$ & \multicolumn{2}{c|}{} &  \\
        $\hat D_6$ & \multicolumn{2}{c|}{} & \multicolumn{2}{c|}{$(\ell_1-q)\cdot v_1$} & $(\ell_1-q)\cdot v_1$ \\
        \hline
        $D_7$ & \multicolumn{2}{c|}{$(\ell_1-\bell_1)^2$} & \multicolumn{2}{c|}{$(\bell_1-\ell_1)^2$} & $(\bell_1-\ell_1)^2$ \\
        $D_8$ & \multicolumn{2}{c|}{$(\ell_2-\bell_1)^2$} & \multicolumn{2}{c|}{$(\bell_2-\ell_1)^2$} & $(\bell_2-\ell_1+q)^2$ \\
        \hline
        $D_9$ & \multicolumn{2}{c|}{$(\ell_1-\ell_2)^2$} & \multicolumn{2}{c|}{$(\bell_1-\bell_2)^2$} & $(\bell_2-\bell_1+q)^2$ \\
        $D_{10}$ & \multicolumn{2}{c|}{$(\ell_1-q)^2$} & \multicolumn{2}{c|}{$(\bell_1-q)^2$} & $(\bell_1-q)^2$ \\
        $D_{11}$ & \multicolumn{2}{c|}{$(\ell_2-q)^2$} & \multicolumn{2}{c|}{$(\bell_2-q)^2$} & $\bell_2^2$ \\
        $D_{12}$ & \multicolumn{2}{c|}{$(\bell_1-q)^2$} & \multicolumn{2}{c|}{$(\ell_1-q)^2$} & $(\ell_1-q)^2$ \\
        $D_{13}$ & \multicolumn{2}{c|}{$\ell_1^2$} & \multicolumn{2}{c|}{$\bell_1^2$} & $\bell_1^2$ \\
        $D_{14}$ & \multicolumn{2}{c|}{$\ell_2^2$} & \multicolumn{2}{c|}{$\bell_2^2$} &  \\
        $D_{15}$ & \multicolumn{2}{c|}{$\bell_1^2$} & \multicolumn{2}{c|}{$\ell_1^2$} & $\ell_1^2$ \\
    \end{tabular}
    \caption{
        Off-shell (inverse) propagators of the 1SF Mondrian families M1 and M2, the \oneSFBar{} Mondrian families $\overbar{\rm M1}$ and $\overbar{\rm M2}$, and the \oneSFBar{} non-planar family $\overbar{\rm NP}$.
        The 1SF families have two unbarred momenta $\ell_{1,2}$ and the $\overbar{\rm 1SF}$ families have two barred momenta $\bar{\ell}_{1,2}$.
        The propagators with subscripts 4--8 are active and, depending on labelling, have either a retarded or an advanced $\iZero$ prescription.
    }
    \label{tab:families}
\end{table}

\begin{figure}[t]
  \centering
  \begin{subfigure}[b]{0.45\textwidth}
    \centering
    \begin{tikzpicture}
      \coordinate (t1) at (0,1.2);    \coordinate (t2) at (.6,1.2);   \coordinate (t3) at (1.2,1.2);
      \coordinate (t4) at (1.75,1.2); \coordinate (t5) at (2.75,1.2);
      \coordinate (m1) at (.6,.6);
      \coordinate (b1) at (1.75,0);   \coordinate (b2) at (2.75,0);
      \drawDottedLine{t1}{t3}
      \drawDottedLine{b1}{b2} \drawDottedLineR{b2}{.5}
      \drawBHLine{t3}{t4} \drawBHLine{t4}{t5}
      \drawRadiativeCurvedGravitonLine{t1}{m1}{-90}{180}
      \drawRadiativeCurvedGravitonLine{t3}{m1}{-90}{0}
      \drawGravitonLine{t2}{m1}
      \drawGravitonLine{t4}{b1} \drawGravitonLine{t5}{b2}
      \drawDirectedBHLineR{t5}{.5}
      \foreach \v in {t1,t2,t3,t4,t5,m1,b1,b2} {\drawPerturbativeVertex{\v}}
    \end{tikzpicture}
    \caption{}
    \label{fig:wqft2-a}
  \end{subfigure}
  \hfill
  \begin{subfigure}[b]{0.45\textwidth}
    \centering
    \begin{tikzpicture}
      \coordinate (t1) at (0,1.2);    \coordinate (t2) at (1,1.2);
      \coordinate (t3) at (1.55,1.2); \coordinate (t4) at (2.75,1.2);
      \coordinate (m1) at (0,.6);     \coordinate (m2) at (1,.6);
      \coordinate (b1) at (0,0);      \coordinate (b2) at (1,0);
      \drawDottedLine{t1}{t2}
      \drawDottedLine{t3}{t4}
      \drawDottedLine{b1}{b2} \drawDottedLineR{b2}{.5}
      \drawBHLine{t2}{t3}
      \drawGravitonLine{t1}{m1} \drawGravitonLine{m1}{b1}
      \drawActiveGravitonLine{t2}{m2} \drawGravitonLine{m2}{b2}
      \drawActiveGravitonLine{m1}{m2}
      \drawRadiativeCurvedGravitonLine{t3}{t4}{-90}{-90}
      \drawDirectedBHLineR{t4}{.5}
      \foreach \v in {t1,t2,t3,t4,m1,m2,b1,b2} {\drawPerturbativeVertex{\v}}
    \end{tikzpicture}
    \caption{}
    \label{fig:wqft2-b}
  \end{subfigure}
  \caption{
    Two diagrams whose contribution to $\Delta p^\mu_i$ or $\Delta b_i^\mu$ vanish.
    The first \subref{fig:wqft2-a} is scaleless for any outgoing energy $\omega$ while the second \subref{fig:wqft2-b} is scaleless only on-shell.
    This is still true for the contribution to $\Delta b_i^\mu$ where the external line is still put on shell after the derivative is evaluated.
    Cutting the worldline in the middle of \subref{fig:wqft2-a}, the left-most part with the red radiative graviton is non-interacting, i.e., WL2 is not part of this subdiagram.
    A similar cut is possible for \subref{fig:wqft2-b}.
  }
  \label{fig:wqft-diagrams-2}
\end{figure}
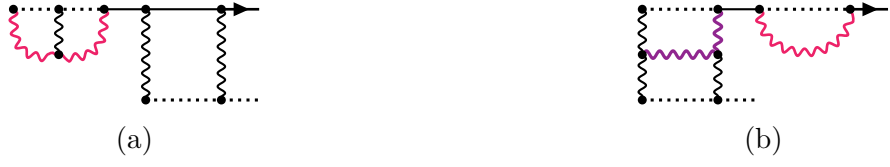

In this counting of diagrams we have omitted a class of graphs that are scaleless and therefore trivially vanish.
This class of graphs is characterised by having a propagating WL that, if cut, divides the graph into an interacting half and a non-interacting half.
Two examples are given in \fig{fig:wqft-diagrams-2}.
Namely, the dynamic scale $|q|$ comes only from interactions between the two WLs.

The next step is to label the momenta in each diagram such that every diagram admits a representation in terms of a small number of integral families.
At 1SF, these families are the planar-1 (M1) and planar-2 (M2) Mondrian families,\footnote{
    The name is an allusion to the similarity of the associated diagrams with the works of Dutch painter Piet Mondrian.
}
\begin{equation}\label{eq:off-shellFamily}
    I_{F,\{\nu\}}^{\{\sigma\}}(|q|,\gamma,\omega_{\rm IR})
    =
    \int_{\ell_1,\ell_2,\bell_1}
    \frac{
        \dd^{(\nu_1-1)}(D_{F,1})
        \dd^{(\nu_2-1)}(D_{F,2})
        \dd^{(\nu_3-1)}(D_{F,3})
    }{
        \prod_{i=4}^{15}
        \big[D_{F,i}^{\sigma_i}\big]^{\nu_i}
        \big[\hat D_{F,5}^{\hat\sigma_5}\big]^{\hat\nu_5}
    }
    \, ,
\end{equation}
where $F=\rm M1,M2$ and we list the (inverse) propagators $D_{F,i}$ in \tab{tab:families}.
The causality prescriptions of the propagators are controlled by the labels $\sigma_i=\pm 1$, corresponding to retarded ($+1$) and advanced prescriptions ($-1$).
Due to the off-shell nature of our computation, we are forced to introduce the hatted propagator $\hat D_{F,5}^{\hat\sigma_5}$.
Once the on-shell limit is taken though, it maps to its unhatted cousin $\hat D_{F,5}^{\hat\sigma_5} \to D_{F,5}^{\sigma_5}$.\footnote{
    In the on-shell limit, it is possible to use partial fractions to map all integrals to the M1 family \cite{Driesse:2024xad}; in this work, we carry out this mapping at the level of master integrals.
}

At \oneSFBar{}, we have the corresponding Mondrian families plus a non-planar family~\cite{Driesse:2026qiz}:
\begin{align}
    I_{\bar F,\{\nu\}}^{\{\sigma\}}(|q|,\gamma,\omega_{\rm IR})
    =
    \int_{\bell_1,\bell_2,\ell_1}
    \frac{
        \dd^{(\nu_1-1)}(D_{\bar F,1})
        \dd^{(\nu_2-1)}(D_{\bar F,2})
        \dd^{(\nu_3-1)}(D_{\bar F,3})
    }{
        \prod_{i=4}^{15}\big[D_{\bar F,i}^{\sigma_i}\big]^{\nu_i}
        [\hat D_{\bar F,6}^{\hat\sigma_6}]^{\hat\nu_6}
    }
    \, ,
\end{align}
for $\bar F=\rm\overbar{M1},\overbar{M2},\rm \overbar{NP}$, whose (inverse) propagators can be found as well in \tab{tab:families}.
We again have on shell that $\hat D_{\bar F,6}^{\hat\sigma_6} \to D_{\bar F,6}^{\sigma_6}$.
If every diagram is to be routed such that solely the total transferred momentum $q^\mu$ flows out of the outgoing line, i.e. $\omega_{\rm IR}=q\cdot v_1$, then the Mondrian families no longer suffice at \oneSFBar{}.
This is the reason for introducing the $\overbar{\rm NP}$ family. Still, in the on-shell limit it maps onto the $\overbar{\rm M1}$ family by the shift $\bell_2\to\bell_2-q$.

With the integral labellings in place, we insert the Feynman rules using the optimised gauge of \cite{Driesse:2024feo}.
The integrand is then reduced to scalar integrals by replacing loop momenta $\ell^\mu$ carrying a free index by
\begin{align}
    \ell^\mu
    \to
    (\ell\cdot e_i)\check e^{i\mu}
    \, ,
    \qquad
    \check e^{i\mu}
    =
    G^{ij}e_j^\mu
    \, ,
    \qquad
    G^{ij}
    =
    (G^{-1})_{ij}
    \, ,
    \qquad
    G_{ij}
    =
    e_i\cdot e_j
    \, ,
\end{align}
where summation on the latin indices is implied.
The basis vectors are $e_i^\mu = v_i^\mu$ for $i=1,2$ and $e_3^\mu = q^\mu$.
On shell, $q^\mu$ is orthogonal to both velocities, so the dual basis reduces to $q^\mu/q^2$ and the usual dual velocities introduced in \eqn{eq:dual-vels}.
Off shell, however, $q^\mu$ is no longer orthogonal to the velocities, and the dual basis receives corrections,
\begin{subequations}
\begin{align}
    \check e^{1\mu}
    &=
    \check v^{1\mu}
    +
    \frac{
        \omega_{\rm IR}
        q^\mu
    }{
        q^2(\gamma^2-1)
    }
    +
    \mathcal{O}(\omega_{\rm IR}^2)
    \, ,
    \\
    \check e^{2\mu}
    &=
    \check v^{2\mu}
    -
    \frac{
        \omega_{\rm IR}
        \gamma
        q^\mu
    }{
        q^2(\gamma^2-1)
    }
    +
    \mathcal{O}(\omega_{\rm IR}^2)
    \, ,
    \\
    \check e^{3\mu}
    &=
    \frac{q^\mu}{q^2}
    +
    \frac{
        \omega_{\rm IR}
        (v_1^\mu-\gamma v_2^\mu)
    }{
        q^2(\gamma^2-1)
    }
    +
    \mathcal{O}(\omega_{\rm IR}^2)
    \, .
\end{align}
\end{subequations}
Note that we keep the notation $\check v^{i\mu}$ reserved for the on-shell dual vectors.
This simple tensor reduction assumes that our integrals are at most rank one which is indeed the case for the spinless off-shell integrand.

The remaining task is then to take the derivative of the deflection one-point function,
\begin{align}
    \frac{\d}{\d \omega_{\rm IR}}
    \Big[
    \i \omega_{\rm IR}^2
    \,
    \begin{tikzpicture}[baseline={(0,-.1)}]
        \coordinate (cL) at (-1,0);
        \coordinate (cM) at (0,0);
        \coordinate (cMP) at (-.1,-.1);
        \drawDirectedBHLine{cL}{cM}
        \drawPerturbativeVertex{cM}
        \draw[line width=.8pt,fill=white] (cL) circle (.23cm) ;
        \draw[pattern=north east lines] (cL) circle (.23cm) ;
        \node[below right=.1cm, font=\footnotesize] at (cL) {$\ \ \omega_{\rm IR}$} ;
        \node[above right=.1cm, font=\scriptsize] at (cMP) {$\mu$} ;
    \end{tikzpicture}
    \Big]
    =
    \frac{\d}{\d \omega_{\rm IR}}
    \int_q
    \dd(\omega_{\rm IR}-q\cdot v_1)
    \dd(q\cdot v_2)
    e^{\i q\cdot b}
    \mathcal{I}^\mu(q)
    \, ,
\end{align}
where $\mathcal{I}^\mu(q)$ is the sum of diagrams in $q$-space. By virtue of the first delta function, we may regard the integrand as independent of $\omega_{\rm IR}$.
In taking the derivative, one is helped by the rewriting,
\begin{align}
    \frac{\d}{\d \omega_{\rm IR}}
    \dd(\omega_{\rm IR}-q\cdot v_1)
    =
    -u^\mu\frac\pat{\pat q^\mu}
    \dd(\omega_{\rm IR}-q\cdot v_1)
    \, ,
    \qquad
    u\cdot v_1
    =
    1
    \, ,
\end{align}
where a convenient choice for the directional derivative is $u^\mu=\check v^{1\mu}$.
With this choice, $\check v^{1\mu}\pat/\pat q^\mu[\dd(q\cdot v_2)e^{\i q\cdot b}]=0$, assuming that $b\cdot \check v^1=0$.
It is not surprising that $\Delta b_i^\mu$ depends on a choice of $b\cdot \check v^1=0$ since it is not invariant under proper time translations.
A partial integration and taking the limit then leads to
\begin{align}\label{eq:integration-pick-up-here}
    \frac{\d}{\d \omega_{\rm IR}}
    \Big[
    \i \omega_{\rm IR}^2
    \,
    \begin{tikzpicture}[baseline={(0,-.1)}]
        \coordinate (cL) at (-1,0);
        \coordinate (cM) at (0,0);
        \coordinate (cMP) at (-.1,-.1);
        \drawDirectedBHLine{cL}{cM}
        \drawPerturbativeVertex{cM}
        \draw[line width=.8pt,fill=white] (cL) circle (.23cm) ;
        \draw[pattern=north east lines] (cL) circle (.23cm) ;
        \node[below right=.1cm, font=\footnotesize] at (cL) {$\ \ \omega_{\rm IR}$} ;
        \node[above right=.1cm, font=\scriptsize] at (cMP) {$\mu$} ;
    \end{tikzpicture}
    \Big]\Big|_{\omega_{\rm IR}\to0}
    =
    \int_q
    \dd(q\cdot v_1)
    \dd(q\cdot v_2)
    e^{\i q\cdot b}
    \,
    \check v^{1\nu}
    \frac{\pat\mathcal{I}^\mu(q)}{\pat q^\nu}
    \, .
\end{align}
In conclusion, the derivative acts only on the $q$-space sum of diagrams.
Since this sum is built from integrals of the families introduced above, so is its derivative, and we may thus easily take the on-shell limit, which in particular sends the hatted propagators to their unhatted partners.

\subsection{Elimination of worldline energies and insertion of dotted lines}
With the integrand in hand, the natural next step is to consider its integrals.
Instead, we will continue dwelling a bit on the formal structure of WQFT Feynman diagrams.
In the following, we will discuss how one may integrate out all worldline energies at the level of the diagrams which gives a formal meaning to the dotted line in WQFT Feynman diagrams.
These formal steps highlight several properties of the PM expansion from WQFT such as how, and why, we may always choose a basis of ``passive'' momenta in our loop integrals which dot to zero either with $v_1^\mu$ or with $v_2^\mu$.

The first goal will be to integrate out the worldline energies.
We will employ the dotted line for this purpose and, in this way, it acquires a life of its own and particular rules associated with it.
In contrast, in ``conventional'' WQFT Feynman diagrams, the dotted line is often used mainly to trace each black hole through the diagram.
A conventional WQFT diagram may therefore be drawn entirely without dotted lines as
\begin{align}\label{eq:exampleDiagramHej}
    \begin{tikzpicture}[baseline={(0,-.1)}]
        \coordinate (t1) at (0,.65);
        \coordinate (t2) at (.55,.65);
        \coordinate (t3) at (1.1,.65);
        \coordinate (t4) at (1.65,.65);
        \coordinate (out) at (2.1,.65);
        \coordinate (b1) at (0,-.65);
        \coordinate (b2) at (.55,-.65);
        \coordinate (b3) at (1.1,-.65);
        \coordinate (b4) at (1.65,-.65);
        \drawBHLine{t1}{t2}
        \drawBHLine{t2}{t3}
        \drawBHLine{b3}{b4}
        \drawDirectedBHLine{t4}{out}
        \drawGravitonLine{t1}{b1}
        \drawGravitonLine{t2}{b2}
        \drawActiveGravitonLine{t3}{b3}
        \drawActiveGravitonLine{b4}{t4}
        \foreach \v in {t1,t2,t3,t4,b1,b2,b3,b4} {\drawMarkedVertex{\v}}
    \end{tikzpicture}
    \ ,
\end{align}
as long as it is made clear what background parameters each WL fluctuation corresponds to (in this case by drawing them above or below the gravitational interactions). 
In this diagram, we use square vertices to represent the conventional, energy-conserving vertices as provided in \eqn{eq:hejFarvel}.

In order to integrate out the worldline energies, we focus on each subdiagram in which only worldlines propagate and refer to each such subdiagram as a ``touch''. 
An example is
\begin{align}
    \begin{tikzpicture}[baseline={(0,-.4)}]
        \coordinate (a) at (0,0);
        \coordinate (aB) at (0,-1);
        \coordinate (b) at (1.3,0);
        \coordinate (bB) at (1.3,-1);
        \coordinate (c) at (2.6,0);
        \coordinate (cB) at (2.6,-1);
        \coordinate (T) at (-.65,.65);
        \drawDirectedBHLine{a}{b}
        \drawDirectedBHLine{b}{c}
        \drawDirectedGravitonLine{aB}{a}
        \drawDirectedGravitonLine{bB}{b}
        \drawDirectedGravitonLine{c}{cB}
        \foreach \v in {a,b,c} {\drawMarkedVertex{\v}}
        \node at ($(aB)!.5!(a)$) [left=.08cm] {\footnotesize $k_1$};
        \node at ($(bB)!.5!(b)$) [left=.08cm] {\footnotesize $k_2$};
        \node at ($(cB)!.5!(c)$) [left=.08cm] {\footnotesize $k_3$};
        \node at ($(a)!.5!(b)$) [above=.08cm] {\footnotesize $\omega_1$};
        \node at ($(b)!.5!(c)$) [above=.08cm] {\footnotesize $\omega_2$};
    \end{tikzpicture}
    \ .
\end{align}
The example graph in \eqn{eq:exampleDiagramHej} has in total two such touches on WL1 and three on WL2 including the simple one-point emissions without any worldline propagation.
Note that in this subgraph we omit distinguishing between active and potential gravitons (and simply let all gravitons be black).

Integrating out the energies of the worldlines in a touch fixes each of them in terms of the graviton momenta coming into the touch.
We may thus regard the worldlines as carrying (linear combinations of) the graviton momenta $k_i^\mu$.
It also leaves behind a phase and a single delta function, which we depict as an open-ended dotted line
\begin{align}\label{eq:open-dotted-line-rule}
    \begin{tikzpicture}[baseline={(0,-.1)}]
        \coordinate (x) at (0,0);
        \coordinate (y) at (1.3,0);
        \drawPerturbativeVertex{x}
        \drawDottedLine{x}{y}
        \node at ($(x)!.5!(y)$) [above=.08cm] {\footnotesize $K$};
    \end{tikzpicture}
    \;=\;
    m_j e^{-\i K\cdot b_j}
    \dd(K\cdot v_j)
    \, ,
\end{align}
where $K^\mu$ is the total momentum flowing into the touch through its connecting gravitons.
It is also convenient to pull out an overall factor of mass which allows one to omit it in WL propagators and vertices; the round vertices and propagation between them are therefore independent of mass.
As the dotted line carries no energy, it may be placed freely on any vertex of the touch so long as momentum conservation is observed.
Consider as an example the three possible placements in the following touch:
\begin{align}\label{eq:touch-example}
        \begin{tikzpicture}[baseline={(0,-.4)}]
        \coordinate (a) at (0,0);
        \coordinate (aB) at (0,-1);
        \coordinate (b) at (1.3,0);
        \coordinate (bB) at (1.3,-1);
        \coordinate (c) at (2.6,0);
        \coordinate (cB) at (2.6,-1);
        \coordinate (T) at (-.65,.65);
        \drawDirectedBHLine{a}{b}
        \drawDirectedBHLine{b}{c}
        \drawDirectedGravitonLine{aB}{a}
        \drawDirectedGravitonLine{bB}{b}
        \drawDirectedGravitonLine{c}{cB}
        \foreach \v in {a,b,c} {\drawMarkedVertex{\v}}
        \node at ($(aB)!.5!(a)$) [right=.08cm] {\footnotesize $k_1$};
        \node at ($(bB)!.5!(b)$) [left=.08cm] {\footnotesize $k_2$};
        \node at ($(cB)!.5!(c)$) [left=.08cm] {\footnotesize $k_3$};
        \node at ($(a)!.5!(b)$) [above=.08cm] {\footnotesize $\omega_1$};
        \node at ($(b)!.5!(c)$) [above=.08cm] {\footnotesize $\omega_2$};
    \end{tikzpicture}
    \,=\,
    \begin{tikzpicture}[baseline={(0,-.4)}]
        \coordinate (a) at (0,0);
        \coordinate (aB) at (0,-1);
        \coordinate (b) at (1.3,0);
        \coordinate (bB) at (1.3,-1);
        \coordinate (c) at (2.6,0);
        \coordinate (cB) at (2.6,-1);
        \coordinate (T) at (-.65,.65);
        \drawDirectedBHLine{a}{b}
        \drawDirectedBHLine{b}{c}
        \drawDottedLine{a}{T}
        \drawDirectedGravitonLine{aB}{a}
        \drawDirectedGravitonLine{bB}{b}
        \drawDirectedGravitonLine{c}{cB}
        \foreach \v in {a,b,c} {\drawPerturbativeVertex{\v}}
        \node at ($(aB)!.5!(a)$) [right=.08cm] {\footnotesize $k_1$};
        \node at ($(bB)!.5!(b)$) [left=.08cm] {\footnotesize $k_2$};
        \node at ($(cB)!.5!(c)$) [left=.08cm] {\footnotesize $k_3$};
        \node at ($(a)!.5!(b)$) [above=.08cm] {\footnotesize $k_3{-}k_2$};
        \node at ($(b)!.5!(c)$) [above=.08cm] {\footnotesize $k_3$};
        \node at ($(a)!.5!(T)$) [below left=-.08cm] {\footnotesize $K$};
    \end{tikzpicture}
    \,=\,
    \begin{tikzpicture}[baseline={(0,-.4)}]
        \coordinate (a) at (0,0);
        \coordinate (aB) at (0,-1);
        \coordinate (b) at (1.3,0);
        \coordinate (bB) at (1.3,-1);
        \coordinate (c) at (2.6,0);
        \coordinate (cB) at (2.6,-1);
        \coordinate (T) at (1.3,.9);
        \drawDirectedBHLine{a}{b}
        \drawDirectedBHLine{b}{c}
        \drawDottedLine{b}{T}
        \drawDirectedGravitonLine{aB}{a}
        \drawDirectedGravitonLine{bB}{b}
        \drawDirectedGravitonLine{c}{cB}
        \foreach \v in {a,b,c} {\drawPerturbativeVertex{\v}}
        \node at ($(aB)!.5!(a)$) [right=.08cm] {\footnotesize $k_1$};
        \node at ($(bB)!.5!(b)$) [left=.08cm] {\footnotesize $k_2$};
        \node at ($(cB)!.5!(c)$) [left=.08cm] {\footnotesize $k_3$};
        \node at ($(a)!.5!(b)$) [above=.08cm] {\footnotesize $k_1$};
        \node at ($(b)!.5!(c)$) [above right=.08cm] {\footnotesize $k_3$};
        \node at ($(b)!.5!(T)$) [above right] {\footnotesize $K$};
    \end{tikzpicture}
    \,=\,
    \begin{tikzpicture}[baseline={(0,-.4)}]
        \coordinate (a) at (0,0);
        \coordinate (aB) at (0,-1);
        \coordinate (b) at (1.3,0);
        \coordinate (bB) at (1.3,-1);
        \coordinate (c) at (2.6,0);
        \coordinate (cB) at (2.6,-1);
        \coordinate (T) at (3.25,.65);
        \drawDirectedBHLine{a}{b}
        \drawDirectedBHLine{b}{c}
        \drawDottedLine{c}{T}
        \drawDirectedGravitonLine{aB}{a}
        \drawDirectedGravitonLine{bB}{b}
        \drawDirectedGravitonLine{c}{cB}
        \foreach \v in {a,b,c} {\drawPerturbativeVertex{\v}}
        \node at ($(aB)!.5!(a)$) [right=.08cm] {\footnotesize $k_1$};
        \node at ($(bB)!.5!(b)$) [left=.08cm] {\footnotesize $k_2$};
        \node at ($(cB)!.5!(c)$) [left=.08cm] {\footnotesize $k_3$};
        \node at ($(a)!.5!(b)$) [above=.08cm] {\footnotesize $k_1$};
        \node at ($(b)!.5!(c)$) [above=.08cm] {\footnotesize $k_1{+}k_2$};
        \node at ($(c)!.5!(T)$) [below right=-.08cm] {\footnotesize $K$};
    \end{tikzpicture}
    ,
\end{align}
where $K^\mu=k_1^\mu+k_2^\mu-k_3^\mu$.
In the first diagram, solid lines have energies associated with them.
Instead, in the three next diagrams, the solid lines have momenta associated with them and there is a single insertion of the dotted line.
Using the Feynman rule in \eqn{eq:open-dotted-line-rule} together with the following rule,
\begin{align}
    \begin{tikzpicture}[baseline={(0,-.1)}]
        \coordinate (x) at (-.6,0);
        \coordinate (y) at (.6,0);
        \drawPerturbativeVertex{x}
        \drawPerturbativeVertex{y}
        \drawDirectedBHLine{x}{y}
        \node at ($(x)!.5!(y)$) [above=.08cm] {\footnotesize $k$};
    \end{tikzpicture}
    \;&=\;
    -\frac{\i\eta^{\mu\nu}}{
        (k\cdot v_j+\iZero)^{2}}
    \, ,
\end{align}
all diagrams in \eqn{eq:touch-example} evaluate to the same expression.
We note that this rule differs from the one given in eq.~\eqref{WLprops} (which is labelled by an energy) by the omission of the mass.
Importantly, we define the round vertex rules without energy-conserving delta functions, the exponential phase or masses. These elements are now all taken care of by the dotted line.

It is a simple consequence of the above equations that
\begin{align}\label{eq:connectingFriends}
    \begin{tikzpicture}[baseline={(0,-.4)}]
        \coordinate (x1) at (0,0);
        \coordinate (x1B) at (0,-1);
        \coordinate (T) at (0,.9);
        \coordinate (x2) at (1.3,0);
        \coordinate (x2B) at (1.3,-1);
        \coordinate (y) at (2.6,0);
        \drawDottedLine{x1}{T}
        \drawDottedLine{x2}{y}
        \drawDirectedGravitonLine{x1B}{x1}
        \drawDirectedGravitonLine{x2B}{x2}
        \drawPerturbativeVertex{x1}
        \drawPerturbativeVertex{x2}
        \node at ($(x1)!.5!(T)$) [right=.04cm] {\footnotesize $K_1$};
        \node at ($(x2)!.5!(y)$) [above=.08cm] {\footnotesize $K_2$};
        \node at ($(x1B)!.5!(x1)$) [right=.08cm] {\footnotesize $K_1$};
        \node at ($(x2B)!.5!(x2)$) [right=.08cm] {\footnotesize $K_2$};
    \end{tikzpicture}
    \;=\;
    \begin{tikzpicture}[baseline={(0,-.4)}]
        \coordinate (x1) at (0,0);
        \coordinate (x1B) at (0,-1);
        \coordinate (x2) at (1.3,0);
        \coordinate (x2B) at (1.3,-1);
        \coordinate (y) at (3,0);
        \drawDottedLine{x1}{x2}
        \drawDottedLine{x2}{y}
        \drawDirectedGravitonLine{x1B}{x1}
        \drawDirectedGravitonLine{x2B}{x2}
        \drawPerturbativeVertex{x1}
        \drawPerturbativeVertex{x2}
        \node at ($(x1)!.5!(x2)$) [above=.08cm] {\footnotesize $K_1$};
        \node at ($(x2)!.5!(y)$) [above=.08cm] {\footnotesize $K_1{+}K_2$};
        \node at ($(x1B)!.5!(x1)$) [right=.08cm] {\footnotesize $K_1$};
        \node at ($(x2B)!.5!(x2)$) [right=.08cm] {\footnotesize $K_2$};
    \end{tikzpicture}
    \; ,
\end{align}
In this equation, each round interaction represents the simplest kind of worldline touch, i.e., a one-point interaction.
It shows how one may connect individual touches into a larger touch by combining them with the dotted line and imposing four-momentum conservation.
In order for this to work, the ``propagator'' of a dotted line must take the form,
\begin{align}
    \begin{tikzpicture}[baseline={(0,-.1)}]
        \coordinate (x) at (-.6,0);
        \coordinate (y) at (.6,0);
        \drawPerturbativeVertex{x}
        \drawPerturbativeVertex{y}
        \drawDottedLine{x}{y}
        \node at ($(x)!.5!(y)$) [above=.08cm] {\footnotesize $k$};
    \end{tikzpicture}
    \;&=\;
    m_j\dd(k\cdot v_j)
\end{align}
It represents propagation along the unperturbed background trajectory.
We note that this aligns with notation for WQFT loop integrals in refs.~\cite{Driesse:2024xad,Driesse:2024feo,benThesis} and is in direct correspondence with the velocity cuts of \cite{Bjerrum-Bohr:2021wwt,Bjerrum-Bohr:2021din}.

The logic of \eqn{eq:connectingFriends} remains true with pairs of more complex worldline touches such as the one considered in \eqn{eq:touch-example}.
One special case of the rules arises if the touch contains the (amputated) outgoing line. Then, instead of the dotted line, one has
\begin{align}
    \begin{tikzpicture}[baseline={(0,-.1)}]
        \coordinate (x) at (0,0);
        \coordinate (y) at (1.3,0);
        \drawPerturbativeVertex{x}
        \drawDirectedBHLine{x}{y}
        \node at ($(x)!.5!(y)$) [above=.08cm] {\footnotesize $K$};
    \end{tikzpicture}
    \;=\;
    m_j
    e^{-\i K\cdot b_j}
    \dd(\omega-K\cdot v_j)
    \, ,
\end{align}
with the outgoing energy $\omega$.
In this case, one may still combine subdiagrams in the same manner as \eqn{eq:connectingFriends}:
\begin{align}
    \begin{tikzpicture}[baseline={(0,-.4)}]
        \coordinate (x1) at (0,0);
        \coordinate (x1B) at (0,-1);
        \coordinate (T) at (0,.9);
        \coordinate (x2) at (1.3,0);
        \coordinate (x2B) at (1.3,-1);
        \coordinate (y) at (2.6,0);
        \drawDottedLine{x1}{T}
        \drawDirectedBHLine{x2}{y}
        \drawDirectedGravitonLine{x1B}{x1}
        \drawDirectedGravitonLine{x2B}{x2}
        \drawPerturbativeVertex{x1}
        \drawPerturbativeVertex{x2}
        \node at ($(x1)!.5!(T)$) [right=.04cm] {\footnotesize $K_1$};
        \node at ($(x2)!.5!(y)$) [above=.08cm] {\footnotesize $K_2$};
        \node at ($(x1B)!.5!(x1)$) [right=.08cm] {\footnotesize $K_1$};
        \node at ($(x2B)!.5!(x2)$) [right=.08cm] {\footnotesize $K_2$};
    \end{tikzpicture}
    \;=\;
    \begin{tikzpicture}[baseline={(0,-.4)}]
        \coordinate (x1) at (0,0);
        \coordinate (x1B) at (0,-1);
        \coordinate (x2) at (1.3,0);
        \coordinate (x2B) at (1.3,-1);
        \coordinate (y) at (3,0);
        \drawDottedLine{x1}{x2}
        \drawDirectedBHLine{x2}{y}
        \drawDirectedGravitonLine{x1B}{x1}
        \drawDirectedGravitonLine{x2B}{x2}
        \drawPerturbativeVertex{x1}
        \drawPerturbativeVertex{x2}
        \node at ($(x1)!.5!(x2)$) [above=.08cm] {\footnotesize $K_1$};
        \node at ($(x2)!.5!(y)$) [above=.08cm] {\footnotesize $K_1{+}K_2$};
        \node at ($(x1B)!.5!(x1)$) [right=.08cm] {\footnotesize $K_1$};
        \node at ($(x2B)!.5!(x2)$) [right=.08cm] {\footnotesize $K_2$};
    \end{tikzpicture}
    \; ,
\end{align}

Returning to the initial example in \eqn{eq:exampleDiagramHej}, one would then integrate out the worldline energies diagrammatically as,
\begin{align}\label{eq:threeGraphsMoveRightToLeft}
    \begin{tikzpicture}[baseline={(0,-.1)}]
        \coordinate (t1) at (0,.65);
        \coordinate (t2) at (.55,.65);
        \coordinate (t3) at (1.1,.65);
        \coordinate (t4) at (1.65,.65);
        \coordinate (out) at (2.1,.65);
        \coordinate (b1) at (0,-.65);
        \coordinate (b2) at (.55,-.65);
        \coordinate (b3) at (1.1,-.65);
        \coordinate (b4) at (1.65,-.65);
        \drawBHLine{t1}{t2}
        \drawBHLine{t2}{t3}
        \drawBHLine{b3}{b4}
        \drawDirectedBHLine{t4}{out}
        \drawGravitonLine{t1}{b1}
        \drawGravitonLine{t2}{b2}
        \drawActiveGravitonLine{t3}{b3}
        \drawActiveGravitonLine{b4}{t4}
        \foreach \v in {t1,t2,t3,t4,b1,b2,b3,b4} {\drawMarkedVertex{\v}}
    \end{tikzpicture}
    \;=\;
    \begin{tikzpicture}[baseline={(0,-.1)}]
        \coordinate (t1) at (0,.65);
        \coordinate (t2) at (.55,.65);
        \coordinate (t3) at (1.1,.65);
        \coordinate (t4) at (1.65,.65);
        \coordinate (out) at (2.1,.65);
        \coordinate (b1) at (0,-.65);
        \coordinate (b2) at (.55,-.65);
        \coordinate (b3) at (1.1,-.65);
        \coordinate (b4) at (1.65,-.65);
        \drawDottedLine{t3}{$(t3)+(0,.55)$}
        \drawDottedLine{b1}{$(b1)-(0,.55)$}
        \drawDottedLine{b2}{$(b2)-(0,.55)$}
        \drawDottedLine{b4}{$(b4)-(0,.55)$}
        \drawBHLine{t1}{t2}
        \drawBHLine{t2}{t3}
        \drawBHLine{b3}{b4}
        \drawDirectedBHLine{t4}{out}
        \drawGravitonLine{t1}{b1}
        \drawGravitonLine{t2}{b2}
        \drawActiveGravitonLine{t3}{b3}
        \drawActiveGravitonLine{b4}{t4}
        \foreach \v in {t1,t2,t3,t4,b1,b2,b3,b4} {\drawPerturbativeVertex{\v}}
    \end{tikzpicture}
    \;=\;
    \begin{tikzpicture}[baseline={(0,-.1)}]
        \coordinate (t1) at (0,.65);
        \coordinate (t2) at (.55,.65);
        \coordinate (t3) at (1.1,.65);
        \coordinate (t4) at (1.65,.65);
        \coordinate (out) at (2.1,.65);
        \coordinate (b1) at (0,-.65);
        \coordinate (b2) at (.55,-.65);
        \coordinate (b3) at (1.1,-.65);
        \coordinate (b4) at (1.65,-.65);
        \drawDottedLine{t3}{t4}
        \drawDottedLine{b1}{b3}
        \drawDottedLine{b4}{out|-b4}
        \drawBHLine{t1}{t2}
        \drawBHLine{t2}{t3}
        \drawBHLine{b3}{b4}
        \drawDirectedBHLine{t4}{out}
        \drawGravitonLine{t1}{b1}
        \drawGravitonLine{t2}{b2}
        \drawActiveGravitonLine{t3}{b3}
        \drawActiveGravitonLine{b4}{t4}
        \foreach \v in {t1,t2,t3,t4,b1,b2,b3,b4} {\drawPerturbativeVertex{\v}}
    \end{tikzpicture}
    \ .
\end{align}
In the first graph, each solid line is assigned an energy.
In contrast, in the second and third graphs each solid line and each dotted line is assigned a four-momentum with four-momentum conservation imposed at each (round) vertex.
The third graph clearly shows the loop integral structure and may be read as a ``standard'' three-loop Feynman graph.
Each manipulation going from left to right in \eqn{eq:threeGraphsMoveRightToLeft} is not unique and the ``classical loops'' can therefore be ``closed'' in many different manners.
Any choice is legitimate and leads to the same result when vertex rules are inserted.

\subsection{Four-momentum conserving WQFT Feynman rules}
Let us summarise this formulation of the WQFT Feynman rules, starting with the propagators:
\begin{subequations}\label{eq:new-rules}
\begin{align}
    \begin{tikzpicture}[baseline={(0,-.1)}]
        \coordinate (x) at (-.6,0);
        \coordinate (y) at (.6,0);
        \drawPerturbativeVertex{x}
        \drawPerturbativeVertex{y}
        \drawDirectedBHLine{x}{y}
        \node at ($(x)!.5!(y)$) [above=.08cm] {\footnotesize $k$};
        \node at (x) [below left] {\footnotesize $j$};
    \end{tikzpicture}
    \;&=\;
    -\frac{\i\eta^{\mu\nu}}{(k\cdot v_j+\iZero)^{2}}
    \, ,
    &
    \begin{tikzpicture}[baseline={(0,-.1)}]
        \coordinate (x) at (-.6,0);
        \coordinate (y) at (.6,0);
        \drawPerturbativeVertex{x}
        \drawPerturbativeVertex{y}
        \drawDottedLine{x}{y}
        \node at ($(x)!.5!(y)$) [above=.08cm] {\footnotesize $k$};
        \node at (x) [below left] {\footnotesize $j$};
    \end{tikzpicture}
    \;&=\;
    m_j\dd(k\cdot v_j)
    \, ,
    \\
    \begin{tikzpicture}[baseline={(0,-.1)}]
        \coordinate (x) at (-.6,0);
        \coordinate (y) at (.6,0);
        \drawPerturbativeVertex{x}
        \drawDirectedBHLine{x}{y}
        \node at ($(x)!.5!(y)$) [above=.08cm] {\footnotesize $k$};
        \node at (x) [below left] {\footnotesize $j$};
    \end{tikzpicture}
    \;&=\;
    m_j
    e^{-\i k\cdot b_j}
    \dd(\omega-k\cdot v_j)
    \, ,
    &
    \begin{tikzpicture}[baseline={(0,-.1)}]
        \coordinate (x) at (-.6,0);
        \coordinate (y) at (.6,0);
        \drawPerturbativeVertex{x}
        \drawDottedLine{x}{y}
        \node at ($(x)!.5!(y)$) [above=.08cm] {\footnotesize $k$};
        \node at (x) [below left] {\footnotesize $j$};
    \end{tikzpicture}
    \;&=\;
    m_j
    e^{-\i k\cdot b_j}
    \dd(k\cdot v_j)
    \, .
\end{align}
\end{subequations}
Each propagator (external or internal) is assigned a four-momentum $k^\mu$ which is taken to flow towards the right.
The index $j$ traces which of the worldlines the rule is applied to and is usually left implicit by drawing WL1 above and WL2 below the gravitational interactions.

The two simplest vertex rules read,
\begin{subequations}\label{eq:FeynmanRulesWL}
\begin{align}
    \begin{tikzpicture}[baseline={(0,-.6)}]
        \coordinate (v) at (0,0);
        \coordinate (gOut) at (0,-1);
        \drawPerturbativeVertex{v}
        \drawGravitonLine{v}{gOut}
        \node at (gOut) [below] {\footnotesize $\mn$};
        \node at ($(v)!.5!(gOut)$) [left=.08cm] {\footnotesize $k$};
    \end{tikzpicture}
    \;&=\;
    -
    \frac{\i\kappa}2
    v_j^\mu v_j^\nu
    \, , 
    \\
    \begin{tikzpicture}[baseline={(0,-.6)}]
        \coordinate (v) at (0,0);
        \coordinate (out) at (1,0);
        \coordinate (gOut) at (0,-1);
        \drawBHLine{v}{out}
        \drawPerturbativeVertex{v}
        \drawGravitonLine{v}{gOut}
        \node at (gOut) [below] {\footnotesize $\mn$};
        \node at (out) [right] {\footnotesize $\rho$};
        \node at ($(v)!.5!(gOut)$) [left=.08cm] {\footnotesize $k$};
        \node at ($(v)!.5!(out)$) [above=.08cm] {\footnotesize $p$};
    \end{tikzpicture}
    \;&=\;
    \frac{\kappa}2
    \big(
        v_j^\mu v_j^\nu k_\rho
        +
        2(p\cdot v_j)v_j^{(\mu}\delta^{\nu)}_{\rho}
    \big)
    \, ,
\end{align}
\end{subequations}
with all momenta outgoing.
The general $n$-point vertex~\cite{Mogull:2020sak} is,
\begin{align}
    &
    \begin{tikzpicture}[baseline={([yshift=-.5ex]current bounding box.center)},yscale=-1.15,xscale=1.15]
        \drawCurvedGravitonLine{.6,0}{1.3,.7}{90}{180}
        \drawBHLineR{.6,0}{-.6}
        \drawCurvedBHLine{.6,0}{0,.7}{150}{0}
        \drawPerturbativeVertex{.6,0}
        \node[rotate=90] at (0,.35) {\footnotesize \dots} ;
        \node at (1.05,.7) [below=2pt] {\footnotesize $k$} ;
        \node at (1.3,.7) [right=1pt] {\footnotesize $\mu\nu$} ;
        \node at (.3,0) [above=1pt] {\footnotesize $p_1$} ;
        \node at (.3,.64) [below=1pt] {\footnotesize $p_n$} ;
        \node at (0,0) [left=-2pt] {\footnotesize $\rho_1$} ;
        \node at (0,.7) [left=-2pt] {\footnotesize $\rho_n$} ;
    \end{tikzpicture}
    =
    \i^{n-1}
    \kappa 
    \bigg[
        \frac{1}{2}
            \bigg(
            \prod_{i=1}^n k_{\rho_i}
            \bigg)
            v_j^\mu v_j^\nu
    +\sum_{i=1}^n
    (v_j\cdot p_i)
        \bigg(
        \prod_{r\neq i}^n k_{\rho_r}
        \bigg)
        v_j^{(\mu}\delta^{\nu)}_{\rho_i}
        \\*
        &\hspace{8cm}
    +\sum_{i<r}^n
    (v_j\cdot p_i)
    (v_j\cdot p_r)
        \bigg(
        \prod_{l\neq i,r}^n k_{\rho_l}
        \bigg)
        \delta^{(\mu}_{\rho_i}\delta^{\nu)}_{\rho_r}
    \bigg].
    \nn
\end{align}
Again, all lines, solid and wiggly, are assigned a four-momentum on which the vertex rules depend.

To evaluate a diagram with these rules, one first assembles all tree-level WQFT diagrams ``as usual'' without any dotted lines and energy-conserving square interactions on the WLs.
Then, all loops are closed with dotted lines whereby, each square energy-conserving vertex is turned into a round four-momentum conserving vertex.
Indeed, these are the steps exemplified in \eqn{eq:threeGraphsMoveRightToLeft}.

Let us comment a bit on the general structure of WQFT diagrams from this perspective:
At $n$PM order, a graph will always have $n-1$ loops after turning all square interactions into four-momentum conserving round ones.
Each loop will be composed of exactly one dotted line and otherwise gravitons and solid lines.
Further the self-force order of a graph is determined by the number of dotted lines on each of the WLs.
Indeed, all of this formalises patterns known from work on WQFT~\cite{Driesse:2026qiz,Driesse:2024feo,Haddad:2025cmw,Driesse:2024xad}.

\begin{figure}[t]
    \centering
    \captionsetup[sub]{skip=0pt}
    \subcaptionbox{\label{fig:example-labellings-a}}{%
        \begin{tikzpicture}[baseline={(0,0)}]
            \coordinate (t1) at (0,1);
            \coordinate (v2) at (1.3,.4);
            \coordinate (t3) at (2.6,1);
            \coordinate (t4) at (3.9,1);
            \coordinate (out) at (4.8,1);
            \coordinate (b1) at (0,-1);
            \coordinate (b2) at (1.3,-1);
            \coordinate (b3) at (2.6,-1);
            \coordinate (b4) at (3.9,-1);
            \coordinate (qe) at (4.8,-1);
            \drawDottedLine{t3}{t4}
            \drawDottedLine{b1}{b2}
            \drawDottedLine{b2}{b3}
            \drawDottedLine{b4}{qe}
            \drawBHLine{t1}{t3}
            \drawCurvedBHLine{v2}{t3}{0}{180}
            \drawBHLine{b3}{b4}
            \drawDirectedBHLine{t4}{out}
            \drawGravitonLine{t1}{b1}
            \drawGravitonLine{v2}{b2}
            \drawActiveGravitonLine{t3}{b3}
            \drawActiveGravitonLine{b4}{t4}
            \foreach \v in {t1,v2,t3,t4,b1,b2,b3,b4} {\drawPerturbativeVertex{\v}}
            \begin{scope}[transparency group, opacity=.4]
                \draw[loopContour=.72, sharp corners]
                    (1.175,.4) -- (1.175,-.775)
                    .. controls (1.175,-.83) and (1.13,-.875) .. (1.075,-.875)
                    -- (.225,-.875)
                    .. controls (.17,-.875) and (.125,-.83) .. (.125,-.775)
                    -- (.125,.775)
                    .. controls (.125,.83) and (.17,.875) .. (.225,.875)
                    -- (1.869,.875)
                    .. controls (1.969,.875) and (1.969,.875) .. (1.891,.813)
                    .. controls (1.715,.671) and (1.538,.525) .. (1.3,.525)
                    .. controls (1.231,.525) and (1.175,.469) .. (1.175,.4)
                    -- cycle;
                \draw[loopContour]
                    (1.425,.284) .. controls (1.885,.346) and (2.121,.795) .. (2.475,.862)
                    -- (2.475,-.875) -- (1.425,-.875) -- cycle;
                \draw[loopContour]
                    (2.725,.875) -- (3.775,.875) -- (3.775,-.875) -- (2.725,-.875) -- cycle;
                \draw[loopContour=.11]
                    (4.75,-.875) -- (4.025,-.875) -- (4.025,.875) -- (4.75,.875);
            \end{scope}
            \node at (.6,0) {\scriptsize$\ell_1$};
            \node at (1.9,-.35) {\scriptsize$\ell_2$};
            \node at (3.25,0) {\scriptsize$\bar\ell_1$};
            \node at (4.4,0) {\scriptsize$q$};
            \path (2.4,-1.85); 
        \end{tikzpicture}%
    }
    \hfill
    \subcaptionbox{\label{fig:example-labellings-b}}{%
        \begin{tikzpicture}[baseline={(0,0)}]
            \coordinate (t1) at (0,1);
            \coordinate (v2) at (1.3,.4);
            \coordinate (t3) at (2.6,1);
            \coordinate (t4) at (3.9,1);
            \coordinate (out) at (4.8,1);
            \coordinate (b1) at (0,-1);
            \coordinate (b2) at (1.3,-1);
            \coordinate (b3) at (2.6,-1);
            \coordinate (b4) at (3.9,-1);
            \coordinate (qe) at (2.6,-1.6);
            \drawDottedLine{t3}{t4}
            \drawDottedLine{b2}{b3}
            \drawCurvedDottedLine[1]{b1}{b3}{-50}{-130}
            \drawDottedLine{b3}{qe}
            \drawBHLine{t1}{t3}
            \drawCurvedBHLine{v2}{t3}{0}{180}
            \drawBHLine{b3}{b4}
            \drawDirectedBHLine{t4}{out}
            \drawGravitonLine{t1}{b1}
            \drawGravitonLine{v2}{b2}
            \drawActiveGravitonLine{t3}{b3}
            \drawActiveGravitonLine{b4}{t4}
            \foreach \v in {t1,v2,t3,t4,b1,b2,b3,b4} {\drawPerturbativeVertex{\v}}
            \begin{scope}[transparency group, opacity=.4]
                \draw[loopContour]
                    (.125,.875) -- (2.475,.875) -- (2.475,-.954)
                    .. controls (1.875,-1.625) and (.725,-1.625) .. (.125,-.954) -- cycle;
                \draw[loopContour]
                    (1.425,.284) .. controls (1.841,.344) and (2.056,.692) .. (2.35,.821)
                    -- (2.35,-.875) -- (1.425,-.875) -- cycle;
                \draw[loopContour]
                    (2.725,.875) -- (3.775,.875) -- (3.775,-.875) -- (2.725,-.875) -- cycle;
                \draw[loopContour=.25]
                    (2.725,-1.6) -- (2.725,-1.125) -- (4.025,-1.125) -- (4.025,.875) -- (4.75,.875);
            \end{scope}
            \node at (.6,0) {\scriptsize$\ell_1$};
            \node at (1.9,-.35) {\scriptsize$\ell_2$};
            \node at (3.25,0) {\scriptsize$\bar\ell_1$};
            \node at (4.4,0) {\scriptsize$q$};
            \path (2.4,-1.85); 
        \end{tikzpicture}%
    }
    \hfill
    \subcaptionbox{\label{fig:example-labellings-c}}{%
        \begin{tikzpicture}[baseline={(0,0)}]
            \coordinate (tl) at (0,1);
            \coordinate (tm) at (1.6,1);
            \coordinate (tr) at (3.2,1);
            \coordinate (out) at (4.0,1);
            \coordinate (g1) at (0,.15);
            \coordinate (r) at (3.2,.15);
            \coordinate (g3) at (1.6,-.45);
            \coordinate (bl) at (0,-1);
            \coordinate (bm) at (1.6,-1);
            \coordinate (qe) at (4.0,-1);
            \drawCurvedDottedLine[1]{tl}{tr}{50}{130}
            \drawDottedLine{tm}{tr}
            \drawDottedLine{bl}{bm}
            \drawDottedLine{bm}{qe}
            \drawDirectedBHLine{tr}{out}
            \drawGravitonLine{tl}{g1}
            \drawGravitonLine{g1}{bl}
            \drawActiveGravitonLine{tr}{r}
            \drawGravitonLine{g3}{bm}
            \drawGravitonLine{tm}{g3}
            \drawShadowLine{g1}{r}
            \drawActiveGravitonLine{r}{g3}
            \drawActiveGravitonLine{g1}{r}
            \foreach \v in {tl,tm,tr,g1,r,g3,bl,bm} {\drawPerturbativeVertex{\v}}
            \begin{scope}[transparency group, opacity=.4]
                \draw[loopContour=.22]
                    (.126,.954) .. controls (.876,1.804) and (2.324,1.804) .. (3.074,.954)
                    -- (3.075,.275) -- (.125,.275) -- cycle;
                \draw[loopContour]
                    (1.85,.75) -- (1.85,-.089) -- (2.95,.323) -- (2.95,.75) -- cycle;
                \draw[loopContour]
                    (.125,.025) -- (2.511,.025) -- (1.475,-.363) -- (1.475,-.875) -- (.125,-.875) -- cycle;
                \draw[loopContour=.19]
                    (3.95,-.875) -- (1.725,-.875) -- (1.725,-.537) -- (3.325,.063)
                    -- (3.325,.875) -- (3.95,.875);
            \end{scope}
            \node at (.8,.65) {\scriptsize$\bar\ell_1$};
            \node at (2.4,.47) {\scriptsize$\bar\ell_2$};
            \node at (.75,-.45) {\scriptsize$\ell_1$};
            \node at (3.65,-.25) {\scriptsize$q$};
            \path (2,-1.85); 
        \end{tikzpicture}%
    }
        \caption{
            Possible ways of closing the classical loops of the diagram in \fig{fig:sample-diagrams-c} are shown in \subref{fig:example-labellings-a} and \subref{fig:example-labellings-b}, while a single way of doing it for the diagram in \fig{fig:sample-diagrams-j} is shown in \subref{fig:example-labellings-c}.
            The routing in \subref{fig:example-labellings-a} belongs to the M2 family, while \subref{fig:example-labellings-b} fits none of our families.
            The routing in \subref{fig:example-labellings-c} belongs to the $\overbar{\rm NP}$ family.
        }
    \label{fig:example-labellings}
\end{figure}
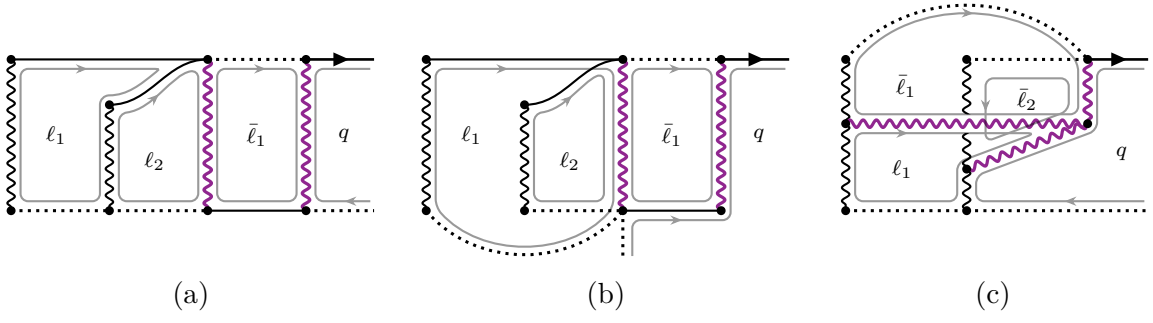

The dotted lines clearly exhibit one of the simplifications of the classical limit:
Loop integration variables may always be chosen in $D-1$ dimensions instead of $D$ dimensions.
Namely, consider a diagram at $n$PM order and $m\text{SF}=\overbar{m{\rm SF}}$ order where $\bar m=n-m-1$ and the SF order is generally labelled by the lesser of $m$ and $\bar m$.
In this case there must be $m$ dotted lines on WL1 and $\bar m$ dotted lines on WL2.
One may choose the loop momenta that flow through these dotted lines as integration variables and one then ends up with $\{\ell_i,\bar\ell_j\}$ with $1\le i\le \bar m$ and $1\le j\le m$ that satisfy,
\begin{align}\label{eq:potentialCon}
    \ell_i\cdot v_2=0
    \ ,
    \hspace{3cm}
    \bar\ell_j\cdot v_1=0
    \ .
\end{align}
This choice of loop momenta at 4PM 1SF and $\overbar{1\rm SF}$ is depicted in \fig{fig:example-labellings}.
Namely, the dotted lines in those diagrams are each assigned a single four-momentum in contrast to graviton and deflection lines which, generally, are assigned linear combinations of four-momenta.
The integral families in \tab{tab:families} have been chosen in exactly this manner.

With this labelling choice one may also straightforwardly categorise momenta into active and passive or radiative and potential ones.
Namely, the integration variables themselves are potential due to \eqn{eq:potentialCon}.
Any linear combination of unbarred momenta is still potential and the same is true for any linear combination of barred momenta.
Active graviton momenta, then, are always linear combinations of at least one barred and one unbarred momentum.

Whether a Mondrian labelling is possible is equally visible.
In \fig{fig:example-labellings-a}, every line of the diagram separates two loops and thus carries the difference of their momenta.
The curved deflection line, in particular, gives rise to the (inverse) propagator $(\ell_1-\ell_2)\cdot v_1$ of M2. In \fig{fig:example-labellings-b}, on the other hand, the third graviton is bordered by three loops and thus falls outside the Mondrian family.
The diagram in \fig{fig:sample-diagrams-j}, finally, admits no Mondrian labelling off shell, but may be mapped to $\overbar{\rm M1}$ after the on-shell limit is taken.

It is interesting to note that after closing the classical loops, the ``one-point functions'' with one outgoing solid line are now, labelling-wise, a two-point function with a single four-momentum $q^\mu$ flowing through the graph.
Another example would be the waveform which is given from graphs with one outgoing graviton line (i.e. \fig{fig:waveformkappa3}).
Each of the two WLs adds an external dotted line after closing the classical loops.
Labelling-wise, the waveform is therefore a three-point function which depends on two external momenta $q_1$ and $q_2$.
Those momenta are exactly the ones of \eqn{eq:paraFour}.

This perspective of the $\langle z^\mu(\omega)\rangle$ being a two-point function and $\langle h_\mn(k)\rangle$ being a three-point function is most clear if one works in $q$-space instead of $b$-space.
Namely, conventionally in WQFT diagrams, one integrates over all loop momenta.
Due to the phase on the external lines in \eqn{eq:new-rules}, integrations on the momentum transfers are Fourier transforms.
To work in $q$-space, one simply omits the phases of external lines and omits an integration over the external momenta.

\section{Three-loop integration}
\label{sec:integrals}
Having constructed the integrand, we now proceed with integrating it.
We describe the on-shell integral family, as well as the reduction of the integrals to a set of master integrals through integration-by-parts (IBP) identities and the canonicalisation of the resulting system of differential equations (DE).
Finally, some details on the boundary matching in the PN limit are given.

The 1SF planar-2 family, together with the \oneSFBar{} planar-1, planar-2 and non-planar families, is easily mapped to the 1SF planar-1 family once the on-shell limit has been taken (for details on this, see ref. \cite{Driesse:2024xad}).
We hence confine our attention to the 1SF planar-1 family throughout this section.

\subsection{On-shell integral family}
The on-shell 1SF planar-1 family reads
\begin{equation}\label{eq:M1-onshell}
    I^{\{\sigma\}}_{\{\nu\}}(|q|,\gamma,\epsilon)
    =
    \int_{\ell_1,\ell_2,\bell_1}
    \frac{
        \dd^{(\nu_1-1)}(D_1)\,
        \dd^{(\nu_2-1)}(D_2)\,
        \dd^{(\nu_3-1)}(D_3)
    }{
        \prod_{i=4}^{15}\big[D_{i}^{\sigma_i}\big]^{\nu_i}
    }
    \, ,
\end{equation}
where superscripts on delta functions signify derivatives.
The worldline propagators are
\begin{align}\label{eq:M1-wl-props}
\begin{aligned}
    D_1
    &=
    \ell_1\cdot v_2
    \, ,
    &\hspace{2.5cm}
    D_4^{\sigma_4}
    &=
    \ell_1\cdot v_1+\sigma_4 \iZero
    \, ,
    \\
    D_2
    &=
    \ell_2\cdot v_2
    \, ,
    &\hspace{2.5cm}
    D_5^{\sigma_5}
    &=
    \ell_2\cdot v_1+\sigma_5 \iZero
    \, ,
    \\
    D_3
    &=
    \bell_1\cdot v_1
    \, ,
    &\hspace{2.5cm}
    D_6^{\sigma_6}
    &=
    \bell_1\cdot v_2+\sigma_6 \iZero
    \, ,
\end{aligned}
\end{align}
where the first column is associated with the dotted line and the second with the deflection lines.
The dotted line indices are naturally restricted to $\nu_{1,2,3}\geq1$.
The two active gravitons, i.e., the gravitons that can go on shell within the integration region, are
\begin{align}\label{eq:M1-active-props}
\begin{aligned}
    D_7^{\sigma_7}
    =
    (\ell_1^0-\bell_1^0 + \sigma_7 \iZero)^2
    -
    (\vct{\ell}_1-\vct{\bell}_1)^2
    \, ,
    \\
    D_8^{\sigma_8}
    =
    (\ell_2^0 - \bell_1^0+\sigma_8 \iZero)^2
    -
    (\vct{\ell}_2 - \vct{\bell}_1)^2
    \, ,
\end{aligned}
\end{align}
and the remaining passive graviton propagators are
\begin{align}\label{eq:M1-passive-props}
    D_9
    &=
    (\ell_1-\ell_2)^2
    \, ,
    &\quad
    D_{10}
    &=
    (\ell_1-q)^2
    \, ,
    &\quad
    D_{11}
    &=
    (\ell_2-q)^2
    \, ,
    &\quad
    D_{12}
    &=
    (\bell_1-q)^2
    \, ,
    \notag\\
    D_{13}
    &=
    \ell_1^2
    \, ,
    &\quad
    D_{14}
    &=
    \ell_2^2
    \, ,
    &\quad
    D_{15}
    &=
    \bell_1^2
    \, .
\end{align}
Recall that $\sigma_i$ marks the pole prescription, which determines the causality of the process and must be kept track of throughout all steps of the calculation other than the IBP reduction.

Ignoring pole prescriptions, the 1SF integrand contains 38\,797 integrals of the form \eqref{eq:M1-onshell} with distinct indices $\{\nu\}$.
There are a few observations we can make to decrease this number before IBP reduction.
First, we note that many of the integrals are related by the symmetries.
These are
\begin{align}
    \text{12-swap}:
    &\quad
    (\ell_1^{\mu}\leftrightarrow \ell_2^{\mu})
    \, ,
    \nonumber\\
    \text{Sign flip}:
    &\quad
    (
        \ell_1^\mu\rightarrow -\ell_1^\mu
        \, , \
        \ell_2^\mu\rightarrow -\ell_2^\mu
        \, , \
        \bar{\ell}_1^\mu\rightarrow -\bar{\ell}_1^\mu
        \, , \
        q^\mu\rightarrow-q^\mu
    )
    \, ,
    \\
    \text{$q$-shift}:
    &\quad
    (
        \ell_1^\mu\rightarrow\ell_1^\mu-q^\mu
        \, , \
        \ell_2^\mu\rightarrow \ell_2^\mu-q^\mu
        \, , \
        \bar{\ell}_1^\mu\rightarrow \bar{\ell}_1^\mu-q^\mu
        \, , \
        q^\mu\rightarrow-q^\mu
    )
    \, .
    \notag
    \label{eq:symmetries}
\end{align}
Each transformation is understood as a simultaneous replacement.
Since the integrals depend on $q^\mu$ only through the modulus $|q|$, the replacement $q^\mu\to-q^\mu$ is harmless, and the shift by $q^\mu$ leaves the linear propagators invariant only because of the on-shell condition $q\cdot v_1=q\cdot v_2=0$.
One can also understand these symmetries diagrammatically using the tools introduced in the integrand section. In particular, the $q$-shift corresponds to moving the dotted lines from the right of the diagram to the left.
Note that symmetry replacements may change the $\sigma_i$. 
Applying these symmetries, the number of integrals, ignoring pole prescriptions, reduces to 28\,145.

A further simplification is offered by the fact that we only need the 4PM worldline offset shift along the $b^\mu$-direction, meaning we need only work with integrals with odd parity under $v_i^\mu\to-v_i^\mu$.
Such odd-parity integrals have an odd number of worldline propagators including delta functions, which, in terms of a condition on the indices $\{\nu\}$, corresponds to keeping integrals for which the condition $\sum_{i=1}^6 \nu_i=2 m+1$ for $m\in \mathbb{Z}$ is true.

\subsection{IBPs and the canonical DE}
Computing such a large number of integrals directly is unfeasible.
We therefore take advantage of IBP relations, which allow us to reduce all integrals appearing in our integrand to a significantly smaller number of master integrals. IBPs are centred around the dimensional regularisation identity
\begin{align}
\int_k \frac{\partial}{\partial k^\mu}\frac{w^\mu \mathcal{N}(|q|,\gamma,\epsilon)}{\prod_{i=1}^n  D_{i}^{\nu_i}}=0\ ,
\end{align}
where, in the setting of this work, $w^\mu$ is any vector in the span of $\{v_1^\mu, v_2^\mu, q^\mu,\ell_1^\mu, \ell_2^\mu, \bar{\ell}_1^\mu\}$, while $k^\mu$ can be any of our loop momenta.
These relations lead to a system of equations which allows the reduction to master integrals.

On a practical level, one uses codes that implement the IBP reduction algorithmically. For our problem, we use \texttt{Kira 3.0} \cite{Maierhofer:2017gsa, Klappert:2020nbg, Lange:2025fba} to generate the IBP relations and carry out the ensuing reduction of our integrals.
For IBPs, the pole prescription is irrelevant and we may thus disregard them when doing the reduction.
They however re-enter the computation when performing the boundary matching.

One is helped in the reduction and the preparation of input files for \texttt{Kira 3.0} by implementing the organising principle of top sectors.
After acting with the symmetries above, we find fifteen such top sectors, four of which are irreducible (see \fig{fig:red-top-sectors} and \fig{fig:irr-top-sectors}).
To each of the top sectors we assign seeds adapted precisely to the integrals appearing in the concrete integrand we are dealing with.
Additionally, if a top sector contains a subsector with particularly high seeds, it is favourable to reduce this sector independently of its top sector, such that one avoids running the whole top sector with seeds that, for the most part, are too high.
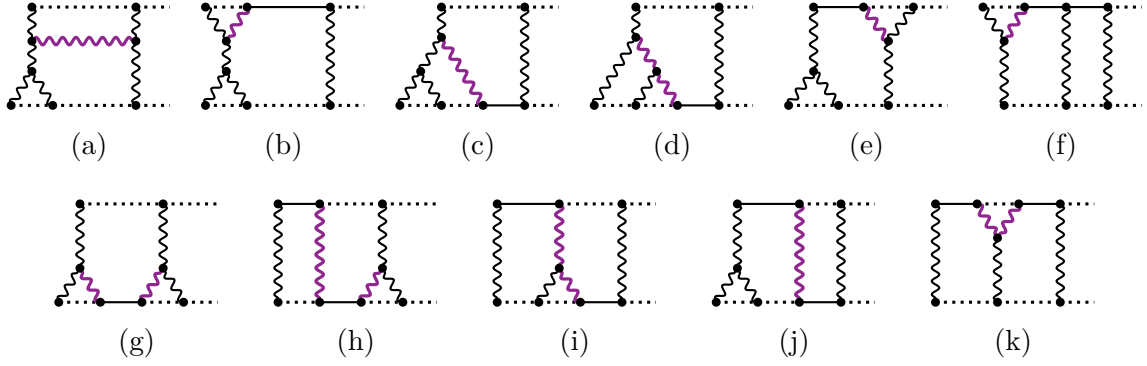
\begin{figure}[t]
    \centering
    \begin{subfigure}[b]{0.15\textwidth}
        \centering
        \begin{tikzpicture}[baseline={(0,-.1)}]
            \coordinate (b1) at (0.000,-0.650);
            \coordinate (b2) at (0.550,-0.650);
            \coordinate (b3) at (1.650,-0.650);
            \coordinate (g1) at (0.275,-0.200);
            \coordinate (g2) at (0.275,0.200);
            \coordinate (g3) at (1.650,0.200);
            \coordinate (t1) at (0.275,0.650);
            \coordinate (t2) at (1.650,0.650);
            \drawDottedLine{t1}{t2}
            \drawDottedLineR{t2}{0.45}
            \drawDottedLine{b1}{b2}
            \drawDottedLine{b2}{b3}
            \drawDottedLineR{b3}{0.45}
            \drawGravitonLine{b2}{g1}
            \drawActiveGravitonLine{g2}{g3}
            \drawGravitonLine{g1}{g2}
            \drawGravitonLine{b3}{g3}
            \drawGravitonLine{b1}{g1}
            \drawGravitonLine{t1}{g2}
            \drawGravitonLine{t2}{g3}
            \foreach \v in {b1,b2,b3,g1,g2,g3,t1,t2} {\drawPerturbativeVertex{\v}}
        \end{tikzpicture}
        \caption{}
        \label{fig:red-top-sectors-a}
    \end{subfigure}
    \hfill
    \begin{subfigure}[b]{0.15\textwidth}
        \centering
        \begin{tikzpicture}[baseline={(0,-.1)}]
            \coordinate (b1) at (0.000,-0.650);
            \coordinate (b2) at (0.550,-0.650);
            \coordinate (b3) at (1.650,-0.650);
            \coordinate (g1) at (0.275,-0.200);
            \coordinate (g2) at (0.275,0.200);
            \coordinate (t1) at (0.000,0.650);
            \coordinate (t2) at (0.550,0.650);
            \coordinate (t3) at (1.650,0.650);
            \drawDottedLine{t1}{t2}
            \drawBHLine{t2}{t3}
            \drawDottedLineR{t3}{0.45}
            \drawDottedLine{b1}{b2}
            \drawDottedLine{b2}{b3}
            \drawDottedLineR{b3}{0.45}
            \drawGravitonLine{b2}{g1}
            \drawActiveGravitonLine{t2}{g2}
            \drawGravitonLine{g1}{g2}
            \drawGravitonLine{t3}{b3}
            \drawGravitonLine{b1}{g1}
            \drawGravitonLine{t1}{g2}
            \foreach \v in {b1,b2,b3,g1,g2,t1,t2,t3} {\drawPerturbativeVertex{\v}}
        \end{tikzpicture}
        \caption{}
        \label{fig:red-top-sectors-b}
    \end{subfigure}
    \hfill
    \begin{subfigure}[b]{0.15\textwidth}
        \centering
        \begin{tikzpicture}[baseline={(0,-.1)}]
            \coordinate (b1) at (0.000,-0.650);
            \coordinate (b2) at (0.550,-0.650);
            \coordinate (b3) at (1.100,-0.650);
            \coordinate (b4) at (1.650,-0.650);
            \coordinate (g1) at (0.275,-0.200);
            \coordinate (g2) at (0.550,0.250);
            \coordinate (t1) at (0.550,0.650);
            \coordinate (t2) at (1.650,0.650);
            \drawDottedLine{t1}{t2}
            \drawDottedLineR{t2}{0.45}
            \drawDottedLine{b1}{b2}
            \drawDottedLine{b2}{b3}
            \drawBHLine{b3}{b4}
            \drawDottedLineR{b4}{0.45}
            \drawGravitonLine{b2}{g1}
            \drawActiveGravitonLine{b3}{g2}
            \drawGravitonLine{g1}{g2}
            \drawGravitonLine{b1}{g1}
            \drawGravitonLine{t1}{g2}
            \drawGravitonLine{t2}{b4}
            \foreach \v in {b1,b2,b3,b4,g1,g2,t1,t2} {\drawPerturbativeVertex{\v}}
        \end{tikzpicture}
        \caption{}
        \label{fig:red-top-sectors-c}
    \end{subfigure}
    \hfill
    \begin{subfigure}[b]{0.15\textwidth}
        \centering
        \begin{tikzpicture}[baseline={(0,-.1)}]
            \coordinate (b1) at (0.000,-0.650);
            \coordinate (b2) at (0.550,-0.650);
            \coordinate (b3) at (1.100,-0.650);
            \coordinate (b4) at (1.650,-0.650);
            \coordinate (g1) at (0.550,0.250);
            \coordinate (g2) at (0.825,-0.200);
            \coordinate (t1) at (0.550,0.650);
            \coordinate (t2) at (1.650,0.650);
            \drawDottedLine{t1}{t2}
            \drawDottedLineR{t2}{0.45}
            \drawDottedLine{b1}{b2}
            \drawDottedLine{b2}{b3}
            \drawBHLine{b3}{b4}
            \drawDottedLineR{b4}{0.45}
            \drawGravitonLine{b2}{g2}
            \drawActiveGravitonLine{g1}{g2}
            \drawGravitonLine{b1}{g1}
            \drawActiveGravitonLine{b3}{g2}
            \drawGravitonLine{t1}{g1}
            \drawGravitonLine{t2}{b4}
            \foreach \v in {b1,b2,b3,b4,g1,g2,t1,t2} {\drawPerturbativeVertex{\v}}
        \end{tikzpicture}
        \caption{}
        \label{fig:red-top-sectors-d}
    \end{subfigure}
    \hfill
    \begin{subfigure}[b]{0.15\textwidth}
        \centering
        \begin{tikzpicture}[baseline={(0,-.1)}]
            \coordinate (b1) at (0.00,-0.65);
            \coordinate (b2) at (0.66,-0.65);
            \coordinate (b3) at (1.32,-0.65);
            \coordinate (g1) at (0.33,-0.20);
            \coordinate (g2) at (1.32,0.20);
            \coordinate (t1) at (0.33,0.65);
            \coordinate (t2) at (0.99,0.65);
            \coordinate (t3) at (1.65,0.65);
            \drawBHLine{t1}{t2}
            \drawDottedLine{t2}{t3}
            \drawDottedLineR{t3}{0.45}
            \drawDottedLine{b1}{b2}
            \drawDottedLine{b2}{b3}
            \drawDottedLineR{b3}{0.78}
            \drawGravitonLine{b2}{g1}
            \drawActiveGravitonLine{t2}{g2}
            \drawGravitonLine{t1}{g1}
            \drawGravitonLine{b3}{g2}
            \drawGravitonLine{b1}{g1}
            \drawGravitonLine{t3}{g2}
            \foreach \v in {b1,b2,b3,g1,g2,t1,t2,t3} {\drawPerturbativeVertex{\v}}
        \end{tikzpicture}
        \caption{}
        \label{fig:red-top-sectors-e}
    \end{subfigure}
    \hfill
    \begin{subfigure}[b]{0.15\textwidth}
        \centering
        \begin{tikzpicture}[baseline={(0,-.1)}]
            \coordinate (b1) at (0.28,-0.65);
            \coordinate (b2) at (1.10,-0.65);
            \coordinate (b3) at (1.65,-0.65);
            \coordinate (g1) at (0.28,0.20);
            \coordinate (t1) at (0.00,0.65);
            \coordinate (t2) at (0.55,0.65);
            \coordinate (t3) at (1.10,0.65);
            \coordinate (t4) at (1.65,0.65);
            \drawDottedLine{t1}{t2}
            \drawBHLine{t2}{t3}
            \drawBHLine{t3}{t4}
            \drawDottedLineR{t4}{0.45}
            \drawDottedLine{b1}{b2}
            \drawDottedLine{b2}{b3}
            \drawDottedLineR{b3}{0.45}
            \drawGravitonLine{t3}{b2}
            \drawActiveGravitonLine{t2}{g1}
            \drawGravitonLine{b1}{g1}
            \drawGravitonLine{t4}{b3}
            \drawGravitonLine{t1}{g1}
            \foreach \v in {b1,b2,b3,g1,t1,t2,t3,t4} {\drawPerturbativeVertex{\v}}
        \end{tikzpicture}
        \caption{}
        \label{fig:red-top-sectors-f}
    \end{subfigure}
    \\[1em]
    \hspace*{\fill}
    \begin{subfigure}[b]{0.15\textwidth}
        \centering
        \begin{tikzpicture}[baseline={(0,-.1)}]
            \coordinate (b1) at (0.00,-0.65);
            \coordinate (b2) at (0.55,-0.65);
            \coordinate (b3) at (1.10,-0.65);
            \coordinate (b4) at (1.65,-0.65);
            \coordinate (g1) at (0.28,-0.20);
            \coordinate (g2) at (1.38,-0.20);
            \coordinate (t1) at (0.28,0.65);
            \coordinate (t2) at (1.38,0.65);
            \drawDottedLine{t1}{t2}
            \drawDottedLineR{t2}{0.72}
            \drawDottedLine{b1}{b2}
            \drawBHLine{b2}{b3}
            \drawDottedLine{b3}{b4}
            \drawDottedLineR{b4}{0.45}
            \drawActiveGravitonLine{b2}{g1}
            \drawGravitonLine{b1}{g1}
            \drawActiveGravitonLine{b3}{g2}
            \drawGravitonLine{b4}{g2}
            \drawGravitonLine{t1}{g1}
            \drawGravitonLine{t2}{g2}
            \foreach \v in {b1,b2,b3,b4,g1,g2,t1,t2} {\drawPerturbativeVertex{\v}}
        \end{tikzpicture}
        \caption{}
        \label{fig:red-top-sectors-g}
    \end{subfigure}
    \hfill
    \begin{subfigure}[b]{0.15\textwidth}
        \centering
        \begin{tikzpicture}[baseline={(0,-.1)}]
            \coordinate (b1) at (0.00,-0.65);
            \coordinate (b2) at (0.55,-0.65);
            \coordinate (b3) at (1.10,-0.65);
            \coordinate (b4) at (1.65,-0.65);
            \coordinate (g1) at (1.38,-0.20);
            \coordinate (t1) at (0.00,0.65);
            \coordinate (t2) at (0.55,0.65);
            \coordinate (t3) at (1.38,0.65);
            \drawBHLine{t1}{t2}
            \drawDottedLine{t2}{t3}
            \drawDottedLineR{t3}{0.72}
            \drawDottedLine{b1}{b2}
            \drawBHLine{b2}{b3}
            \drawDottedLine{b3}{b4}
            \drawDottedLineR{b4}{0.45}
            \drawActiveGravitonLine{t2}{b2}
            \drawGravitonLine{t1}{b1}
            \drawActiveGravitonLine{b3}{g1}
            \drawGravitonLine{b4}{g1}
            \drawGravitonLine{t3}{g1}
            \foreach \v in {b1,b2,b3,b4,g1,t1,t2,t3} {\drawPerturbativeVertex{\v}}
        \end{tikzpicture}
        \caption{}
        \label{fig:red-top-sectors-h}
    \end{subfigure}
    \hfill
    \begin{subfigure}[b]{0.15\textwidth}
        \centering
        \begin{tikzpicture}[baseline={(0,-.1)}]
            \coordinate (b1) at (0.00,-0.65);
            \coordinate (b2) at (0.55,-0.65);
            \coordinate (b3) at (1.10,-0.65);
            \coordinate (b4) at (1.65,-0.65);
            \coordinate (g1) at (0.82,-0.20);
            \coordinate (t1) at (0.00,0.65);
            \coordinate (t2) at (0.82,0.65);
            \coordinate (t3) at (1.65,0.65);
            \drawBHLine{t1}{t2}
            \drawDottedLine{t2}{t3}
            \drawDottedLineR{t3}{0.45}
            \drawDottedLine{b1}{b2}
            \drawDottedLine{b2}{b3}
            \drawBHLine{b3}{b4}
            \drawDottedLineR{b4}{0.45}
            \drawGravitonLine{b2}{g1}
            \drawActiveGravitonLine{t2}{g1}
            \drawGravitonLine{t1}{b1}
            \drawActiveGravitonLine{b3}{g1}
            \drawGravitonLine{t3}{b4}
            \foreach \v in {b1,b2,b3,b4,g1,t1,t2,t3} {\drawPerturbativeVertex{\v}}
        \end{tikzpicture}
        \caption{}
        \label{fig:red-top-sectors-i}
    \end{subfigure}
    \hfill
    \begin{subfigure}[b]{0.15\textwidth}
        \centering
        \begin{tikzpicture}[baseline={(0,-.1)}]
            \coordinate (b1) at (0.00,-0.65);
            \coordinate (b2) at (0.55,-0.65);
            \coordinate (b3) at (1.10,-0.65);
            \coordinate (b4) at (1.65,-0.65);
            \coordinate (g1) at (0.28,-0.20);
            \coordinate (t1) at (0.28,0.65);
            \coordinate (t2) at (1.10,0.65);
            \coordinate (t3) at (1.65,0.65);
            \drawBHLine{t1}{t2}
            \drawDottedLine{t2}{t3}
            \drawDottedLineR{t3}{0.45}
            \drawDottedLine{b1}{b2}
            \drawDottedLine{b2}{b3}
            \drawBHLine{b3}{b4}
            \drawDottedLineR{b4}{0.45}
            \drawGravitonLine{b2}{g1}
            \drawActiveGravitonLine{t2}{b3}
            \drawGravitonLine{t1}{g1}
            \drawGravitonLine{b1}{g1}
            \drawGravitonLine{t3}{b4}
            \foreach \v in {b1,b2,b3,b4,g1,t1,t2,t3} {\drawPerturbativeVertex{\v}}
        \end{tikzpicture}
        \caption{}
        \label{fig:red-top-sectors-j}
    \end{subfigure}
    \hfill
    \begin{subfigure}[b]{0.15\textwidth}
        \centering
        \begin{tikzpicture}[baseline={(0,-.1)}]
            \coordinate (b1) at (0.00,-0.65);
            \coordinate (b2) at (0.82,-0.65);
            \coordinate (b3) at (1.65,-0.65);
            \coordinate (g1) at (0.82,0.20);
            \coordinate (t1) at (0.00,0.65);
            \coordinate (t2) at (0.55,0.65);
            \coordinate (t3) at (1.10,0.65);
            \coordinate (t4) at (1.65,0.65);
            \drawBHLine{t1}{t2}
            \drawDottedLine{t2}{t3}
            \drawBHLine{t3}{t4}
            \drawDottedLineR{t4}{0.45}
            \drawDottedLine{b1}{b2}
            \drawDottedLine{b2}{b3}
            \drawDottedLineR{b3}{0.45}
            \drawGravitonLine{b2}{g1}
            \drawActiveGravitonLine{t2}{g1}
            \drawGravitonLine{t1}{b1}
            \drawActiveGravitonLine{t3}{g1}
            \drawGravitonLine{t4}{b3}
            \foreach \v in {b1,b2,b3,g1,t1,t2,t3,t4} {\drawPerturbativeVertex{\v}}
        \end{tikzpicture}
        \caption{}
        \label{fig:red-top-sectors-k}
    \end{subfigure}
    \hspace*{\fill}
    \caption{The eleven reducible top sectors of the 1SF planar-1 family.}
    \label{fig:red-top-sectors}
\end{figure}
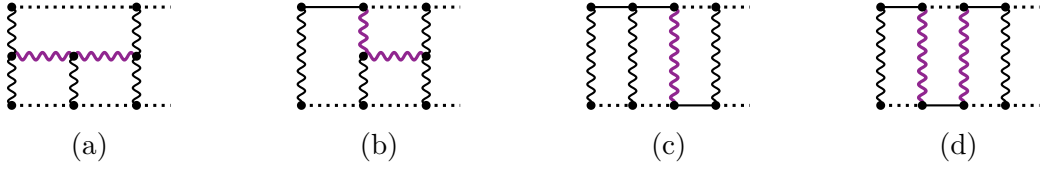
\begin{figure}[t]
    \centering
    \begin{subfigure}[b]{0.24\textwidth}
        \centering
        \begin{tikzpicture}[baseline={(0,-.1)}]
            \coordinate (b1) at (0.00,-0.65);
            \coordinate (b2) at (0.82,-0.65);
            \coordinate (b3) at (1.65,-0.65);
            \coordinate (g1) at (0.00,0.00);
            \coordinate (g2) at (0.82,0.00);
            \coordinate (g3) at (1.65,0.00);
            \coordinate (t1) at (0.00,0.65);
            \coordinate (t2) at (1.65,0.65);
            \drawDottedLine{t1}{t2}
            \drawDottedLineR{t2}{0.45}
            \drawDottedLine{b1}{b2}
            \drawDottedLine{b2}{b3}
            \drawDottedLineR{b3}{0.45}
            \drawGravitonLine{b2}{g2}
            \drawActiveGravitonLine{g1}{g2}
            \drawGravitonLine{b1}{g1}
            \drawActiveGravitonLine{g2}{g3}
            \drawGravitonLine{b3}{g3}
            \drawGravitonLine{t1}{g1}
            \drawGravitonLine{t2}{g3}
            \foreach \v in {b1,b2,b3,g1,g2,g3,t1,t2} {\drawPerturbativeVertex{\v}}
        \end{tikzpicture}
        \caption{}
        \label{fig:irr-top-sectors-a}
    \end{subfigure}
    \hfill
    \begin{subfigure}[b]{0.24\textwidth}
        \centering
        \begin{tikzpicture}[baseline={(0,-.1)}]
            \coordinate (b1) at (0.00,-0.65);
            \coordinate (b2) at (0.82,-0.65);
            \coordinate (b3) at (1.65,-0.65);
            \coordinate (g1) at (0.82,0.00);
            \coordinate (g2) at (1.65,0.00);
            \coordinate (t1) at (0.00,0.65);
            \coordinate (t2) at (0.82,0.65);
            \coordinate (t3) at (1.65,0.65);
            \drawBHLine{t1}{t2}
            \drawDottedLine{t2}{t3}
            \drawDottedLineR{t3}{0.45}
            \drawDottedLine{b1}{b2}
            \drawDottedLine{b2}{b3}
            \drawDottedLineR{b3}{0.45}
            \drawGravitonLine{b2}{g1}
            \drawActiveGravitonLine{t2}{g1}
            \drawGravitonLine{t1}{b1}
            \drawActiveGravitonLine{g1}{g2}
            \drawGravitonLine{b3}{g2}
            \drawGravitonLine{t3}{g2}
            \foreach \v in {b1,b2,b3,g1,g2,t1,t2,t3} {\drawPerturbativeVertex{\v}}
        \end{tikzpicture}
        \caption{}
        \label{fig:irr-top-sectors-b}
    \end{subfigure}
    \hfill
    \begin{subfigure}[b]{0.24\textwidth}
        \centering
        \begin{tikzpicture}[baseline={(0,-.1)}]
            \coordinate (b1) at (0.00,-0.65);
            \coordinate (b2) at (0.55,-0.65);
            \coordinate (b3) at (1.10,-0.65);
            \coordinate (b4) at (1.65,-0.65);
            \coordinate (t1) at (0.00,0.65);
            \coordinate (t2) at (0.55,0.65);
            \coordinate (t3) at (1.10,0.65);
            \coordinate (t4) at (1.65,0.65);
            \drawBHLine{t1}{t2}
            \drawBHLine{t2}{t3}
            \drawDottedLine{t3}{t4}
            \drawDottedLineR{t4}{0.45}
            \drawDottedLine{b1}{b2}
            \drawDottedLine{b2}{b3}
            \drawBHLine{b3}{b4}
            \drawDottedLineR{b4}{0.45}
            \drawGravitonLine{t2}{b2}
            \drawGravitonLine{t1}{b1}
            \drawActiveGravitonLine{t3}{b3}
            \drawGravitonLine{t4}{b4}
            \foreach \v in {b1,b2,b3,b4,t1,t2,t3,t4} {\drawPerturbativeVertex{\v}}
        \end{tikzpicture}
        \caption{}
        \label{fig:irr-top-sectors-c}
    \end{subfigure}
    \hfill
    \begin{subfigure}[b]{0.24\textwidth}
        \centering
        \begin{tikzpicture}[baseline={(0,-.1)}]
            \coordinate (b1) at (0.00,-0.65);
            \coordinate (b2) at (0.55,-0.65);
            \coordinate (b3) at (1.10,-0.65);
            \coordinate (b4) at (1.65,-0.65);
            \coordinate (t1) at (0.00,0.65);
            \coordinate (t2) at (0.55,0.65);
            \coordinate (t3) at (1.10,0.65);
            \coordinate (t4) at (1.65,0.65);
            \drawBHLine{t1}{t2}
            \drawDottedLine{t2}{t3}
            \drawBHLine{t3}{t4}
            \drawDottedLineR{t4}{0.45}
            \drawDottedLine{b1}{b2}
            \drawBHLine{b2}{b3}
            \drawDottedLine{b3}{b4}
            \drawDottedLineR{b4}{0.45}
            \drawActiveGravitonLine{t2}{b2}
            \drawGravitonLine{t1}{b1}
            \drawActiveGravitonLine{t3}{b3}
            \drawGravitonLine{t4}{b4}
            \foreach \v in {b1,b2,b3,b4,t1,t2,t3,t4} {\drawPerturbativeVertex{\v}}
        \end{tikzpicture}
        \caption{}
        \label{fig:irr-top-sectors-d}
    \end{subfigure}
    \caption{
        The four top sectors of the planar-1 family to which, up to symmetries, all our integrals can be reduced.
        The active gravitons $D_7$ and $D_8$ are highlighted.
    }
    \label{fig:irr-top-sectors}
\end{figure}
The most demanding sector in our reduction is the sector associated with the integral
\begin{align}
    I_{1,1,1,1,0,0,0,1,1,1,0,0,0,1,0}
    \;=\;\;\;
    \begin{tikzpicture}[baseline={(0,-.1)}]
        \coordinate (b1) at (0.00,-0.65);
        \coordinate (b2) at (0.82,-0.65);
        \coordinate (b3) at (1.65,-0.65);
        \coordinate (b4) at (2,-0.65);
        \coordinate (t1) at (0.00,0.65);
        \coordinate (t2) at (0.82,0.65);
        \coordinate (t3) at (1.65,0.65);
        \coordinate (t4) at (2,0.65);
        \drawDottedLine{t1}{t2}
        \drawBHLine{t2}{t3}
        \drawDottedLine{b1}{b2}
        \drawDottedLine{b2}{b3}
        \drawGravitonLine{t1}{b1}
        \drawGravitonLine{t2}{b2}
        \drawGravitonLine{t3}{b3}
        \drawDottedLine{t3}{t4}
        \drawDottedLine{b3}{b4}
        \drawRadiativeCurvedGravitonLine{t1}{t2}{-85}{-95}
        \foreach \v in {b1,b2,b3,t1,t2,t3} {\drawPerturbativeVertex{\v}}
    \end{tikzpicture}
    \; ,
\end{align}
which contains representatives requiring seeds $s=7$ and $d=5$, where $s$ is the maximal number of numerators and $d$ the maximal number of dots.

We emphasise that one needs higher seeds for the offset shift than for the impulse, as taking the derivative with respect to $\omega_{\rm IR}$ generically raises the maximal power of propagators.
Choosing the minimal working seeds, however, this demanding sector runs in a few hours on a laptop.
Finally, after IBP reduction we land on 75 odd master integrals.
\begin{figure}[t]
    \centering
    \begin{minipage}[c]{.5\textwidth}
        \centering
        \includegraphics[width=.9\textwidth]{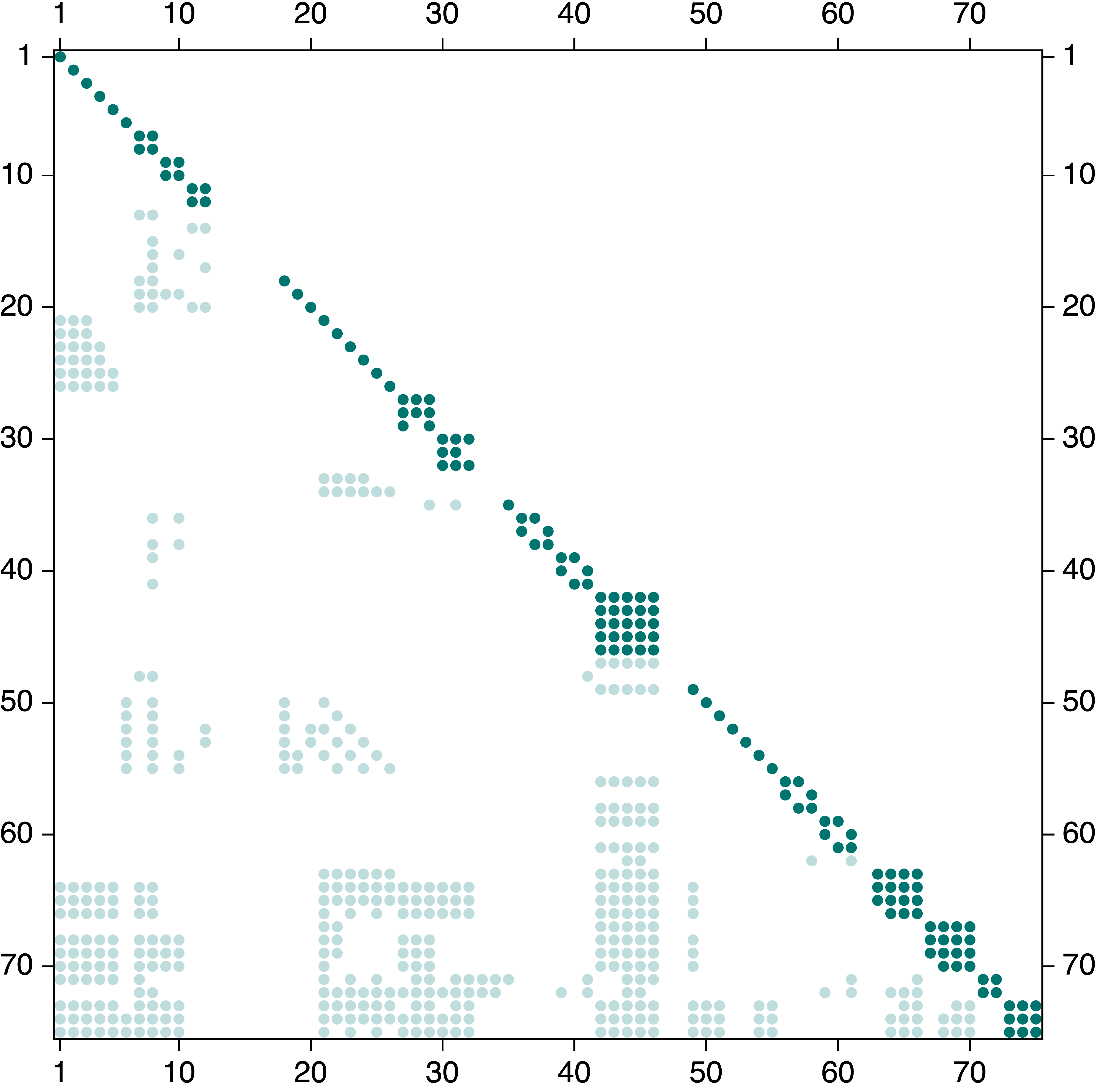}
    \end{minipage}
    \caption{
        Density plot of the canonical differential equation matrix $M(x)$, with non-zero entries represented by dots.
        The diagonal blocks correspond to sectors; the sectors with large diagonal blocks and many off-diagonal entries are computationally the most intense to canonicalise.
    }
    \label{fig:deMatrix}
\end{figure}

We now focus on solving the master integrals, to which end we use the method of differential equations.
This involves acting with a derivative with respect to the variable $\gamma$ on the vector $\vec I(\gamma;\eps)$, whose components are the master integrals.
IBP reducing this derivative, a DE for $\vec I(\gamma;\eps)$ can then be constructed.
To avoid square roots in the DE matrix, it is more convenient to use the variable $x=\gamma-\sqrt{\gamma^2-1}$, in terms of which we obtain the DE
\begin{align}
    \frac{\d}{\d x}\vec{I}(x;\eps)
    =
    A(x,\eps)\vec{I}(x;\eps)
    \, ,
\end{align}
where $A(x,\eps)$ is in block-diagonal form with 41 diagonal blocks of which the largest comprises five master integrals.
Using the package \texttt{CANONICA} \cite{Meyer:2017joq}, we then find a transformation matrix $T(x;\eps)$ that brings the entire matrix $A(x,\eps)$ into canonical form.
The original basis $\vec I(x;\eps)$ is then related to the canonical basis $\Ican(x;\eps)$ through the relation
\begin{align}
    \vec I(x;\eps)
    =
    T(x;\eps)\Ican(x;\eps)
    \, ,
\end{align}
and the resulting canonical DE is given by
\begin{align}
    \frac{\d}{\d x}\Ican(x;\eps)
    =
    A_{\rm c}(x,\eps)\Ican(x;\eps)
    =
    \eps M(x)\Ican(x;\eps)
    \, ,
\end{align}
where the matrix $M(x)$ can be expanded in terms of numerical matrices and our five letters:
\begin{align}
    M(x)
    =
    \sum_{x_i}
    \frac{
        M_{x_i}
    }{
        x-x_i
    }
    \, ,
    \qquad
    x_i\in\{
        0,1,-1,\i,-\i
    \}
    \, .
\end{align}
See \fig{fig:deMatrix} for a density plot of $M(x)$. The canonical DE admits a formal, well studied solution in terms of a path ordered exponential. Denoting the boundary integrals by $\Jcan(\eps)$, we have
\begin{align}\label{eq:path-ordered-exp}
    \Ican(x;\eps)
    =
    \mathcal{P}\exp\Big[
        \eps\int_1^x\d x'\,
        M(x')
    \Big]\Jcan(\eps)
    \, .
\end{align}
We employ \texttt{PolyLogTools} \cite{Duhr:2019tlz} for the integration of the canonical DE.

\subsection{Boundary matching}
We fix the integration constants $\Jcan(\eps)$ of \eqn{eq:path-ordered-exp} in the PN limit $\gamma\to1$, which corresponds to $x\to1$ and thus explains the choice of basepoint in \eqn{eq:path-ordered-exp}.
Following Wasow's method \cite{WASOW1966378}, we first determine the leading behaviour of the canonical integrals as $x\to1$.
Expanding \eqn{eq:path-ordered-exp}, we get
\begin{align}
    \Ican(x;\eps)
    \sim
    \exp\Big[
        \eps\int_1^x\d x'\,
        \frac{M_1}{x'-1}
    \Big]
    \Jcan(\eps)
    =
    (1-x)^{\eps M_1}
    \Jcan(\eps)
    \, ,
\end{align}
where we made use of the shuffle-regulated integral representation
\begin{align}
    \log(1-x)
    =
    -
    \int_1^x
    \frac{\d x'}{1-x'}
    \, .
\end{align}
The matrix exponential is controlled by the eigenvalues $\lambda=-4,-2,0,2,6,8$ of $M_1$:
\begin{align}
    (1-x)^{\eps M_1}
    =
    \sum_{\lambda}
    \sum_{k\geq0}
    \frac{
        (1-x)^{\eps\lambda}
        \eps^k\log^k(1-x)
    }{k!}
    M_{1,\lambda,k}
    \, ,
\end{align}
where $M_{1,\lambda,k}$ are numerical matrices.
The presence of logarithms originates in the failure of $M_1$ to be diagonalisable.
Which of the scalings $(1-x)^{\eps\lambda}$ are physical may be ascertained from the PN limit, in which gravitons in the dynamical region are either potential (P) or radiative (R), cf. \eqn{eq:scalingsDyn}.
Our integrals contain at most two active gravitons, so the occuring integration regions are PP, PR, RP, and RR.
They are characterised by the following scalings:
\begin{align}
    \text{PP} : (1-x)^0
    \, ,
    \qquad
    \text{PR/RP} : (1-x)^{-2\eps}
    \, ,
    \qquad
    \text{RR} : (1-x)^{-4\eps}
    \, .
\end{align}
Immediately then, we obtain the constraints $M_{1,\lambda,k}\Jcan(\eps)=0$ for $\lambda=2,6,8$ and any $k$, as well as for $\lambda=-4,-2,0$ and $k>0$, as the leading terms of the region integrals are free of logarithms.
The constants are then split into their PP, PR/RP, and RR contributions,
\begin{align}
    \Jcan(\eps)
    =
    \Jcan{}_{,\rm PP}(\eps)
    +
    \Jcan{}_{,\rm PR/RP}(\eps)
    +
    \Jcan{}_{,\rm RR}(\eps)
    \, ,
\end{align}
where in each region the scalings of the other regions are set to zero.
In particular, since we are concerned with dissipative physics, we discard $\Jcan{}_{,\rm PP}(\eps)$ and set the scalings of the PP region to zero always.
This leaves only a few constants to be fixed within each region.
To do this, we need the asymptotic expansion of the canonical integrals for $x\to 1$ beyond leading order.
We will not dwell on this here beyond noting that it can be obtained by making the ansatz
\begin{figure}[t]
    \centering
    \begin{subfigure}[b]{0.3\textwidth}
        \centering
        \begin{tikzpicture}
            \coordinate (t1) at (0,1.3);   \coordinate (t2) at (.6,1.3); \coordinate (t3) at (1.1,1.3);
            \coordinate (t4) at (1.6,1.3); \coordinate (t5) at (2.2,1.3);
            \coordinate (m1) at (1.1,.65);
            \coordinate (b1) at (0,0);     \coordinate (b2) at (2.2,0);
            \drawDottedLine{t2}{t4} \drawDottedLineR{t5}{.35}
            \drawDottedLine{b1}{b2} \drawDottedLineR{b2}{.35}
            \drawBHLine{t1}{t2} \drawBHLine{t4}{t5}
            \drawGravitonLine{t1}{b1} \drawGravitonLine{t5}{b2}
            \drawRadiativeCurvedGravitonLine{t2}{m1}{-90}{180}
            \drawRadiativeCurvedGravitonLine{t4}{m1}{-90}{0}
            \drawGravitonLine{t3}{m1}
            \foreach \v in {t1,t2,t3,t4,t5,m1,b1,b2} {\drawPerturbativeVertex{\v}}
        \end{tikzpicture}
        \caption{}
        \label{fig:boundary-RR}
    \end{subfigure}
    \hfill
    \begin{subfigure}[b]{0.3\textwidth}
        \centering
        \begin{tikzpicture}
            \coordinate (t1) at (0,1.3);   \coordinate (t2) at (.8,1.3);  \coordinate (t3) at (2.2,1.3);
            \coordinate (b1) at (0,0);     \coordinate (b2) at (.8,0);    \coordinate (b3) at (2.2,0);
            \drawBHLine{t1}{t2} \drawDottedLine{t2}{t3} \drawDottedLineR{t3}{.35}
            \drawDottedLine{b1}{b3} \drawDottedLineR{b3}{.35}
            \drawGravitonLine{t1}{b1}
            \drawGravitonLine{t2}{b2}
            \drawGravitonLine{t3}{b3}
            \drawRadiativeCurvedGravitonLine{t2}{t3}{-90}{-90}
            \foreach \v in {t1,t2,t3,b1,b2,b3} {\drawPerturbativeVertex{\v}}
        \end{tikzpicture}
        \caption{}
        \label{fig:boundary-PR1}
    \end{subfigure}
    \hfill
    \begin{subfigure}[b]{0.3\textwidth}
        \centering
        \begin{tikzpicture}
            \coordinate (t1) at (.4,1.3);  \coordinate (t2) at (1.1,1.3); \coordinate (t3) at (2.2,1.3);
            \coordinate (b1) at (0,0);     \coordinate (b2) at (.8,0);    \coordinate (b3) at (2.2,0);
            \drawBHLine{t1}{t2} \drawDottedLine{t2}{t3} \drawDottedLineR{t3}{.35}
            \drawDottedLine{b1}{b3} \drawDottedLineR{b3}{.35}
            \drawGravitonLine{t1}{b1}
            \drawGravitonLine{t1}{b2}
            \drawGravitonLine{t3}{b3}
            \drawRadiativeCurvedGravitonLine{t2}{t3}{-90}{-90}
            \foreach \v in {t1,t2,t3,b1,b2,b3} {\drawPerturbativeVertex{\v}}
        \end{tikzpicture}
        \caption{}
        \label{fig:boundary-PR2}
    \end{subfigure}
    \caption{
        In \subref{fig:boundary-RR} we show the RR integral family, which we evaluate for generic indices.
        In \subref{fig:boundary-PR1} and \subref{fig:boundary-PR2} we display the two PR master integrals.
    }
    \label{fig:boundary-integrals}
\end{figure}
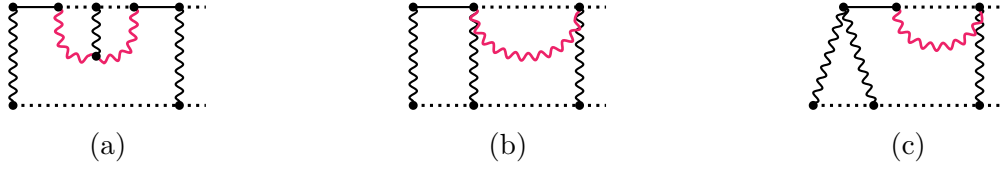
\begin{align}
    \Ican(x;\eps)
    =
    U(x;\eps)
    (1-x)^{\eps M_1}
    \Jcan(\eps)
    \, ,
    \qquad
    U(x;\eps)
    =
    1
    +
    \sum_{n=1}^\infty
    (1-x)^nU^{(n)}(\eps)
    \, ,
\end{align}
and using the DE to solve for the $U^{(n)}(\eps)$.

We employ the method of regions to obtain the asymptotic expansion of the integrals \cite{Beneke:1997zp,Smirnov:2012gma,Becher:2014oda}.
It is most straightforwardly applied, however, in a basis with no derivatives on the delta functions.
We therefore rotate from the IBP basis to a basis
\begin{align}
    \vec I_{\rm sp}(x;\eps)
    =
    T_{\rm sp}(x;\eps)\vec I(x;\eps)
\end{align}
which contains scalar products in the numerator instead.
With this in place, we have
\begin{align}
    \vec I_{\rm sp}(x;\eps)
    =
    T_{\rm sp}(x;\eps)
    T(x;\eps)
    U(x;\eps)
    (1-x)^{\eps M_1}
    \Jcan(\eps)
    \, .
\end{align}
In essence, we solve this equation for the constants $\Jcan(\eps)$ in each region.
On the left-hand side, we insert the leading term of the region expansion of each master integral, while on the right-hand side, we expand to high enough order in $(1-x)$ such that we get enough equations for the remaining constants.

The boundary integrals needed are shown in \fig{fig:boundary-integrals}.
In the RR region, the expansion maps our master integrals onto a simple RR integral family which admits a closed-form solution for generic indices.
In the PR region, the expansion similarly maps our master integrals onto a simpler PR integral family, which we in this case IBP-reduce to two odd-parity master integrals.
The RP region follows from the PR one by a 12-swap.

\section{Results}\label{section:Results}
In this section we present our main results which can also be found in the ancillary file submitted along with this article.
The central result of this work is the mechanical loss of angular momentum $|L|_{\rm loss}$ at the 4PM order.
As discussed, it is convenient to parametrise this loss in terms of $(E|L|)_{\rm loss}$ and $E_{\rm loss}$ whose 4PM contributions we expand schematically in their mass dependence as follows,
\begin{align}
    (E|L|)_{\rm loss}^{\rm(4PM)}
    &=
    \frac{G^4 M^6\nu^2}{|b|^3}
        \Big(
        \vec{c}^{\,\,(\rm 1 SF)}(\gamma)
        +
        \nu
        \vec{c}^{\,\,(\overbar{\rm 1 SF})}(\gamma)
        \Big)
        \cdot
        \vec{f}(\gamma)
    \ ,
    \\
    E_{\rm loss}^{\rm(4PM)}
    &=
    \frac{G^4 M^6 \nu^2}{E |b|^4}
    \,
        \Big(
        \vec{d}^{\,\,(\rm 1 SF)}(\gamma)
        +
        \nu
        \vec{d}^{\,\,(\overbar{\rm 1 SF})}(\gamma)
        \Big)
        \cdot
        \vec{f}(\gamma)
    \ .
\end{align}
Here, $\vec f(\gamma)$ is a vector spanning the relevant function space, while $\vec{c}(\gamma)$ and $\vec{d}(\gamma)$ are the corresponding coefficient vectors, whose entries are rational functions with possible poles at $\gamma=0$ and $\gamma=1$. Recall that we already introduced this schematic form in \eqn{eq:nuparamet} in \secref{sec:WEFTWQFT}.
The angular momentum loss is easily obtained from the previous two results, through
\begin{align}\label{eq:lalala}
     |L|^{(4\rm PM)}_{\rm loss}
    =
    \frac1{E}
    (E|L|)^{(4\rm PM)}_{\rm loss}
    -
    \frac{|L|}{E}
     E^{(4\rm PM)}_{\rm loss}\ .
\end{align}
The function space vector $\vec f(\gamma)$ is given explicitly as
\begin{equation}
    \vec{f}(\gamma)
    =
    \begin{pmatrix}
        1
        \\[6pt]
        \dfrac{\arccosh(\gamma)}{\sqrt{\gamma^2-1}}
        \\[12pt]
        \arccosh^2(\gamma)
        \\[6pt]
        \dfrac{1}{2\sqrt{\gamma^2-1}}
        \left[
            \operatorname{Li}_2\!\left(\sqrt{\tfrac{\gamma-1}{\gamma+1}}\right)
            -
            \operatorname{Li}_2\!\left(-\sqrt{\tfrac{\gamma-1}{\gamma+1}}\right)
        \right]
        \\[12pt]
        \dfrac{1}{\sqrt{\gamma^2-1}}
        \Big[
            -\arccosh(\gamma) \, \log\!\big(2(1+\gamma)\big)
            -
            \operatorname{Li}_2\!\left(-x^{-1}\right)
            +
            \operatorname{Li}_2(-x)
        \Big]
        \\[12pt]
        \dfrac{1}{\sqrt{\gamma^2-1}}
        \Big[
            -2 \arccosh(\gamma) \, \log(2\gamma)
            +
            \tfrac{1}{2}
            \Big(
                \operatorname{Li}_2\!\left(-x^{2}\right)
                -
                \operatorname{Li}_2\!\left(-x^{-2}\right)
            \Big)
        \Big]
    \end{pmatrix}
    \ ,
\end{equation}
where the final two functions are written in terms of $x=\gamma-\sqrt{\gamma^2-1}$ instead of $\gamma$ for brevity.
Note that we have chosen all basis functions to be parity-even under flipping the sign of the relative velocity $v\to -v$.
As we shall see, the latter three functions only enter in the dynamical part of the R contributions.
Instead the RR contributions and, in fact, the whole static part, are spanned entirely by $\arccosh(\gamma)$.

We distinguish R and RR contributions as well as dynamical and static ones in the coefficient functions:
\begin{align}\label{eq:labelingConventions}
    \vec{c}^{\,\,(n{\rm SF})}
    &=
    \vec{c}_{\rm(dyn)\mathrm{R}}^{\,\,(n\mathrm{SF})}
    +
    \vec{c}_{\rm(dyn)\mathrm{RR}}^{\,\,(n\mathrm{SF})}
    +
    \vec{c}_{\rm(stat)\mathrm{R}}^{\,\,(n\mathrm{SF})}
    +
    \vec{c}_{\rm(stat)\mathrm{RR}}^{\,\,(n\mathrm{SF})}
    \ ,
    \\
    \vec{d}^{\,\,(n{\rm SF})}
    &=
    \vec{d}_{\rm(dyn)\mathrm{R}}^{\,\,(n\mathrm{SF})}
    +
    \vec{d}_{\rm(dyn)\mathrm{RR}}^{\,\,(n\mathrm{SF})}
    \ .
\end{align}
Naturally, the $\vec{d}(\gamma)$ coefficients associated with $E_{\rm loss}$ have no static region.
In this equation we can have $n=1$ or $n=\bar 1$ and we provide all non-zero coefficients of $E_{\rm loss}$ in \app{app:results}.
We expand any coefficient as,
\begin{equation}
    \vec X
    =
    (X^1,X^2,X^3,X^4,X^5,X^6)
    \ .
\end{equation}

\subsubsection*{Static coefficients of $(E|L|)_{\rm loss}$}

The static RR region is particularly simple.
First, it has only a $\overbar{\rm 1SF}$ contribution.
Second, this contribution combines into a square of $\mathcal{I}(\gamma)$.
We find,
\begin{align}
    \vec{c}_{\rm(stat)\mathrm{RR}}^{\,\,(1\mathrm{SF})}
    &=0
    \\
    \vec{c}_{\rm(stat)\mathrm{RR}}^{\,\,(\overbar{1\mathrm{SF}})}
    \cdot \vec{f}
    &=
    \frac{2\,(2\gamma^2-1)^2}{\sqrt{\gamma^2-1}}
    \mathcal{I}^2(\gamma)
    \ .
\end{align}
The R static coefficients are spanned by the first three functions only:
\begin{subequations}
\begin{align}
    c_{\rm(stat)\mathrm{R}}^{(1\mathrm{SF})\,1}
    &=
    -\frac{4 \, (16 - 473\gamma^{2} + 1080\gamma^{4} - 528\gamma^{6} - 272\gamma^{8} + 192\gamma^{10})}
          {15 \, (\gamma^{2}-1)^{3}}
    \ ,
    \\
    c_{\rm(stat)\mathrm{R}}^{(1\mathrm{SF})\,2}
    &=
    \frac{4 \, \gamma \, (2\gamma^{2}-3)(-1 + 16\gamma^{2} - 32\gamma^{4} + 16\gamma^{6})}
          {(\gamma^{2}-1)^{3}}
    \ ,
    \\
    c_{\rm(stat)\mathrm{R}}^{(1\mathrm{SF})\,3}
    &=
    0
    \ ,
    \\
    c_{\rm(stat)\mathrm{R}}^{(\overbar{1\mathrm{SF}})\,1}
    &=
    \frac{8}{45 \, (\gamma^{2}-1)^{3}}
    \Big(48 - 2345\gamma - 1419\gamma^{2} + 7565\gamma^{3} + 3240\gamma^{4} - 9840\gamma^{5} \nonumber \\
        &- 1584\gamma^{6} + 5930\gamma^{7} - 816\gamma^{8} - 1220\gamma^{9} + 576\gamma^{10}
    \Big)
    \ ,\\
    c_{\rm(stat)\mathrm{R}}^{(\overbar{1\mathrm{SF}})\,2}
    &= \frac{8}{3 \, (\gamma^{2}-1)^{3}}
    \Big( 57 - 9\gamma + 171\gamma^{2} + 150\gamma^{3} - 818\gamma^{4} - 384\gamma^{5}
        \nonumber \\
        &+ 1008\gamma^{6} + 336\gamma^{7} - 504\gamma^{8} - 96\gamma^{9} + 80\gamma^{10}
    \Big)
    \ ,
    \\
    c_{\rm(stat)\mathrm{R}}^{(\overbar{1\mathrm{SF}})\,3}
    &=
    -\frac{16 \, \gamma \, (2\gamma^{2}-3)(-3 - 12\gamma^{2} + 4\gamma^{4})}
          {(\gamma^{2}-1)^{2}}
    \ .
\end{align}
The $\overbar{1{\rm SF}}$ coefficients are generally slightly more complicated than the 1SF counterparts.

All other coefficients are zero:
\begin{align}
    c_{\rm(stat)\mathrm{R}}^{(n\mathrm{SF})\,i}
    =
    0
    \qquad
    \text{for }i\ge4
    \ .
\end{align}
\end{subequations}
We collect the remaining coefficients which contribute to $(E|L|)_{\rm loss}$ in \tabs{tab:cdynRR}{tab:cR}.
\newpage
\subsubsection*{Dynamical RR coefficients of $(E|L|)_{\rm loss}$}
\begin{table}[h]
\centering
\small
\setlength{\fboxsep}{0pt}
\fbox{%
\begin{minipage}{\dimexpr\textwidth-2\fboxrule\relax}
\setlength{\abovedisplayskip}{3pt}\setlength{\belowdisplayskip}{3pt}%
\setlength{\abovedisplayshortskip}{3pt}\setlength{\belowdisplayshortskip}{3pt}%
\setlength{\jot}{1.5pt}%
\vspace{4pt}
\centerline{\textbf{Coefficients at 1SF}}
\vspace{4pt}
\hrule
\vspace{5pt}
\begin{align*}
c_{\rm(dyn)\mathrm{RR}}^{(1\mathrm{SF})\,1} &= \frac{16\,\gamma\,(804-745\gamma^{2}+48\gamma^{4}-92\gamma^{6})}{15\,(\gamma^{2}-1)^{3/2}}\\[2pt]
c_{\rm(dyn)\mathrm{RR}}^{(1\mathrm{SF})\,2} &= -\frac{16\,(116+353\gamma^{2}-420\gamma^{4}-28\gamma^{6}-16\gamma^{8})}{5\,(\gamma^{2}-1)^{3/2}}\\[2pt]
c_{\rm(dyn)\mathrm{RR}}^{(1\mathrm{SF})\,3} &= -\frac{96\,\gamma\,(5+6\gamma^{2})}{(\gamma^{2}-1)^{3/2}}
\end{align*}
\vspace{2pt}
\hrule
\vspace{4pt}
\centerline{\textbf{Coefficients at $\overline{\text{1SF}}$}}
\vspace{4pt}
\hrule
\vspace{5pt}
\begin{align*}
c_{\rm(dyn)\mathrm{RR}}^{(\overbar{1\mathrm{SF}})\,1} &= \frac{16}{15\,(\gamma^{2}-1)^{3/2}}\Bigl(-680-1608\gamma-305\gamma^{2}+1490\gamma^{3}+330\gamma^{4}\\
&\qquad-96\gamma^{5}+640\gamma^{6}+184\gamma^{7}\Bigr)\\[2pt]
c_{\rm(dyn)\mathrm{RR}}^{(\overbar{1\mathrm{SF}})\,2} &= -\frac{16}{15\,(\gamma^{2}-1)^{3/2}}\Bigl(-696-2695\gamma-2118\gamma^{2}+1180\gamma^{3}+2520\gamma^{4}\\
&\qquad+940\gamma^{5}+168\gamma^{6}+560\gamma^{7}+96\gamma^{8}\Bigr)\\[2pt]
c_{\rm(dyn)\mathrm{RR}}^{(\overbar{1\mathrm{SF}})\,3} &= \frac{64\,(3+15\gamma+29\gamma^{2}+18\gamma^{3}+7\gamma^{4}+2\gamma^{6})}{(\gamma^{2}-1)^{3/2}}
\end{align*}
\vspace{2pt}
\end{minipage}}
\caption{Coefficients $c_{\rm(dyn)\mathrm{RR}}^{(1\mathrm{SF})\,i}$ and $c_{\rm(dyn)\mathrm{RR}}^{(\overbar{1\mathrm{SF}})\,i}$ of the basis functions in the dynamical RR contribution for $i\le3$. 
The dynamical RR coefficients vanish for $i\ge4$.}
\label{tab:cdynRR}
\end{table}
\newpage
\subsubsection*{Dynamical R coefficients of $(E|L|)_{\rm loss}$}
\begin{table}[H]
\centering
\small
\setlength{\fboxsep}{0pt}
\fbox{%
\begin{minipage}{\dimexpr\textwidth-2\fboxrule\relax}
\setlength{\abovedisplayskip}{3pt}\setlength{\belowdisplayskip}{3pt}%
   \setlength{\abovedisplayshortskip}{3pt}\setlength{\belowdisplayshortskip}{3pt}%
   \setlength{\jot}{1.5pt}%
\vspace{4pt}
\centerline{\textbf{Coefficients at 1SF}}
\vspace{4pt}
\hrule
\vspace{5pt}
\begin{align*}
c_{\rm(dyn)\mathrm{R}}^{(1\mathrm{SF})\,1} &= \frac{1}{960\,\gamma^{6}(\gamma^{2}-1)^{3}}\Bigl(-300+760\gamma^{2}+3960\gamma^{4}+210313\gamma^{6}+647777\gamma^{8}\\
&\qquad-735123\gamma^{10}-235029\gamma^{12}+21114\gamma^{14}+94208\gamma^{16}\Bigr)\\[4pt]
c_{\rm(dyn)\mathrm{R}}^{(1\mathrm{SF})\,2} &= -\frac{1}{320\,\gamma^{7}(\gamma^{2}-1)^{3}}\Bigl(100-320\gamma^{2}-1160\gamma^{4}+37840\gamma^{6}+494101\gamma^{8}\\
&\qquad-113271\gamma^{10}-624547\gamma^{12}+122647\gamma^{14}+70786\gamma^{16}+16384\gamma^{18}\Bigr)\\[4pt]
c_{\rm(dyn)\mathrm{R}}^{(1\mathrm{SF})\,3} &= \frac{96\,\gamma^{2}(2\gamma^{2}-3)(6\gamma^{2}+5)}{(\gamma^{2}-1)^{3}}\\[4pt]
c_{\rm(dyn)\mathrm{R}}^{(1\mathrm{SF})\,4} &= \frac{3\,(1961-9376\gamma+17784\gamma^{2}-17088\gamma^{3}+5605\gamma^{4}+2400\gamma^{5}-350\gamma^{6})}{32\,(\gamma^{2}-1)}\\[4pt]
c_{\rm(dyn)\mathrm{R}}^{(1\mathrm{SF})\,5} &= \frac{3\,\gamma\,(293+534\gamma^{2}-75\gamma^{4})}{2\,(\gamma^{2}-1)}\\[4pt]
c_{\rm(dyn)\mathrm{R}}^{(1\mathrm{SF})\,6} &= \frac{3\,\gamma\,(667+618\gamma^{2}+75\gamma^{4})}{4\,(\gamma^{2}-1)}
\end{align*}
\vspace{2pt}
\hrule
\vspace{4pt}
\centerline{\textbf{Coefficients at $\overline{\text{1SF}}$}}
\vspace{4pt}
\hrule
\vspace{5pt}
\begin{align*}
c_{\rm(dyn)\mathrm{R}}^{(\overbar{1\mathrm{SF}})\,1} &= \frac{1}{100800\,\gamma^{7}(\gamma^{2}-1)^{3}}\Bigl(55125+63000\gamma-375375\gamma^{2}-159600\gamma^{3}+1884050\gamma^{4}\\
&\qquad-831600\gamma^{5}-9175250\gamma^{6}-44165730\gamma^{7}-352360351\gamma^{8}-136033170\gamma^{9}\\
&\qquad+477970819\gamma^{10}+154375830\gamma^{11}-161339956\gamma^{12}+49356090\gamma^{13}\\
&\qquad+165085834\gamma^{14}-4433940\gamma^{15}-122551296\gamma^{16}-19783680\gamma^{17}\Bigr)\\[4pt]
c_{\rm(dyn)\mathrm{R}}^{(\overbar{1\mathrm{SF}})\,2} &= \frac{1}{6720\,\gamma^{8}(\gamma^{2}-1)^{3}}\Bigl(3675+4200\gamma-27475\gamma^{2}-13440\gamma^{3}+141960\gamma^{4}-48720\gamma^{5}\\
&\qquad-693350\gamma^{6}+1589280\gamma^{7}+9338139\gamma^{8}+20752242\gamma^{9}+30628579\gamma^{10}\\
&\qquad-4757382\gamma^{11}-51962758\gamma^{12}-26230974\gamma^{13}+5150002\gamma^{14}+5151174\gamma^{15}\\
&\qquad+2584364\gamma^{16}+2973012\gamma^{17}+4890624\gamma^{18}+688128\gamma^{19}\Bigr)\\[4pt]
c_{\rm(dyn)\mathrm{R}}^{(\overbar{1\mathrm{SF}})\,3} &= -\frac{16\,\gamma\,(2\gamma^{2}-3)(14+60\gamma+83\gamma^{2}+72\gamma^{3}+6\gamma^{4}+4\gamma^{6})}{(\gamma^{2}-1)^{3}}\\[4pt]
c_{\rm(dyn)\mathrm{R}}^{(\overbar{1\mathrm{SF}})\,4} &= -\frac{6447-27792\gamma+40884\gamma^{2}-28016\gamma^{3}+15227\gamma^{4}-7200\gamma^{5}+1050\gamma^{6}}{16\,(\gamma^{2}-1)}\\[4pt]
c_{\rm(dyn)\mathrm{R}}^{(\overbar{1\mathrm{SF}})\,5} &= \frac{141-7032\gamma-3117\gamma^{2}-12816\gamma^{3}-397\gamma^{4}+1800\gamma^{5}+525\gamma^{6}}{8\,(\gamma^{2}-1)}\\[4pt]
c_{\rm(dyn)\mathrm{R}}^{(\overbar{1\mathrm{SF}})\,6} &= -\frac{3597+16008\gamma+6611\gamma^{2}+14832\gamma^{3}-6541\gamma^{4}+1800\gamma^{5}+525\gamma^{6}}{16\,(\gamma^{2}-1)}
\end{align*}
\vspace{2pt}
\end{minipage}}
\caption{Coefficients $c_{\rm(dyn)\mathrm{R}}^{(1\mathrm{SF})\,i}$ and $c_{\rm(dyn)\mathrm{R}}^{(\overbar{1\mathrm{SF}})\,i}$ of the basis functions $f^i(\gamma)$ in the R region.}
\label{tab:cR}
\end{table}

\subsection{Checks against existing literature}
Although the complete 4PM angular momentum loss has not been available before, several of its ingredients have been computed independently, and we use them to validate our result.
First, the static contribution to the angular momentum loss, which originates from linear memory, and the complete RR contribution, comprising both its static and its dynamical part, have been obtained in refs.~\cite{Heissenberg:2024umh,Heissenberg:2025ocy}.
We find exact agreement with both results.
Furthermore, in our quest to obtain $|L|_{\rm loss}$, we have also independently computed $E_{\rm loss}$ and the $v_i^\mu$-components of the impulse as by-products, and we find agreement with ref.~\cite{Dlapa:2022lmu}.
Finally, our complete result for $|L|_{\rm loss}$ agrees with the independent computation of ref.~\cite{GHMR}.

A further check is provided by the PN expansion, i.e.\ the expansion around $\gamma\to1$ or, equivalently, around small relative velocities $v=\sqrt{\gamma^2-1}/\gamma\to0$.
In PN literature, it is conventional to expand around $v\gamma=\sqrt{\gamma^2-1}$.
The PN expansion of the angular momentum loss then reads 
\begin{align}\label{eq:DeltaLPN}
\frac{|L|^{\rm (4PM)}_{\rm loss}}{|L|} &= \nu\left(\frac{G M^2 \nu}{|L|}\right)^4\biggl(
\frac{176}{5}\,(v\gamma)
+ \left(\frac{8144}{105} - \frac{1072}{9}\,\nu\right) (v\gamma)^3
+ \frac{448}{5}\,(v\gamma)^4 \\
&+ \left(-\frac{93664}{1575} - \frac{247724}{1575}\,\nu + \frac{11056}{45}\,\nu^2\right) (v\gamma)^5
+ \left(\frac{1184}{21} - \frac{431936}{1575}\,\nu\right) (v\gamma)^6 \nonumber\\
&+ \left(-\frac{4955072}{121275} + \frac{1459694}{11025}\,\nu + \frac{67432}{315}\,\nu^2 - \frac{6224}{15}\,\nu^3\right) (v\gamma)^7 \nonumber\\
&+ \left(-\frac{13736}{315} - \frac{126736}{1575}\,\nu + \frac{857152}{1575}\,\nu^2\right) (v\gamma)^8 \nonumber\\
&+ \left(-\frac{29857664}{1576575} + \frac{38220257}{485100}\,\nu - \frac{1406603}{6615}\,\nu^2 - \frac{113356}{525}\,\nu^3 + \frac{28208}{45}\,\nu^4\right) (v\gamma)^9 \nonumber\\
&+\left(\frac{724868}{17325} + \frac{3100472}{121275}\,\nu + \frac{4544}{315}\,\nu^2 - \frac{472256}{525}\,\nu^3\right)(v\gamma)^{10}\nonumber \\
&+ \mathcal{O}((v\gamma)^{11})\biggr)
\ ,
\nonumber
\end{align}
where $|L| = M^2 \nu \sqrt{\gamma^2-1}\,|b|/E$ is the initial angular momentum of the two-body system.
We have checked the first six terms, i.e.\ up to relative 3PN order, against the known PN results of refs.~\cite{Bini:2021gat,Bini:2022enm}, and the 1SF terms up to 7PN order against ref.~\cite{Geralico:2025rof}, finding agreement throughout.

Our calculation directly yields $(E|L|)_{\rm loss}$ rather than $|L|_{\rm loss}$.
That is because following \eqn{eq:lalala} and given $E_{\rm loss}$, the two carry the same information.
The combination $(E|L|)_{\rm loss}$ is, however, the simpler quantity.
It follows directly from the fact that changes in the worldline parameters are polynomial in the masses, whereas $L^\mu$ carries a factor $1/E$ with $E=M\sqrt{1+2\nu(\gamma-1)}$.
The quantity $(E|L|)_{\rm loss}$ therefore consists of only a 1SF and a $\overbar{\rm 1SF}$ part, while $|L|_{\rm loss}$ picks up inverse powers of $E$, both directly and through $E_{\rm loss}$. This can also be seen directly in its PN expansion, which reads
\begin{align}\label{eq:DeltaELPN}
\frac{(E|L|)^{\rm (4PM)}_{\rm loss}}{E|L|} &= \nu\left(\frac{G M}{|b|\, v\gamma}\right)^4\biggl(
\frac{176}{5}\,(v\gamma)
+ \left(\frac{8144}{105} - \frac{208}{15}\,\nu\right) (v\gamma)^3
+ \frac{448}{5}\,(v\gamma)^4\\
&+ \left(-\frac{93664}{1575} + \frac{988}{63}\,\nu\right) (v\gamma)^5
+ \left(\frac{1184}{21} - \frac{13312}{525}\,\nu\right) (v\gamma)^6\nonumber\\
&+ \left(-\frac{4955072}{121275} + \frac{5458}{1575}\,\nu\right) (v\gamma)^7
+ \left(-\frac{13736}{315} - \frac{208}{225}\,\nu\right) (v\gamma)^8\nonumber\\
&+ \left(-\frac{29857664}{1576575} + \frac{14275463}{485100}\,\nu\right) (v\gamma)^9
+ \left(\frac{724868}{17325} + \frac{849608}{121275}\,\nu\right) (v\gamma)^{10}\nonumber\\
&+ \mathcal{O}((v\gamma)^{11})\biggr)\nonumber
\, .
\end{align}
Every coefficient is linear in $\nu$, reflecting the 1SF and $\overbar{\rm 1SF}$ structure, and the $\nu$-independent terms agree with those of \eqn{eq:DeltaLPN}.

\section{Conclusions}
\label{sec:conclusions}
In this paper, we compute the loss of mechanical angular momentum $|L|_{\rm loss}$ of a gravitational two-body system at 4PM order using the formalism set up in ref.~\cite{Jakobsen:2026ayw}.
The approach relates the angular momentum loss to the \emph{worldline offset parameter shifts} $\Delta b_i^\mu$ and impulses $\Delta p_i^\mu$, both derived from the same WQFT one-point function of the worldline deflection.
The impulse, a well-studied observable, is obtained by putting the one-point function on shell, i.e., setting the outgoing energy $\omega_{\rm IR}$ to zero.
The offset parameter shift emerges at the first subleading order in the expansion in small $\omega_{\rm IR}$, and is the new ingredient computed here.

Taking the outgoing energy $\omega_{\rm IR}$ to zero is more subtle for the offset shift than for the impulse.
Namely, one introduces in this limit a static region in which graviton momenta scale with $\omega_{\rm IR}$.
This region is absent from the impulse and is tied to gravitational memory, both linear and nonlinear.
In this paper, we identify diagrammatically which static-region contributions correspond to the former and the latter.
We find, in agreement with ref.~\cite{Heissenberg:2024umh},
that the nonlinear memory contributions vanish at 4PM order due to symmetries of the leading-order waveform.
The diagrams contributing to the linear memory factorise into products of lower-order impulses and static correlators, analogously to the 2PM and 3PM static offset parameter shifts \cite{Jakobsen:2026ayw}.

In the dynamical region, where graviton momenta scale with the momentum transferred from BH2 to BH1, the offset parameter shift involves the same diagrams as the impulse, but these must be generated off shell to capture the subleading in $\omega_{\rm IR}$ behaviour.

To this end, we reformulate the WQFT Feynman rules such that four-momentum is conserved at every vertex, which grants the dotted line a Feynman rule of its own and makes the classical loops of each diagram manifest~\cite{Driesse:2024xad}.
With the loops in plain sight, the mapping to integral families is greatly eased.
The appearing integrals are reduced to master integrals by means of tensor reduction, symmetry relations, and IBP identities, where only integrals of odd parity are needed.
We then derive a differential equation for the masters, bring it to canonical form and solve for the canonical integrals in terms of polylogarithms. Due to our restriction to odd-parity integrals, no elliptic functions appear, in contrast to the even-parity integrals entering the conservative dynamics~\cite{Bern:2021yeh,Dlapa:2021vgp}.
The boundary conditions are fixed in the PN limit, where we must consider the RR region, with two radiative gravitons, and the PR region, with one.
The latter introduces a divergence in the 4PM offset parameter shift, which cancels in the angular momentum loss against the conservative 1PM offset shift combined with the $\mathcal{O}(\eps)$ part of the 3PM impulse, so that the final result is finite.

The final result, expressed in a basis of polylogarithms, is provided in sec.~\ref{section:Results} and in the ancillary files.
It agrees with the known post-Newtonian results to relative 3PN order~\cite{Bini:2022enm} and at 1SF to 7PN order~\cite{Geralico:2025rof}.
Its RR and static-region contributions reproduce known 4PM results~\cite{Heissenberg:2024umh,Heissenberg:2025ocy}, and the complete result agrees with the independent computation of ref.~\cite{GHMR}.

The 4PM angular momentum loss also informs higher-order calculations: through linear response, it fixes the contribution to the 5PM impulse that is linear in radiation reaction~\cite{Bini:2012ji,Damour:2020tta,Bini:2021gat,Jakobsen:2022zsx}.
It is further a missing ingredient in the construction of a fully 5PM-informed EOB Hamiltonian.

Two extensions follow naturally from this work.
The first is the 5PM order, where the dissipative part of the 1SF impulse already involves Calabi--Yau periods~\cite{Driesse:2024feo}.
It remains to be seen which functions survive in the offset parameter shift at that order, and whether divergences in the 5PM dissipative sector still cancel against conservative terms.
A general explanation of this cancellation would clarify whether it is a structural property of the angular momentum loss.
The second extension is spin, for which WQFT is well suited: its bosonic-oscillator formulation describes spinning bodies to all orders in spin~\cite{Haddad:2024ebn}, and it has already yielded the radiated angular momentum from the full scattering trajectory at 2PM to linear order in spin~\cite{Mogull:2025cfn}.
Together with the high-order spin results at $\mathcal{O}(G^2)$~\cite{Alessio:2025flu}, the result at $\mathcal{O}(G^3 S^1)$~\cite{Jakobsen:2023hig, Heissenberg:2023uvo} and a future non-spinning 5PM result, repeating the present computation at  $\mathcal{O}(G^3 S^2)$ and  $\mathcal{O}(G^4 S)$ would give the angular momentum loss of spinning black holes at physical 5PM order.

\newpage

\section{Acknowledgements}
We thank Mathias Driesse, Carlo Heissenberg, Jitze Hoogeveen, Gustav Mogull, Jan Plefka, Benjamin Sauer and Johann Usovitsch for useful discussions.
We are grateful to Alessandro Georgoudis, Carlo Heissenberg, Gustav Mogull and Rodolfo Russo for coordinating the release of our papers.
AI tools were used for coding and text editing.
Every step of the computation has been human verified.
This work was funded by the Deutsche Forschungsgemeinschaft (DFG, German Research Foundation) Projektnummer 417533893/GRK2575 “Rethinking Quantum Field Theory” (LB, CJE, KS).

\appendix
\section{Expansion coefficients of $E_{\rm loss}$}\label{app:results}
\subsubsection*{Dynamical RR coefficients of $E_{\rm loss}$}
\begin{table}[H]
\centering
\small
\setlength{\fboxsep}{0pt}
\fbox{%
\begin{minipage}{\dimexpr\textwidth-2\fboxrule\relax}
\setlength{\abovedisplayskip}{3pt}\setlength{\belowdisplayskip}{3pt}%
\setlength{\abovedisplayshortskip}{3pt}\setlength{\belowdisplayshortskip}{3pt}%
\setlength{\jot}{1.5pt}%
\vspace{4pt}
\centerline{\textbf{Coefficients at 1SF}}
\vspace{4pt}
\hrule
\vspace{5pt}
\begin{align*}
d_{\rm(dyn)\mathrm{RR}}^{(1\mathrm{SF})\,1} &= \frac{16\,(1720+4927\gamma^{2}-1484\gamma^{4}+12\gamma^{6})}{45\,(\gamma^{2}-1)}\\[2pt]
d_{\rm(dyn)\mathrm{RR}}^{(1\mathrm{SF})\,2} &= -\frac{16\,\gamma\,(869+496\gamma^{2}-204\gamma^{4}-16\gamma^{6})}{5\,(\gamma^{2}-1)}\\[2pt]
d_{\rm(dyn)\mathrm{RR}}^{(1\mathrm{SF})\,3} &= \frac{32\,(8+51\gamma^{2}+6\gamma^{4}-8\gamma^{6})}{(\gamma^{2}-1)^{2}}
\end{align*}
\vspace{2pt}
\hrule
\vspace{4pt}
\centerline{\textbf{Coefficients at $\overline{\text{1SF}}$}}
\vspace{4pt}
\hrule
\vspace{5pt}
\begin{align*}
d_{\rm(dyn)\mathrm{RR}}^{(\overbar{1\mathrm{SF}})\,1} &= -\frac{32\,(1720+6165\gamma+4927\gamma^{2}+1950\gamma^{3}-1484\gamma^{4}-960\gamma^{5}+12\gamma^{6})}{45\,(\gamma^{2}-1)}\\[2pt]
d_{\rm(dyn)\mathrm{RR}}^{(\overbar{1\mathrm{SF}})\,2} &= -\frac{32}{15\,(\gamma^{2}-1)}\Bigl(-865-2607\gamma-4340\gamma^{2}-1488\gamma^{3}+100\gamma^{4}+612\gamma^{5}\\
&\qquad+320\gamma^{6}+48\gamma^{7}\Bigr)\\[2pt]
d_{\rm(dyn)\mathrm{RR}}^{(\overbar{1\mathrm{SF}})\,3} &= \frac{64\,(-8-40\gamma-51\gamma^{2}-50\gamma^{3}-6\gamma^{4}+6\gamma^{5}+8\gamma^{6}+4\gamma^{7})}{(\gamma^{2}-1)^{2}}
\end{align*}
\vspace{2pt}
\end{minipage}}
\caption{Coefficients $d_{\rm(dyn)\mathrm{RR}}^{(1\mathrm{SF})\,i}$ and $d_{\rm(dyn)\mathrm{RR}}^{(\overbar{1\mathrm{SF}})\,i}$ of the basis functions $f^i(\gamma)$ in the RR region.}
\label{tab:dRR}
\end{table}
\subsubsection*{Dynamical R coefficients of $E_{\rm loss}$}
\begin{table}[H]
\centering
\footnotesize
\setlength{\fboxsep}{0pt}
\fbox{%
\begin{minipage}{\dimexpr\textwidth-2\fboxrule\relax}
\setlength{\abovedisplayskip}{2pt}\setlength{\belowdisplayskip}{2pt}%
\setlength{\abovedisplayshortskip}{2pt}\setlength{\belowdisplayshortskip}{2pt}%
\setlength{\jot}{1.5pt}%
\vspace{3pt}
\centerline{\textbf{Coefficients at 1SF}}
\vspace{3pt}
\hrule
\vspace{4pt}
\begin{align*}
d_{\rm(dyn)\mathrm{R}}^{(1\mathrm{SF})\,1} &= -\frac{1}{1440\,\gamma^{7}\,(\gamma^{2}-1)^{5/2}}\Bigl(1575-8250\gamma^{2}+35710\gamma^{4}-142640\gamma^{6}\\
&\qquad-5560073\gamma^{8}-417302\gamma^{10}+4034092\gamma^{12}-587336\gamma^{14}+6144\gamma^{16}\Bigr)\\[2pt]
d_{\rm(dyn)\mathrm{R}}^{(1\mathrm{SF})\,2} &= -\frac{1}{480\,\gamma^{8}\,(\gamma^{2}-1)^{5/2}}\Bigl(525-3100\gamma^{2}+13690\gamma^{4}-55260\gamma^{6}+816595\gamma^{8}\\
&\qquad+3752006\gamma^{10}-1978290\gamma^{12}-1029342\gamma^{14}+213480\gamma^{16}+24576\gamma^{18}\Bigr)\\[2pt]
d_{\rm(dyn)\mathrm{R}}^{(1\mathrm{SF})\,3} &= \frac{32\,\gamma\,(2\gamma^{2}-3)(-8-51\gamma^{2}-6\gamma^{4}+8\gamma^{6})}{(\gamma^{2}-1)^{7/2}}\\[2pt]
d_{\rm(dyn)\mathrm{R}}^{(1\mathrm{SF})\,4} &= \frac{1}{8\,(\gamma^{2}-1)^{3/2}}\Bigl(36-1209\gamma+4212\gamma^{2}-6422\gamma^{3}+4332\gamma^{4}+1755\gamma^{5}\\
&\qquad-4996\gamma^{6}+2100\gamma^{7}\Bigr)\\[2pt]
d_{\rm(dyn)\mathrm{R}}^{(1\mathrm{SF})\,5} &= \frac{-9-1053\gamma^{2}-1083\gamma^{4}+1249\gamma^{6}}{4\,(\gamma^{2}-1)^{3/2}}\\[2pt]
d_{\rm(dyn)\mathrm{R}}^{(1\mathrm{SF})\,6} &= \frac{-2039-13155\gamma^{2}-1221\gamma^{4}+1823\gamma^{6}}{8\,(\gamma^{2}-1)^{3/2}}
\end{align*}
\vspace{2pt}
\hrule
\vspace{3pt}
\centerline{\textbf{Coefficients at $\overline{\text{1SF}}$}}
\vspace{3pt}
\hrule
\vspace{4pt}
\begin{align*}
d_{\rm(dyn)\mathrm{R}}^{(\overbar{1\mathrm{SF}})\,1} &= \frac{1}{352800\,\gamma^{8}\,(\gamma^{2}-1)^{5/2}}\Bigl(-385875+771750\gamma+1837500\gamma^{2}-4042500\gamma^{3}\\
&\qquad-7188300\gamma^{4}+17497900\gamma^{5}+21241500\gamma^{6}-69893600\gamma^{7}\\
&\qquad-752357266\gamma^{8}-2724435770\gamma^{9}-4066583215\gamma^{10}\\
&\qquad-204477980\gamma^{11}+3666321886\gamma^{12}+1976705080\gamma^{13}\\
&\qquad+558224042\gamma^{14}-287794640\gamma^{15}-318633472\gamma^{16}+3010560\gamma^{17}\Bigr)\\[2pt]
d_{\rm(dyn)\mathrm{R}}^{(\overbar{1\mathrm{SF}})\,2} &= -\frac{1}{3360\,\gamma^{9}\,(\gamma^{2}-1)^{5/2}}\Bigl(3675-7350\gamma-19950\gamma^{2}+43400\gamma^{3}+79800\gamma^{4}\\
&\qquad-191660\gamma^{5}-246540\gamma^{6}+773640\gamma^{7}+222810\gamma^{8}-11432330\gamma^{9}\\
&\qquad-25264989\gamma^{10}-52528084\gamma^{11}-38260656\gamma^{12}+27696060\gamma^{13}\\
&\qquad+48986718\gamma^{14}+14410788\gamma^{15}+81724\gamma^{16}-2988720\gamma^{17}\\
&\qquad-2732032\gamma^{18}-344064\gamma^{19}\Bigr)\\[2pt]
d_{\rm(dyn)\mathrm{R}}^{(\overbar{1\mathrm{SF}})\,3} &= -\frac{64\,\gamma\,(2\gamma^{2}-3)(-8-20\gamma-51\gamma^{2}-25\gamma^{3}-6\gamma^{4}+3\gamma^{5}+8\gamma^{6}+2\gamma^{7})}{(\gamma^{2}-1)^{7/2}}\\[2pt]
d_{\rm(dyn)\mathrm{R}}^{(\overbar{1\mathrm{SF}})\,4} &= -\frac{\gamma\,(\gamma-1)\,(2\gamma^{2}-3)(-329+1207\gamma-1523\gamma^{2}+525\gamma^{3})}{4\,(\gamma^{2}-1)^{3/2}}\\[2pt]
d_{\rm(dyn)\mathrm{R}}^{(\overbar{1\mathrm{SF}})\,5} &= \frac{18+111\gamma+2106\gamma^{2}-555\gamma^{3}+2166\gamma^{4}+1065\gamma^{5}-2498\gamma^{6}-525\gamma^{7}}{4\,(\gamma^{2}-1)^{3/2}}\\[2pt]
d_{\rm(dyn)\mathrm{R}}^{(\overbar{1\mathrm{SF}})\,6} &= \frac{1}{8\,(\gamma^{2}-1)^{3/2}}\Bigl(4078+1041\gamma+26310\gamma^{2}-2773\gamma^{3}+2442\gamma^{4}-1065\gamma^{5}\\
&\qquad-3646\gamma^{6}+525\gamma^{7}\Bigr)
\end{align*}
\vspace{2pt}
\end{minipage}}
\caption{Coefficients $d_{\rm(dyn)\mathrm{R}}^{(1\mathrm{SF})\,i}$ and $d_{\rm(dyn)\mathrm{R}}^{(\overbar{1\mathrm{SF}})\,i}$ of the basis functions $f^i(\gamma)$ in the R region.}
\label{tab:dR}
\end{table}

\bibliographystyle{JHEP}
\bibliography{DJrefs}

\end{document}